\documentclass[11pt,a4paper]{article}

\usepackage{amssymb}
\usepackage[dvips]{graphicx}
\usepackage{bm}

\begin{document}

\title{The NSVZ $\beta$-function for theories regularized by higher covariant derivatives: the all-loop sum of matter and ghost singularities}

\author{
K.V.Stepanyantz\\
{\small{\em Moscow State University,}}\\
{\small{\em Faculty of Physics, Department of Theoretical Physics,}}\\
{\small{\em 119991, Moscow, Russia}}}

\maketitle

\begin{abstract}
The contributions of the matter superfields and of the Faddeev--Popov ghosts to the $\beta$-function of ${\cal N}=1$ supersymmetric gauge theories defined in terms of the bare couplings are calculated in all orders in the case of using the higher covariant derivative regularization. For this purpose we use the recently proved statement that the $\beta$-function in these theories is given by integrals of double total derivatives with respect to the loop momenta. These integrals do not vanish due to singularities of the integrands. This implies that the $\beta$-function beyond the one-loop approximation is given by the sum of the singular contributions, which is calculated in all orders for singularities produced by the matter superfields and by the Faddeev--Popov ghosts. The result is expressed in terms of the anomalous dimensions of these superfields. It coincides with the corresponding part of the new form of the NSVZ equation, which can be reduced to the original one with the help of the non-renormalization theorem for the triple gauge-ghost vertices.
\end{abstract}

\unitlength=1cm

\section{Introduction}
\hspace*{\parindent}

The NSVZ equation \cite{Novikov:1983uc,Jones:1983ip,Novikov:1985rd,Shifman:1986zi} is an important relation between the $\beta$-function of ${\cal N}=1$ supersymmetric gauge theories and the anomalous dimensions of the matter superfileds,

\begin{equation}\label{NSVZ_Old_Equation}
\frac{\beta(\alpha,\lambda)}{\alpha^2} = - \frac{3 C_2 - T(R) + C(R)_i{}^j
\big(\gamma_\phi\big)_j{}^i(\alpha,\lambda)/r}{2\pi(1- C_2\alpha/2\pi)},
\end{equation}

\noindent
where the matter superfields lie in a representation $R$ of a simple gauge group $G$ with dimension $r$. Its generators $T^A$, $A=1,\ldots, r$, satisfy the identities

\begin{equation}
\mbox{tr}(T^A T^B) = T(R) \delta^{AB};\qquad (T^A T^A)_i{}^j = C(R)_i{}^j;\qquad [T^A, T^B] = i f^{ABC} T^C.
\end{equation}

\noindent
The constant $C_2$ is defined by the equation $C_2 \delta^{AB} = f^{ACD} f^{BCD}$. The generators of the fundamental representation $t^A$ are assumed to be normalized by the condition $\mbox{tr}(t^A t^B) = \delta^{AB}/2$.

The NSVZ equation (\ref{NSVZ_Old_Equation}) reduces the number of independent renormalization constants in ${\cal N}=1$ supersymmetric gauge theories and can be considered as a non-renormalization theorem together with the theorems stating the finiteness of the superpotential \cite{Grisaru:1979wc} and of the triple gauge-ghost vertices \cite{Stepanyantz:2016gtk}. For theories with ${\cal N}=2$ supersymmetry the finiteness beyond the one-loop approximation \cite{Grisaru:1982zh,Howe:1983sr,Buchbinder:1997ib} can be derived from the NSVZ relation provided that the quantization is manifestly ${\cal N}=2$ supersymmetric \cite{Shifman:1999mv,Buchbinder:2014wra}. This can be achieved by using the harmonic superspace \cite{Galperin:1984av,Galperin:2001uw} and the corresponding invariant regularization \cite{Buchbinder:2015eva}.

However, the NSVZ equation takes place only for a certain class of the renormalization prescriptions called ``the NSVZ schemes''.\footnote{In the Abelian case this class has been described in Ref. \cite{Goriachuk:2018cac}.} It is well known \cite{Jack:1996vg,Jack:1996cn,Jack:1998uj,Harlander:2006xq,Mihaila:2013wma} that the $\overline{\mbox{DR}}$-scheme is not NSVZ. This implies that with dimensional reduction \cite{Siegel:1979wq} supplemented by modified minimal subtractions \cite{Bardeen:1978yd} Eq. (\ref{NSVZ_Old_Equation}) is not valid starting from the order $O(\alpha^2,\alpha\lambda^2,\lambda^4)$. (The terms of this order are produced by the three-loop $\beta$-function and the two-loop anomalous dimension). Nevertheless, the NSVZ relation can be restored by a proper modification of the subtraction scheme which should be specially tuned in each order of the perturbation theory.

An all-order prescription giving the NSVZ scheme can be obtained in the case of using the higher covariant derivative regularization \cite{Slavnov:1971aw,Slavnov:1972sq} in the supersymmetric version \cite{Krivoshchekov:1978xg,West:1985jx}. It is given by minimal subtractions of logarithms \cite{Shakhmanov:2017wji,Stepanyantz:2017sqg}, when the renormalization constants include only powers of $\ln\Lambda/\mu$, where $\Lambda$ is the dimensionful cut-off parameter and $\mu$ is the renormalization point. The NSVZ scheme constructed in this way is usually called HD+MSL,

\begin{equation}\label{HD+MSL=NSVZ}
\mbox{NSVZ}=\mbox{HD}+\mbox{MSL}.
\end{equation}

The main observation leading to this statement is that the renormalization group functions (RGFs) defined in terms of the bare couplings presumably satisfy the NSVZ equation in all orders in the case of using the higher covariant derivative regularization. Note that these RGFs are independent of a renormalization prescription for a fixed regularization \cite{Kataev:2013eta}, so that this statement is valid in an arbitrary subtraction scheme. For ${\cal N}=1$ supersymmetric gauge theories regularized by dimensional reduction the NSVZ equation for RGFs defined in terms of the bare couplings is not valid starting from the order $O(\alpha_0^2,\alpha_0\lambda_0^2,\lambda_0^4)$, where the dependence on a regularization becomes essential \cite{Aleshin:2016rrr}.

The NSVZ relation for RGFs defined in terms of the bare couplings follows from the factorization of the integrals giving the $\beta$-function into integrals of double total derivatives.\footnote{This is true only in the case of using the higher covariant derivative regularization. For theories regularized by dimensional reduction it is not so \cite{Aleshin:2015qqc}.} The factorization into total and double derivatives was first noted in Refs. \cite{Soloshenko:2003nc} and \cite{Smilga:2004zr}, respectively, in calculating the lowest quantum corrections for ${\cal N}=1$ SQED. Subsequently, the factorization into double total derivatives has been proved in all loops in the Abelian case \cite{Stepanyantz:2011jy,Stepanyantz:2014ima}. This feature of the loop integrals giving the $\beta$-function allows calculating one of them analytically. Note that the result does not vanish, because for a nonsingular function $f(Q^2)$ (where $Q_\mu$ is an Euclidean momentum) rapidly decreasing at infinity

\begin{equation}\label{Total_Derivative_Integral}
I\equiv \int \frac{d^4Q}{(2\pi)^4} \frac{\partial^2}{\partial Q_\mu^2} \Big(\frac{1}{Q^2} f\left(Q^2\right)\hspace*{-1mm}\Big) = \frac{1}{4\pi^2} f(0)
\end{equation}

\noindent
due to the contribution of the singularity at $Q_\mu=0$. In the Abelian case this leads to the relation between the $\beta$-function in a certain order and the anomalous dimension of the matter superfields in the previous order which is equivalent to the NSVZ equation. (Both RGFs in this case are defined in terms of the bare coupling constant and the higher derivatives are used for the regularization.) In all orders the Abelian NSVZ relation \cite{Vainshtein:1986ja,Shifman:1985fi} was derived by this method in Refs. \cite{Stepanyantz:2011jy,Stepanyantz:2014ima}. The similar method can be applied for deriving the exact NSVZ-like equations for the Alder $D$-function in ${\cal N}=1$ SQCD \cite{Shifman:2014cya,Shifman:2015doa} and for the renormalization of the photino mass\footnote{The NSVZ-like equation describing the renormalization group behaviour of the gaugino mass in gauge theories with softly broken supersymmetry was first proposed in Refs. \cite{Hisano:1997ua,Jack:1997pa,Avdeev:1997vx}.} in softly broken ${\cal N}=1$ SQED \cite{Nartsev:2016nym}. In both cases these equations originate from the factorization of the loop integrals into integrals of double total derivatives and for RGFs defined in terms of the bare couplings are valid only with the higher covariant derivative regularization (and are not valid with dimensional reduction \cite{Jack:1997pa,Aleshin:2019yqj}). For RGFs defined in terms of the renormalized couplings Eq. (\ref{HD+MSL=NSVZ}) is also valid \cite{Kataev:2017qvk,Nartsev:2016mvn}, because both definitions of RGFs in the HD+MSL scheme produce the same results up to the renaming of arguments \cite{Kataev:2013eta}.

In the non-Abelian case the factorization of loop integrals giving the $\beta$-function defined in terms of the bare couplings into integrals of (double) total derivatives was confirmed by a large number of explicit calculations made with different versions of the higher derivative regularization, see, e.g., \cite{Pimenov:2009hv,Stepanyantz:2011bz,Stepanyantz:2012zz,Shakhmanov:2017soc,Kazantsev:2018nbl}.
Recently it has been proved in all loops in Ref. \cite{Stepanyantz:2019ihw}. As a fact confirming the correctness of this result one can consider a method for calculating the $\beta$-function proposed on the base of the reasonings used for constructing this proof. This method significantly simplifies the calculations and produces the $\beta$-function (defined in terms of the bare couplings for the theory regularized by higher covariant derivatives) in the form of an integral of double total derivatives. Its application allowed to check independently the (very complicated) calculations of Refs. \cite{Shakhmanov:2017soc,Kazantsev:2018nbl} and also to make some new calculations \cite{Kuzmichev:2019ywn,Stepanyantz:2019lyo}. Exactly as in the Abelian case, the factorization into double total derivatives reduces the calculation of the $\beta$-function to the summation of singular contributions which are produced by integrals similar to (\ref{Total_Derivative_Integral}). At present this sum has not yet been found in all loops. In this paper we find its part, namely, the sum of singularities produced by the matter and ghost superfields.

The paper is organized as follows: In Sect. \ref{Section_Gauge_Theory} we briefly describe the theory under consideration, its regularization by higher covariant derivatives, and quantization. Also we introduce some auxiliary parameters which will be used below for finding the sums of singular contributions. In the next Sect. \ref{Section_Steps} we describe the main steps needed for the all-loop derivation of the non-Abelian NSVZ relation. Sect. \ref{Section_Singularities} is devoted to all-order summing of singular contributions produced by the matter and ghost superfields, which originate from integrals of double total derivatives giving the $\beta$-function. The sum is expressed in terms of the anomalous dimensions of the corresponding superfields and coincides with the relevant terms in the exact NSVZ equation for RGFs defined in terms of the bare couplings.

\section{The theory under consideration}
\hspace*{\parindent}\label{Section_Gauge_Theory}

We consider the massless ${\cal N}=1$ supersymmetric Yang--Mills theory interacting with chiral matter superfields $\phi_i$ formulated in ${\cal N}=1$ superspace, see, e.g., \cite{Gates:1983nr,West:1990tg,Buchbinder:1998qv}. For quantizing this theory we use the background field method \cite{DeWitt:1965jb,Abbott:1980hw,Abbott:1981ke} in the supersymmetric formulation \cite{Grisaru:1982zh,Gates:1983nr} and take into account the necessity of the nonlinear renormalization for the quantum gauge superfield \cite{Piguet:1981fb,Piguet:1981hh,Tyutin:1983rg}. This implies that in the standard classical action one should make the replacement

\begin{equation}\label{Substitution}
e^{2V} \to e^{2\cal {F}(V)} e^{2\bm{V}},
\end{equation}

\noindent
where $\bm{V}$ is the background gauge superfield, and the nonlinear function ${\cal F}(V) = {\cal F}(V)^A t^A$ (or ${\cal F}(V) = {\cal F}(V)^A T^A$) includes an infinite set of parameters. The lowest nonlinear term in this function has been calculated in Refs. \cite{Juer:1982fb,Juer:1982mp},

\begin{equation}
{\cal F}(V)^A = V^A + y_0\, G^{ABCD}\, V^B V^C V^D + \ldots,
\end{equation}

\noindent
where $G^{ABCD} \equiv \big(f^{AKL} f^{BLM} f^{CMN} f^{DNK} + \mbox{permutations of $B$, $C$, and $D$}\big)/6$. The parameters contained in the function ${\cal F}(V)$ are very similar to the gauge fixing parameter $\xi_0$, because instead of introducing the function ${\cal F}(V)$ one can change the gauge fixing condition \cite{Piguet:1981fb}. That is why we will include all of them into a single set $Y_0 \equiv (\xi_0, y_0,\ldots)$. Note that in our notation the bare parameters are denoted by the subscript 0, and the quantum gauge superfield satisfies the constraint $V^+ = e^{-2\bm{V}} V e^{2\bm{V}}$.

After the substitution (\ref{Substitution}) the massless action written in terms of ${\cal N}=1$ superfields takes the form

\begin{eqnarray}\label{Action_Under_Consideration}
&& S = \frac{1}{2e_0^2} \mbox{Re}\,\mbox{tr} \int d^4x\, d^2\theta\, W^a W_a + \frac{1}{4} \int d^4x\, d^4\theta\, \phi^{*i} (e^{2{\cal F}(V)} e^{2\bm{V}})_i{}^j \phi_j\qquad\nonumber\\
&&\qquad\qquad\qquad\qquad\qquad\qquad\qquad\qquad\qquad + \Big(\frac{1}{6} \lambda_0^{ijk} \int d^4x\, d^2\theta\, \phi_i \phi_j \phi_k + \mbox{c.c.}\Big),\qquad
\end{eqnarray}

\noindent
where $e_0$ and $\lambda_0^{ijk}$ are the gauge and Yukawa coupling constants, respectively. The gauge invariant theory is obtained if

\begin{equation}\label{Yukawa_Gauge_Invariance}
\lambda_0^{ijm} (T^A)_m{}^k + \lambda_0^{imk} (T^A)_m{}^j + \lambda_0^{mjk} (T^A)_m{}^i = 0.
\end{equation}

\noindent
We will always assume that this condition is satisfied. The strength of the gauge superfield is given by the expression

\begin{equation}\label{W_Definition}
W_a \equiv \frac{1}{8} \bar D^2 \left(e^{-2\bm{V}} e^{-2{\cal F}(V)}\, D_a \left(e^{2{\cal F}(V)}e^{2\bm{V}}\right)\right),
\end{equation}

\noindent
where $D_a$ and $\bar D_{\dot a}$ are the right and left supersymmetric covariant derivatives, respectively. This expression can be equivalently rewritten in the form

\begin{equation}\label{W_Another_Form}
W_a = \frac{1}{8} e^{-2\bm{V}}\bm{\bar\nabla}^2\left( e^{-2{\cal F}(V)} \bm{\nabla}_a e^{2{\cal F}(V)} \right) e^{2\bm{V}} + \bm{W}_a,
\end{equation}

\noindent
where we have defined the background gauge field strength and the background supersymmetric covariant derivatives as

\begin{equation}\label{Background_Supersymmetric_Derivatives}
\bm{W}_a \equiv \frac{1}{8} \bar D^2\left(e^{-2\bm{V}} D_a e^{2\bm{V}}\right);\qquad\quad
\bm{\nabla}_a \equiv D_a;\qquad\quad \bm{\bar\nabla}_{\dot a} \equiv e^{2\bm{V}} \bar D_{\dot a} e^{-2\bm{V}}.
\end{equation}

The most important ingredient needed for the derivation of the NSVZ $\beta$-function is the higher covariant derivative regularization. Its main idea \cite{Slavnov:1971aw,Slavnov:1972sq} (see also \cite{Faddeev:1980be}) is to add a higher derivative term $S_\Lambda$ to the classical action in such a way that divergences disappear in all orders except for the one-loop approximation. For regularizing the surviving one-loop divergences one should use the Pauli-Villars superfields and insert the corresponding determinants into the generating functional \cite{Slavnov:1977zf}. In this paper, following Refs. \cite{Aleshin:2016yvj,Kazantsev:2017fdc}, we use such a higher derivative term that

\begin{eqnarray}\label{Regularized_Action_Without_G}
&&\hspace*{-6mm} S + S_\Lambda = \frac{1}{2 e_0^2}\mbox{Re}\, \mbox{tr} \int d^4x\, d^2\theta\, W^a \left(e^{-2\bm{V}} e^{-2{\cal F}(V)} \right)_{Adj} R\Big(-\frac{\bar\nabla^2 \nabla^2}{16\Lambda^2}\Big)_{Adj} \left(e^{2{\cal F}(V)}e^{2\bm{V}}\right)_{Adj} W_a \nonumber\\
&&\hspace*{-6mm} + \frac{1}{4} \int d^4x\,d^4\theta\, \phi^{*i} \Big(F\Big(-\frac{\bar\nabla^2 \nabla^2}{16\Lambda^2}\Big) e^{2{\cal F}(V)}e^{2\bm{V}}\Big)_i{}^j \phi_j
+ \Big(\frac{1}{6} \lambda_0^{ijk} \int d^4x\, d^2\theta\, \phi_i \phi_j \phi_k + \mbox{c.c.} \Big),
\end{eqnarray}

\noindent
where the higher derivatives are contained inside the regulator functions $R(x)$ and $F(x)$, which satisfy the condition $R(0)=F(0)=1$ and rapidly increase at infinity. The covariant derivatives are defined as

\begin{equation}\label{Covariant_Derivative_Definition}
\nabla_a = D_a;\qquad \bar\nabla_{\dot a} = e^{2{\cal F}(V)} e^{2\bm{V}} \bar D_{\dot a} e^{-2\bm{V}} e^{-2{\cal F}(V)},
\end{equation}

\noindent
and the subscript $Adj$ means that

\begin{equation}
R(X)_{Adj} Y = \big(1 + r_1 X + r_2 X^2 + \ldots\big)_{Adj} Y \equiv Y + r_1 [X, Y] + r_2 [X,[X,Y]] + \ldots
\end{equation}

Although the expression (\ref{Regularized_Action_Without_G}) allows regularizing all divergences beyond the one-loop approximation, for the all-loop derivation of the NSVZ $\beta$-function we need to introduce some auxiliary parameters, namely, the complex coordinate independent parameter $g$ and the chiral superfield $\mbox{\sl g}(x^\mu,\theta)$.\footnote{Sometimes similar parameters were used earlier, see, e.g., \cite{Ohshima:1999jg,Kraus:2001tg,Kraus:2001kn,Kraus:2002nu,Babington:2005vu}.} The former one relates the original theory corresponding to $g=1$ to the theory in which quantum superfields interact only with the background gauge superfield corresponding to $g\to 0$. This parameter is introduced by the formal substitution $\alpha_0 \to g g^*\alpha_0$, $\lambda_0^{ijk} \to g\lambda_0^{ijk}$, $\lambda_{0ijk}^* \to g^* \lambda^*_{0ijk}$. All terms containing quantum superfields depend on the auxiliary superfield $\mbox{\sl g}$ only through the combination

\begin{equation}\label{Bold_G}
\bm{g} \equiv g + \mbox{\sl g}.
\end{equation}

\noindent
Namely, following Ref. \cite{Stepanyantz:2019lyo}, the regularized action with the auxiliary parameters is defined as

\begin{eqnarray}\label{Regularized_Action_With_G}
&&\hspace*{-7mm} S_{\mbox{\scriptsize reg}}\ \to\ \frac{1}{2 g g^* e_0^2}\,\mbox{Re}\,\mbox{tr} \int d^6x\, \bm{W}^a \bm{W}_a + \frac{1}{e_0^2}\,\mbox{Re}\,\mbox{tr} \int d^8x\, \frac{1}{\bm{g} \bm{g}^*} \Big[ -\frac{1}{4} e^{2\bm{V}} \bm{W}^a e^{-2\bm{V}} e^{-2{\cal F}(V)}
\nonumber\\
&&\hspace*{-7mm} \times \bm{\nabla}_a e^{2{\cal F}(V)}  - \frac{1}{64} e^{-2{\cal F}(V)} \bm{\nabla}^a e^{2{\cal F}(V)}\cdot \bm{\bar\nabla}^2\left(e^{-2{\cal F}(V)} \bm{\nabla}_a e^{2{\cal F}(V)} \right)
+  W^a \left(e^{-2\bm{V}} e^{-2{\cal F}(V)} \right)_{Adj} \vphantom{\frac{\Lambda^2}{\Lambda^2}}\nonumber\\
&&\hspace*{-7mm} \times\,  \frac{\nabla^2}{16\Lambda^2} r\Big(-\frac{\bar\nabla^2 \nabla^2}{16\Lambda^2}\Big)_{Adj} \left(e^{2{\cal F}(V)}e^{2\bm{V}}\right)_{Adj} W_a\Big] + \frac{1}{4} \int d^8x\, \phi^{*i} \Big(F\Big(-\frac{\bar\nabla^2 \nabla^2}{16\Lambda^2}\Big) e^{2{\cal F}(V)}e^{2\bm{V}}\Big)_i{}^j \phi_j\nonumber\\
&&\hspace*{-7mm} + \Big(\frac{1}{6} \lambda_0^{ijk} \int d^6x\, \bm{g}\,\phi_i \phi_j \phi_k + \mbox{c.c.} \Big),
\end{eqnarray}

\noindent
where $r(x) \equiv (R(x)-1)/x$. In the case $g=1$, $\mbox{\sl g}=0$ this expression is reduced to Eq. (\ref{Regularized_Action_Without_G}) with the help of Eq. (\ref{W_Another_Form}) and the relations between the integration measures

\begin{equation}\label{Integration_Measures_Relation}
\int d^4x\, d^4\theta = -\frac{1}{2}\int d^4x\, d^2\theta\, \bar D^2 = -\frac{1}{2}\int d^4x\, d^2\bar\theta\, D^2.
\end{equation}

\noindent
Note that in Eq. (\ref{Regularized_Action_With_G}) we used brief notations for these measures,

\begin{equation}
d^8x \equiv d^4x\, d^4\theta_x;\qquad d^6x \equiv d^4x\, d^2\theta_x;\qquad  d^6\bar x \equiv d^4x\, d^2\bar\theta_x,
\end{equation}

\noindent
which will also be used below.

To quantize the theory under consideration, one should fix a gauge and introduce the Faddeev--Popov and Nielsen--Kallosh ghost superfields. These chiral anticommuting superfields lie in the adjoint representation and are denoted by $\bar c$ (the Faddeev--Popov antighost), $c$ (the Faddeev--Popov ghost), and $b$ (the Nielsen--Kallosh ghost). It is convenient to use the gauge fixing action

\begin{equation}\label{Term_Fixing_Gauge}
S_{\mbox{\scriptsize gf}} = -\frac{1}{16\xi_0 e_0^2}\, \mbox{tr} \int d^8x\,  \bm{\nabla}^2 V \frac{1}{\bm{g}^*}  K\Big(-\frac{\bm{\bar\nabla}^2 \bm{\nabla}^2}{16\Lambda^2}\Big)_{Adj} \frac{1}{\bm{g}} \bm{\bar\nabla}^2 V
\end{equation}

\noindent
invariant under the background gauge transformations. This action also contains higher covariant derivatives inside the regulator function $K(x)$ which satisfies the standard conditions $K(0)=1$, $K(x\to\infty) \to \infty$. The corresponding ghost actions can be written as

\begin{eqnarray}
\label{Ghosts_Faddeev-Popov}
&& S_{\mbox{\scriptsize FP}} = \frac{1}{2} \int
d^8x\, \frac{\partial {\cal F}^{-1}(\widetilde V)^A}{\partial {\widetilde V}^B}\left.\vphantom{\frac{1}{2}}\right|_{\widetilde V = {\cal F}(V)} \left(e^{2\bm{V}}\bar c e^{-2\bm{V}} +
\bar c^+ \right)^A\nonumber\\
&&\qquad\qquad\qquad\qquad \times \left\{\vphantom{\frac{1}{2}}\smash{\Big(\frac{{\cal F}(V)}{1-e^{2{\cal F}(V)}}\Big)_{Adj} c^+
+ \Big(\frac{{\cal F}(V)}{1-e^{-2{\cal F}(V)}}\Big)_{Adj}
\Big(e^{2\bm{V}} c e^{-2\bm{V}}\Big)}\right\}^B;\qquad\\
\label{Ghosts_Nielsen-Kallosh}
&&  S_{\mbox{\scriptsize NK}} = \frac{1}{2}\,\mbox{tr} \int d^8x\,  b^+ \Big(K\Big(-\frac{\bm{\bar\nabla}^2 \bm{\nabla}^2}{16\Lambda^2}\Big) e^{2\bm{V}}\Big)_{Adj} b,
\end{eqnarray}

\noindent
see Ref. \cite{Stepanyantz:2019lyo} for more details. (The dependence of the Nielsen--Kallosh action on the auxiliary parameters was eliminated by a proper redefinition of the superfields $b$ and $b^+$.)
Also we will need some auxiliary sources $\phi_0$, $c_0$, and $\bar c_0$ with the actions $S_{\phi_0}$ and $S_{c_0}$ given below by Eqs. (\ref{S_Phi0}) and (\ref{S_C0}), respectively.

The generating functional of the theory

\begin{eqnarray}\label{Z_Generating_Functional}
&& Z = \int D\mu\, \mbox{Det}(PV,M_\varphi)^{-1}\mbox{Det}(PV,M)^c  \nonumber\\
&&\qquad\qquad\qquad\qquad \times \exp\Big\{i\Big(S_{\mbox{\scriptsize reg}} + S_{\mbox{\scriptsize gf}} + S_{\mbox{\scriptsize FP}} + S_{\mbox{\scriptsize NK}} +S_{\phi_0} + S_{c_0} + S_{\mbox{\scriptsize sources}}\Big)\Big\},\qquad
\end{eqnarray}

\noindent
contains the sources

\begin{equation}
S_{\mbox{\scriptsize sources}} = \int d^8x\, J^A V^A + \Big(\int d^6x\, \Big(j^i \phi_i + j_c^A c^A + \bar j_c^A \bar c^A\Big) + \mbox{c.c.}\Big)
\end{equation}

\noindent
and the Pauli--Villars determinants

\begin{equation}
\mbox{Det}(PV,M_\varphi)^{-1} \equiv \int D\varphi_1\, D\varphi_2\, D\varphi_3\, \exp(iS_\varphi); \qquad \mbox{Det}(PV,M)^{-1} \equiv \int D\Phi\, \exp(iS_\Phi)
\end{equation}

\noindent
needed for regularizing the remaining one-loop divergences \cite{Faddeev:1980be,Slavnov:1977zf}. Here we use the version of the regularization adopted in Ref. \cite{Kazantsev:2017fdc} (a similar variant was proposed earlier in \cite{Aleshin:2016yvj}), when three commuting chiral superfields $\varphi_1$, $\varphi_2$, and $\varphi_3$ in the adoint representation cancel the divergences coming from gauge and ghost loops and the chiral commuting superfield $\Phi_i$ in a certain representation $R_{\mbox{\scriptsize PV}}$ cancels the divergence introduced by a matter loop. Note that the representation $R_{\mbox{\scriptsize PV}}$ should admit the invariant mass term such that
$M^{jk} M^*_{ki} = M^2 \delta_i^j$. (For simple groups considered in this paper it is possible to choose the adjoint representation.)

The one-loop divergences in the supersymmetric case can completely be cancelled if the actions for these Pauli--Villars superfields are written as

\begin{eqnarray}\label{S_varphi}
&&\hspace*{-11mm} S_\varphi = \frac{1}{4} \int d^8x\,\Bigg\{ \varphi_1^{*A} \Big[ \Big(R\Big(-\frac{\bar\nabla^2 \nabla^2}{16\Lambda^2}\Big) e^{2{\cal F}(V)} e^{2\bm{V}}\Big)_{Adj} \varphi_1\Big]_A
+ \varphi_2^{*A} \Big[\Big( e^{2{\cal F}(V)} e^{2\bm{V}}\Big)_{Adj} \varphi_2\Big]_A \nonumber\\
&&\hspace*{-11mm} + \varphi_3^{*A} \Big[\Big(e^{2{\cal F}(V)} e^{2\bm{V}}\Big)_{Adj} \varphi_3\Big]_A
\Bigg\} + \Big(\frac{1}{4} M_\varphi \int d^6x\, \Big((\varphi_1^A)^2 + (\varphi_2^A)^2 + (\varphi_3^A)^2\Big)+\mbox{c.c.}\Big);\\
\label{S_Phi}
&&\hspace*{-11mm} S_\Phi = \frac{1}{4} \int d^8x\, \Phi^{*i} \Big(F\Big(-\frac{\bar\nabla^2 \nabla^2}{16\Lambda^2}\Big) e^{2{\cal F}(V)}e^{2\bm{V}}\Big)_i{}^j \Phi_j
+ \Big(\frac{1}{4} M^{ij} \int d^6x\, \Phi_i \Phi_j + \mbox{c.c.}\Big),\vphantom{\Bigg(}
\end{eqnarray}

\noindent
and the constant $c$ in Eq. (\ref{Z_Generating_Functional}) is equal to $T(R)/T(R_{\mbox{\scriptsize PV}})$. Moreover, we will require that the ratios of the Pauli--Villars superfield masses to the regularization parameter $\Lambda$

\begin{equation}\label{Ratios}
 a_\varphi\equiv M_\varphi/\Lambda;\qquad a\equiv M/\Lambda
\end{equation}

\noindent
do not depend on couplings.

Finally we note that the total action $S_{\mbox{\scriptsize total}}\equiv S_{\mbox{\scriptsize reg}} + S_{\mbox{\scriptsize gf}} + S_{\mbox{\scriptsize FP}} + S_{\mbox{\scriptsize NK}}$ and the Pauli--Villars determinants are invariant under the background gauge transformations

\begin{eqnarray}\label{Background_Gauge_Transformations_Original}
&& e^{2\bm{V}} \to e^{-A^+} e^{2\bm{V}} e^{-A};\qquad V \to  e^{-A^+} V e^{A^+};\qquad c \to e^A c e^{-A};\qquad \bar c \to e^A \bar c e^{-A};\qquad \vphantom{\Big(}\nonumber\\
&& b \to e^A b e^{-A};\qquad\quad \phi_i \to (e^A)_i{}^j \phi_j;\qquad\ \ \Phi_i \to (e^A)_i{}^j \Phi_j;\qquad\ \ \varphi_a \to e^A \varphi_a e^{-A},\qquad\vphantom{\Big(}
\end{eqnarray}

\noindent
parameterized by a chiral superfield $A$ which takes values in the Lie algebra of the gauge group.

\section{The all-loop derivation of the NSVZ $\beta$-function: overview of the main steps.}
\hspace*{\parindent}\label{Section_Steps}

In this section we outline the main steps needed for deriving the NSVZ $\beta$-function in all orders of the perturbation theory. This is done for RGFs defined in terms of the bare couplings in the case of using the higher covariant derivative regularization.

\medskip

1. First, it is necessary to rewrite the NSVZ equation (\ref{NSVZ_Old_Equation}) (for RGFs defined in terms of the bare couplings) in an equivalent form

\begin{eqnarray}\label{NSVZ_New_Equation}
&& \frac{\beta(\alpha_0,\lambda_0, Y_0)}{\alpha_0^2} = - \frac{1}{2\pi}\Big(3 C_2 - T(R) - 2C_2 \gamma_c(\alpha_0,\lambda_0,Y_0)\nonumber\\
&&\qquad\qquad\qquad\qquad\qquad\quad - 2C_2 \gamma_V(\alpha_0,\lambda_0,Y_0) + C(R)_i{}^j \big(\gamma_\phi\big)_j{}^i(\alpha_0,\lambda_0,Y_0)/r\Big) \qquad
\end{eqnarray}

\noindent
using the non-renormalization theorem for the triple gauge-ghost vertices \cite{Stepanyantz:2016gtk}.\footnote{In Ref. \cite{Stepanyantz:2016gtk} this theorem has been derived for an arbitrary value of the gauge fixing parameter $\xi_0$. For the non-supersymmetric Yang--Mills theory \cite{Dudal:2002pq} and for the ${\cal N}=1$ supersymmetric Yang--Mills theory in the component formulation \cite{Capri:2014jqa} similar statements were proved earlier in the Landau gauge $\xi \to 0$.} According to this theorem, the Green functions corresponding to the superdiagrams with two external Faddeev--Popov ghost lines and one external line of the quantum gauge superfield are finite in all orders. As a consequence, the renormalization constants defined by the equations

\begin{equation}
Z_\alpha \equiv \frac{\alpha}{\alpha_0};\qquad V \equiv Z_V Z_\alpha^{-1/2} V_R;\qquad \bar c c = Z_c Z_\alpha^{-1} \bar c_R c_R;\qquad \phi_i = \big(\sqrt{Z_\phi}\big)_i{}^j \big(\phi_R\big)_j,
\end{equation}

\noindent
where the subscript $R$ stands for the renormalized superfields, satisfy the relation $d(Z_\alpha^{-1/2} Z_c Z_V)/d\ln\Lambda = 0$, so that it is possible to choose such a renormalization prescription that

\begin{equation}
Z_\alpha^{-1/2} Z_c Z_V =1.
\end{equation}

\noindent
Below this will be always assumed.

Unlike Eq. (\ref{NSVZ_Old_Equation}), the new form of the NSVZ equation (\ref{NSVZ_New_Equation}) relates the $\beta$-function in a certain order to the anomalous dimensions of the quantum superfields only in the previous order due to the absence of the coupling dependent denominator in the right hand side similarly to the Abelian case \cite{Vainshtein:1986ja,Shifman:1985fi}. Moreover, it has a simple graphical interpretation, which is again analogous to the Abelian case considered in \cite{Smilga:2004zr,Kazantsev:2014yna}. Namely, starting from a supergraph without external lines, one obtains the corresponding part of the $\beta$-function by calculating superdiagrams constructed by attaching two external lines of the background gauge superfield $\bm{V}$ in all possible ways. Eq. (\ref{NSVZ_New_Equation}) relates this part of the $\beta$-function to the contributions to the anomalous dimensions of the quantum superfields which come from superdiagrams produced by all possible cuts of the original supergraph. This qualitative picture has been confirmed by the comparison of the two-loop $\beta$-function with the one-loop anomalous dimensions in Refs. \cite{Shakhmanov:2017wji,Stepanyantz:2019lyo}. In the next order the verification has been made only for certain groups of superdiagrams. Namely, Eq. (\ref{NSVZ_New_Equation}) was checked for the terms of the orders $\alpha\lambda^2$ and $\lambda^4$ \cite{Kazantsev:2018nbl} and for the superdiagrams containing ghost loops \cite{Kuzmichev:2019ywn} (see also Ref. \cite{Kazantsev:2018kjx}).

Thus, it seems that it is Eq. (\ref{NSVZ_New_Equation}) that is obtained in the perturbation theory by summing supergraphs, certainly, in the case of using the higher covariant derivative regularization for RGFs defined in terms of the bare couplings.

\medskip

2. The NSVZ equation (\ref{NSVZ_New_Equation}) can be derived from a certain relation between the Green functions of the theory which is valid with the higher covariant derivative regularization in the limit of the vanishing external momenta \cite{Stepanyantz:2016gtk,Stepanyantz:2019ihw}. Really, the $\beta$-function, defined in terms of the bare couplings can be written as

\begin{equation}\label{Bare_Beta_From_D}
\frac{\beta(\rho\alpha_0,\, \rho\lambda_0 \lambda_0^*, \, Y_0)}{\rho^2 \alpha_0^2} = \left.\frac{d}{d\ln\Lambda} \Big(d^{-1} - \big(g g^*\big)^{-1}\alpha_0^{-1}\Big)\right|_{\alpha,\lambda,Y=\mbox{\scriptsize const};\, p\to 0},
\end{equation}

\noindent
where we have introduced the notation

\begin{equation}
\rho\equiv g g^*.
\end{equation}

\noindent
The function $d^{-1}$ appears in the expression for the part of the effective action corresponding to the two-point Green function of the background gauge superfield $\bm{V} = \bm{V}^A t^A$,

\begin{equation}\label{Two_Point_Function_Background_V}
\Gamma^{(2)}_{\bm{V}} = - \frac{1}{16\pi} \int \frac{d^4p}{(2\pi)^4}\, d^4\theta\, \bm{V}^A(-p,\theta) \partial^2 \Pi_{1/2} \bm{V}^A(p,\theta)\, d^{-1}(\rho\alpha_0,\, \rho\lambda_0 \lambda_0^*,\, Y_0,\,\Lambda/p).
\end{equation}

\noindent
(Certainly here we assume that $\mbox{\sl g} = 0$.) Note that the function $d^{-1}$ depends on couplings only in the combinations $\rho\alpha_0$ and $\rho\lambda_0^{ijk} \lambda^*_{0mnp}$. It is important that the derivative with respect to $\ln\Lambda$ in Eq. (\ref{Bare_Beta_From_D}) should be taken at fixed values of the renormalized couplings (including the parameters of the non-linear renormalization $Y$, see Ref. \cite{Kazantsev:2018kjx}). The limit $p\to 0$ corresponding to the vanishing external momentum is needed in order to get rid of the terms suppressed by powers of $p/\Lambda$.

Similarly, the anomalous dimensions of the quantum superfields can be related to the corresponding Green functions

\begin{eqnarray}
&& \gamma_V(\rho\alpha_0,\, \rho\lambda_0\lambda_0^*,\,Y_0) = \left.\frac{1}{2}\,\frac{d\ln G_V}{d\ln\Lambda}\right|_{\alpha,\lambda,Y = \mbox{\scriptsize const};\ q\to 0};\\
&& \gamma_c(\rho\alpha_0,\,\rho\lambda_0\lambda_0^*,\,Y_0) = \left. \frac{d\ln G_c}{d\ln\Lambda}\right|_{\alpha,\lambda,Y = \mbox{\scriptsize const};\ q\to 0};\\
&& \big(\gamma_\phi\big)_i{}^j(\rho\alpha_0,\,\rho\lambda_0\lambda_0^*,\,Y_0) = \left.\frac{d\big(\ln G_\phi\big)_i{}^j}{d\ln\Lambda}\right|_{\alpha,\lambda,Y = \mbox{\scriptsize const};\ q\to 0}
\end{eqnarray}

\noindent
defined by the equations

\begin{eqnarray}\label{Gamma_V_Quantum_Part}
&&\hspace*{-13mm} \Gamma^{(2)}_V - S_{\mbox{\scriptsize gf}}^{(2)} = -\frac{1}{4 e_0^2 \rho} \int \frac{d^4q}{(2\pi)^4}\, d^4\theta\, V^A(-q,\theta) \partial^2\Pi_{1/2} V^A(q,\theta)\, G_V(\rho\alpha_0,\, \rho\lambda_0 \lambda_0^*,\, Y_0,\, \Lambda/q);\\
\label{Gamma_C_Part}
&&\hspace*{-13mm} \Gamma^{(2)}_c = \frac{1}{4}\int \frac{d^4q}{(2\pi)^4}\, d^4\theta\, \Big(- \bar c^A(-q,\theta) c^{*A}(q,\theta) +  \bar c^{*A}(-q,\theta) c^{A}(q,\theta) \Big) G_c(\rho\alpha_0,\, \rho \lambda_0 \lambda_0^*,\, Y_0,\, \Lambda/q);\\
\label{Gamma_Phi_Part}
&&\hspace*{-13mm} \Gamma^{(2)}_\phi = \frac{1}{4}\int \frac{d^4q}{(2\pi)^4}\, d^4\theta\, \phi^{*i}(-q,\theta) \phi_j(q,\theta) \big(G_\phi\big)_i{}^j(\rho\alpha_0,\, \rho\lambda_0 \lambda_0^*,\, Y_0,\, \Lambda/q).
\end{eqnarray}

\noindent
Then, according to Refs. \cite{Stepanyantz:2016gtk,Stepanyantz:2019ihw}, Eq. (\ref{NSVZ_New_Equation}) can be equivalently rewritten as the relation between the Green functions

\begin{eqnarray}\label{NSVZ_For_Green_Functions}
&& \frac{\partial^2}{\partial g\, \partial g^*} \left.\frac{d}{d\ln\Lambda} \Big(d^{-1} - \big(g g^*\big)^{-1}\alpha_0^{-1}\Big)\right|_{\alpha,\lambda,Y = \mbox{\scriptsize const};\ p\to 0}\nonumber\\
&&\qquad = \frac{1}{2\pi}\, \frac{\partial^2}{\partial g\,\partial g^*} \left.\frac{d}{d\ln\Lambda} \Big(2C_2 \ln G_c + C_2 \ln G_V - \frac{1}{r} C(R)_i{}^j \big(\ln G_\phi\big)_j{}^i\Big)\right|_{\alpha,\lambda,Y = \mbox{\scriptsize const};\ q\to 0},\qquad
\end{eqnarray}

\noindent
where $g$ and $g^*$ are considered as independent variables. The derivatives with respect to these parameters can be removed by the integration

\begin{equation}\label{Rho_Integral}
\int\limits_{+0}^{1} \frac{d\rho}{\rho}\, \int\limits_{+0}^\rho d\rho\, \frac{\partial^2 f(\rho)}{\partial g\,\partial g^*} = f(1) - f(0).
\end{equation}

\noindent
(In the left hand side $f(0) = \beta_{\mbox{\scriptsize 1-loop}}(\alpha_0)/\alpha_0^2$, while in the right hand side $f(0)=0$, see the detailed explanation in Ref. \cite{Stepanyantz:2019ihw}.)

\medskip

3. Next, it is necessary to prove that the left hand side of Eq. (\ref{NSVZ_For_Green_Functions}) is given by integrals of double total derivatives in all loops. In the non-Abelian case this has been done in Ref. \cite{Stepanyantz:2019ihw}. For this purpose, first, it is necessary to extract the left hand side of Eq. (\ref{NSVZ_For_Green_Functions}) from the effective action. To do this, we consider the derivative of the expression

\begin{equation}
\Delta\Gamma^{(2)}_{\bm{V}}\equiv \Gamma^{(2)}_{\bm{V}} - S^{(2)}_{\bm{V}}
\end{equation}

\noindent
with respect to $\ln\Lambda$ and make the formal substitution

\begin{equation}\label{V_To_Theta4}
\bm{V}^A \to \theta^4 v^A,
\end{equation}

\noindent
where $\theta^4 = \theta^a \theta_a \bar\theta^{\dot a} \bar\theta_{\dot a}$, and $v^A$ are functions slowly decreasing at a certain large scale $R\to \infty$. After some calculations it is possible to obtain

\begin{equation}\label{Bare_Beta_From_Effective_Action}
\left.\frac{d\Delta\Gamma^{(2)}_{\bm{V}}}{d\ln\Lambda}\right|_{\alpha,\lambda,Y = \mbox{\scriptsize const};\, \bm{V}= \theta^4 v} = \frac{{\cal V}_4}{2\pi} \cdot \frac{\beta(\rho\alpha_0,\, \rho\lambda_0 \lambda_0^*, \, Y_0)}{\rho^2 \alpha_0^2},
\end{equation}

\noindent
where

\begin{equation}\label{Nu4_Definition}
{\cal V}_4 \equiv \int d^4x\, (v^A)^2 \sim R^4 \to \infty
\end{equation}

\noindent
and all terms suppressed by powers of $1/(\Lambda R)$ should be omitted.

Taking into account that the coordinate independent parameter $g$ and the superfield $\mbox{\sl g}$ enter all terms containing the quantum superfields only in the combination (\ref{Bold_G}) one can equivalently rewrite the left hand side of Eq. (\ref{NSVZ_For_Green_Functions}) in the form

\begin{eqnarray}\label{LHS}
&&\hspace*{-8mm} \frac{\partial^2}{\partial g\, \partial g^*} \left.\frac{d}{d\ln\Lambda} \Big(d^{-1} - \big(g g^*\big)^{-1}\alpha_0^{-1}\Big)\right|_{\alpha,\lambda,Y = \mbox{\scriptsize const};\ p\to 0} = \frac{2\pi}{{\cal V}_4} \cdot \frac{\partial^2}{\partial g\, \partial g^* }\left.\frac{d\Delta\Gamma^{(2)}_{\bm{V}}}{d\ln\Lambda}\right|_{\alpha,\lambda,Y = \mbox{\scriptsize const};\, \bm{V}= \theta^4 v}
\nonumber\\
&&\hspace*{-8mm} = \frac{\pi}{r {\cal V}_4} \int d^8x\, d^8y\, d^6z_1\, d^6\bar z_2\, (\theta^4)_x (v^B)_x\, (\theta^4)_y (v^B)_y \left. \frac{d}{d\ln\Lambda} \frac{\delta^4\Gamma}{\delta\mbox{\sl g}_{z_1} \delta \mbox{\sl g}^*_{z_2} \delta\bm{V}_x^A \delta \bm{V}_y^A} \right|_{\parbox{2.1cm}{\scriptsize $\alpha,\lambda,Y = \mbox{\scriptsize const};$\\ $\mbox{\scriptsize fields}=0;\,\mbox{\scriptsize \sl g}=0$}}.
\end{eqnarray}

\noindent
Note that in our notation the subscript ``fields=0'' does not mean that $\mbox{\sl g}=0$. The latter condition (if needed) is always written separately.

The last equality in Eq. (\ref{LHS}) follows from the fact that all terms in the total action containing quantum superfields depend on $\mbox{\sl g}$ and $g$ only via the combination $\bm{g} = g+\mbox{\sl g}$. This implies that the derivatives with respect to the superfields $\mbox{\sl g}$ and $\mbox{\sl g}^*$ are related to the derivatives with respect to the coordinate independent parameters $g$ and $g^*$ by the identities

\begin{equation}\label{Relation_Between_Derivatives}
\int d^6z_1\, \frac{\delta}{\delta \mbox{\sl g}_{z_1}} = \frac{\partial}{\partial g};\qquad\qquad \int d^6\bar z_2\, \frac{\delta}{\delta \mbox{\sl g}^*_{z_2}} = \frac{\partial}{\partial g^*}.
\end{equation}

\noindent
Also it is necessary to take into account that in the case of a simple gauge group there is the only invariant tensor of the form $I_{AB}$ proportional to $\delta_{AB}$.

The expression (\ref{LHS}) can be transformed with the help of the Slavnov--Taylor identities \cite{Taylor:1971ff,Slavnov:1972fg} for the background gauge invariance, rules for calculating supergraphs, and the identity (first derived in Ref. \cite{Stepanyantz:2014ima})

\begin{equation}
\theta^2 AB \theta^2 + 2(-1)^{P_A+P_B}\theta^a A \theta^2 B \theta_a - \theta^2 A \theta^2 B - A\theta^2 B \theta^2 = O(\theta),
\end{equation}

\noindent
where $A$ and $B$ are the sequences of vertices and propagators connecting certain points of a supergraph. The result can be written as

\begin{eqnarray}\label{Beta_Result}
&& \frac{\partial^2}{\partial g\, \partial g^*}\Big(\frac{\beta(\rho\alpha_0,\, \rho\lambda_0 \lambda_0^*, \, Y_0)}{\rho^2 \alpha_0^2}\Big) = \frac{2\pi}{r {\cal V}_4}\int d^8x\, d^8y\, d^6z_1\, d^6\bar z_2\, \big(\theta^2\big)_{z_1} \big(v^B\big)^2_{z_1}\, \big(\bar\theta^2\big)_{z_2}\,\nonumber\\
&&\qquad\qquad\qquad\qquad \times \big(\bar\theta^{\dot a} (\gamma^\mu)_{\dot a}{}^b \theta_b\big)_x \big(\bar\theta^{\dot c} (\gamma_\mu)_{\dot c}{}^d \theta_d\big)_y \left. \frac{d}{d\ln\Lambda} \frac{\delta^4\Gamma}{\delta\mbox{\sl g}_{z_1} \delta{\mbox{\sl g}}{}^*_{z_2} \delta\bm{V}_x^A \delta\bm{V}_y^A}\right|_{\mbox{\scriptsize fields} = 0;\, \mbox{\scriptsize \sl g}= 0}.\qquad\quad
\end{eqnarray}

An important observation is that this expression {\it formally} vanishes due to the Slavnov--Taylor identity corresponding to the background gauge invariance (\ref{Background_Gauge_Transformations_Original}), see Ref. \cite{Stepanyantz:2019ihw} for the detailed discussion. However, this is not true. In fact, the loop integrals contributing to this expression are integrals of double total derivatives with respect to loop momenta analogous to the integral (\ref{Total_Derivative_Integral}) and do not vanish due to singularities of the integrands. To demonstrate the factorization into double total derivatives, one can {\it formally} rewrite the expression
(\ref{Beta_Result}) in the form

\begin{equation}\label{Double_Variation}
\left. \frac{2\pi}{r {\cal V}_4}\cdot \frac{\partial^2}{\partial b^{\mu B}\, \partial a_\mu^B}\, \bar\delta_b \bar\delta_a\, \frac{d}{d\ln\Lambda} \int d^6z_1\, d^6\bar z_2\, \big(\theta^2\big)_{z_1} \big(v^B\big)_{z_1}^2\, \big(\bar\theta^2\big)_{z_2}\, \frac{\delta^2\Gamma}{\delta\mbox{\sl g}_{z_1} \delta{\mbox{\sl g}}{}^*_{z_2}} \right|_{\mbox{\scriptsize fields}=0;\ \mbox{\scriptsize {\sl g}} = 0} = 0,
\end{equation}

\noindent
where $\bar\delta_a$ denotes the variation of quantum superfields\footnote{In this case the background superfield $\bm{V}$ is not transformed, $\bar\delta_a \bm{V}=0$.} under the background gauge transformations (\ref{Background_Gauge_Transformations_Original}) with the parameters

\begin{equation}\label{Parameters}
A = i a_\mu^B t^B y^\mu;\qquad \mbox{and}\qquad A^+ = -i a_\mu^B t^B (y^\mu)^*.
\end{equation}

\noindent
The variation $\bar\delta_b$ is defined similarly, but $a_\mu^B$ should be replaced by $b_\mu^B$. The vectors $a_\mu^B$ and $b_\mu^B$ are coordinate independent, and $y^\mu = x^\mu + i\bar\theta^{\dot a} (\gamma^\mu)_{\dot a}{}^b \theta_b$ are chiral coordinates. In the momentum representation each of these variations can be interpreted as a certain analog of an integral of double total derivatives \cite{Stepanyantz:2019ihw}. The correctness of this interpretation is confirmed by an algorithm for obtaining the $\beta$-function which has been constructed starting from Eq. (\ref{Double_Variation}) and subsequently verified by some highly nontrivial calculations \cite{Stepanyantz:2019ihw,Kuzmichev:2019ywn,Stepanyantz:2019lyo}.

The formal result (\ref{Double_Variation}) implies that all contributions to the $\beta$-function beyond the one-loop approximation vanish. However, it is not correct, because, as we will see in this paper, in fact, the expressions (\ref{Beta_Result}) and (\ref{Double_Variation}) are not equal. The difference originates from singular contributions, which turn out to be missed in formal transformations.

\medskip

4. Although Eq. (\ref{Double_Variation}) gives the incorrect one-loop result for the $\beta$-function, it is very useful, because it allows to reduce the calculation of the exact $\beta$-function to finding a sum of singular contributions appearing in integrals of double total derivatives. Therefore, to obtain the NSVZ equation one should calculate this sum in all orders. The main purpose of the present paper is to calculate its part coming from singularities produced by the matter superfields and by the Faddeev--Popov ghosts.

To explain, how the singular contributions appear, and what is the difference between Eqs. (\ref{Beta_Result}) and (\ref{Double_Variation}) we again consider a toy integral (\ref{Total_Derivative_Integral}), which can be equivalently rewritten as

\begin{equation}\label{I_Definition}
I = - 2 \int \frac{d^4Q}{(2\pi)^4}\, \frac{Q^\mu}{Q^4}\, \frac{\partial}{\partial Q^\mu} \Big(f(Q^2) - Q^2 f'(Q^2)\Big) = \frac{1}{8\pi^4} \oint\limits_{S^3_\varepsilon} dS\, \frac{1}{Q^3}\, \Big(f(Q^2) - Q^2 f'(Q^2)\Big).
\end{equation}

\noindent
As we have already mentioned, this expression does not vanish due to the surface integral over the small sphere $S^3_\varepsilon$ surrounding the singularity at $Q_\mu = 0$.

However, it is possible to introduce the derivative $\bm{\partial/\partial Q^\mu}$ in such a way that, by definition, the integral of it is always equal to 0. In particular,

\begin{equation}\label{Bold_Integral_Definition}
\bm{I} \equiv - 2 \int \frac{d^4Q}{(2\pi)^4}\,  \frac{\bm{\partial}}{\bm{\partial Q^\mu}}\Big[\frac{Q^\mu}{Q^4}\, \Big(f(Q^2) - Q^2 f'(Q^2)\Big)\Big] \equiv 0,
\end{equation}

\noindent
Then, if we require that

\begin{equation}\label{Identity_With_Delta}
\Big[\frac{\bm{\partial}}{\bm{\partial Q^\mu}},\, \frac{Q^\mu}{Q^4}\Big] = \frac{\bm{\partial}}{\bm{\partial Q^\mu}}\Big(\frac{Q^\mu}{Q^4}\Big) = 2\pi^2 \delta^4(Q),
\end{equation}

\noindent
the original integral $I$ can be calculated in the following way:

\begin{eqnarray}
&& I \equiv -2 \int \frac{d^4Q}{(2\pi)^4}\, \frac{Q^\mu}{Q^4}\, \frac{\bm{\partial}}{\bm{\partial Q^\mu}} \Big(f(Q^2) - Q^2 f'(Q^2)\Big)\nonumber\\
&&\qquad\qquad\qquad\qquad = \bm{I} + 4\pi^2 \int \frac{d^4Q}{(2\pi)^4}\, \delta^4(Q) \Big(f(Q^2) - Q^2 f'(Q^2)\Big) = \frac{1}{4\pi^2} f(0).\qquad
\end{eqnarray}

\noindent
Equivalently, this can be written as

\begin{equation}\label{Singularities_Algorithm}
I = \bm{I} - \mbox{singularities of $\bm{I}$} = - \mbox{singularities of $\bm{I}$}.
\end{equation}

Now we see that the expression (\ref{Beta_Result}) is an analog of the integral $I$, while the expression (\ref{Double_Variation}) is an analog of the vanishing integral $\bm{I}$. Therefore, although they are not equal, the former expression can be obtained by calculating the sum of singular contributions in the latter one using equations analogous to Eq. (\ref{Singularities_Algorithm}). This will done in the next section for the singularities produced by the matter and ghost superfields.

\medskip

5. Finally, it is necessary to construct the all-order prescription giving an NSVZ scheme for RGFs standardly defined in terms of the renormalized couplings. If Eq. (\ref{NSVZ_New_Equation}) is valid in all loops, then it is given by HD+MSL \cite{Stepanyantz:2016gtk}. The proof is made similar to the Abelian case considered in Refs. \cite{Kataev:2013eta,Kataev:2013csa,Kataev:2014gxa}.

\section{The sum of singular contributions}
\label{Section_Singularities}

\subsection{Overview}
\hspace*{\parindent}\label{Subsection_Overview}

The main idea used for calculating the right hand side of Eq. (\ref{Beta_Result}) is that the expression $\bar\theta^{\dot a} (\gamma^\mu)_{\dot a}{}^b \theta_b$ can be considered as a variation of the background gauge superfield under the infinitesimal background gauge transformation (\ref{Background_Gauge_Transformations_Original}) with the parameter (\ref{Parameters}) in the lowest order in $\bm{V}$. In our notation the corresponding variations of various superfields are denoted by $\delta_a$ where the subscript $a$ refers to the parameter $a^A_\mu$ in Eq. (\ref{Parameters}). They should be distinguished from the variations denoted by $\bar\delta_a$ which are nontrivial only for the quantum superfields and, by definition, vanish for the background superfield $\bm{V}$,

\begin{eqnarray}\label{Infinitesimal_Background_Gauge_Transformations}
&&\hspace*{-8mm}  \bar\delta \bm{V} \equiv 0; \qquad \delta \bm{V} = -\Big(\frac{\bm{V}}{1-e^{-2\bm{V}}}\Big)_{Adj} A + \Big(\frac{\bm{V}}{1-e^{2\bm{V}}}\Big)_{Adj} A^+;\qquad\ \delta V =\bar\delta V = -[A^+,V];\nonumber\\
&&\hspace*{-8mm} \delta\phi_i = \bar\delta\phi_i = A_i{}^j \phi_j;\qquad\quad \delta c = \bar\delta c = [A, c];\qquad \delta \bar c =\bar\delta \bar c = [A, \bar c];\qquad\quad  \delta b = \bar\delta b = [A,b];\vphantom{\frac{1}{2}}\nonumber\\
&&\hspace*{-8mm}\qquad\qquad\qquad\qquad \delta\Phi_i = \bar\delta\Phi_i = A_i{}^j \Phi_j;\qquad \delta\varphi_a = \bar\delta\varphi_a = [A,\varphi_a].\vphantom{\frac{1}{2}}
\end{eqnarray}

\noindent
Expanding $\delta \bm{V}$ in powers of $\bm{V}$ and substituting the expressions for $A$ and $A^+$ from Eq. (\ref{Parameters}) we obtain

\begin{equation}\label{Delta_A_V}
\quad \delta_a \bm{V}^A = a_\mu^A\, \bar\theta^{\dot a} (\gamma^\mu)_{\dot a}{}^b \theta_b + f^{ABC} \bm{V}^B a_\mu^C x^\mu + O(\bm{V}^2).\quad
\end{equation}

Due to the background gauge invariance the effective action satisfies the Slavnov--Taylor identity \cite{Taylor:1971ff,Slavnov:1972fg}. It can be obtained in the standard way by making the change of variables in the generating functional coinciding with the transformations (\ref{Background_Gauge_Transformations_Original}) for the quantum superfields (at fixed $\bm{V}$). For the considered symmetry the Slavnov--Taylor identity is reduced to the equation which can be interpreted as the manifest background gauge invariance of the effective action,

\begin{equation}\label{Generating_STI}
0 = \delta\Gamma = \bar\delta\Gamma + \int d^8y\, \delta \bm{V}_y^B \frac{\delta\Gamma}{\delta\bm{V}_y^B},
\end{equation}

\noindent
where the parameter $A$ is unfixed. It is important that the (super)fields in this equation are not set to 0, so that we will call it the generating Slavnov--Taylor identity. However, for the parameter $A$ given by Eq. (\ref{Parameters}) this identity does not actually take place, because the gauge parameter too rapidly grows at the space-time infinity, see Ref. \cite{Stepanyantz:2019ihw}. Thus, in this case Eq. (\ref{Generating_STI}) is valid only formally, and it is necessary to take into account singular contributions.

Nevertheless, if we formally apply Eq. (\ref{Generating_STI}) to the transformations parameterized by $A = i a_\mu^B t^B y^\mu$, then from Eq. (\ref{Beta_Result}) we obtain

\begin{eqnarray}\label{Vanishing_Formal_Result}
&& \frac{\partial^2}{\partial g\, \partial g^*}\Big(\frac{\beta(\rho\alpha_0,\, \rho\lambda_0 \lambda_0^*, \, Y_0)}{\rho^2 \alpha_0^2}\Big) = \mbox{(formally)} =  -\frac{2\pi}{r {\cal V}_4} \cdot \frac{\partial}{\partial a_\mu^A} \bar\delta_a \int d^8x\, d^6z_1\, d^6\bar z_2\, \big(\theta^2\big)_{z_1} \big(v^B\big)^2_{z_1}\, \quad\nonumber\\
&& \times \big(\bar\theta^2\big)_{z_2}\, \big(\bar\theta^{\dot a} (\gamma_\mu)_{\dot a}{}^b \theta_b\big)_x\, \frac{d}{d\ln\Lambda} \frac{\delta^3\Gamma}{\delta\mbox{\sl g}_{z_1} \delta{\mbox{\sl g}}{}^*_{z_2} \delta\bm{V}_x^A}\bigg|_{\mbox{\scriptsize fields} = 0;\, \mbox{\scriptsize \sl g}= 0}  = 0,\qquad\quad
\end{eqnarray}

\noindent
where we took into account that the variation $\bar\delta_a$ vanishes if the quantum (super)fields are set to 0, and the derivative of the second term in Eq. (\ref{Delta_A_V}) with respect to $\bm{V}_x^A$ is proportional to $f^{AAC}=0$. Eq. (\ref{Vanishing_Formal_Result}) is equivalent to the (incorrect) statement that all higher order ($L\ge 2$) contributions to the $\beta$-function vanish.\footnote{The one-loop contribution to the expression $\beta/\rho^2\alpha_0^2$ does not depend on $\rho$ and vanishes after differentiating with respect to $g$ and (or) $g^*$} However, in fact, Eq. (\ref{Vanishing_Formal_Result}) is not valid, because for the background gauge transformations with the parameter (\ref{Parameters}) the generating Slavnov--Taylor identity (\ref{Generating_STI}) is broken by singular contributions, which appear due to the rapid growth of the gauge parameter at infinity,

\begin{equation}\label{Generating_STI_Nonzero}
0 \ne \delta_a\Gamma = \bar\delta_a\Gamma + \int d^8y\, \delta_a \bm{V}_y^B \frac{\delta\Gamma}{\delta\bm{V}_y^B}.
\end{equation}

\noindent
In a certain degree this equation is analogous to Eq. (\ref{Singularities_Algorithm}). Namely, $\bar\delta_a \Gamma$ and $\delta_a\Gamma$ are the analogs of $\bm{I}$ and $I$, respectively, while the last term is given by the sum of singular contributions. This analogy suggests a way of calculating the sum of singular contributions which produce all higher order corrections to the $\beta$-function. Namely, using Eq. (\ref{Generating_STI_Nonzero}) and the last equality in Eq. (\ref{Vanishing_Formal_Result}) we can present the considered expression as a variation $\delta_a$ of a certain derivative of the effective action,

\begin{eqnarray}\label{Beta_From_Singularities}
&& \frac{\partial^2}{\partial g\, \partial g^*}\Big(\frac{\beta(\rho\alpha_0,\, \rho\lambda_0 \lambda_0^*, \, Y_0)}{\rho^2 \alpha_0^2}\Big) = \frac{2\pi}{r {\cal V}_4}\cdot  \frac{\partial}{\partial a_\mu^A} \delta_a \int d^8x\, d^6z_1\, d^6\bar z_2\, \big(\theta^2\big)_{z_1} \big(v^B\big)^2_{z_1}\, \nonumber\\
&&\qquad\qquad\qquad\qquad\qquad\qquad\quad\ \,  \times \big(\bar\theta^2\big)_{z_2}\, \big(\bar\theta^{\dot a} (\gamma_\mu)_{\dot a}{}^b \theta_b\big)_x \left. \frac{d}{d\ln\Lambda} \frac{\delta^3\Gamma}{\delta\mbox{\sl g}_{z_1} \delta{\mbox{\sl g}}{}^*_{z_2} \delta\bm{V}_x^A}\right|_{\mbox{\scriptsize fields} = 0;\, \mbox{\scriptsize \sl g}= 0}.\qquad\quad
\end{eqnarray}

The derivative of the effective action with respect to the background gauge superfield at vanishing quantum fields is equal to the derivative of the generating functional $W = -i\ln Z$ at vanishing sources. The latter one can be presented as  a sum of terms, each of them corresponding to a certain part of the total action,

\begin{eqnarray}\label{Singularities_Parts}
&& \frac{\delta\Gamma}{\delta \bm{V}_x^A}\bigg|_{\mbox{\scriptsize quantum fields} =0} = \Big\langle \frac{\delta S_{\mbox{\scriptsize gauge}}}{\delta\bm{V}_x^A} + \frac{\delta S_{\mbox{\scriptsize gf}}}{\delta\bm{V}_x^A} + \frac{\delta S_{\mbox{\scriptsize matter}}}{\delta\bm{V}_x^A}\nonumber\\
&&\qquad\qquad\qquad\qquad\qquad\ \ + \frac{\delta S_{\mbox{\scriptsize FP}}}{\delta\bm{V}_x^A} + \frac{\delta S_{\mbox{\scriptsize NK}}}{\delta\bm{V}_x^A} + \frac{\delta S_\varphi}{\delta\bm{V}_x^A} - c\,\Big\langle \frac{\delta S_\Phi}{\delta\bm{V}_x^A} \Big\rangle_\Phi \Big\rangle\bigg|_{\mbox{\scriptsize quantum fields} =0},\qquad
\end{eqnarray}

\noindent
where the angular brackets stand for the functional integrations

\begin{eqnarray}\label{Angular_Brackets}
&& \langle B \rangle \equiv \frac{1}{Z} \int D\mu\, B\, \mbox{Det}(PV,M_\varphi)^{-1} \mbox{Det}(PV, M)^c\,\exp\Big\{i \left(S_{\mbox{\scriptsize total}} + S_{\mbox{\scriptsize sources}} \right)\Big\};\qquad\\
&& \langle B \rangle_\Phi \equiv \mbox{Det}(PV,M) \int D\Phi\, B\, \exp(i S_\Phi),
\end{eqnarray}

\noindent
$S_{\mbox{\scriptsize gauge}}$ is a part of $S_{\mbox{\scriptsize reg}}$ which does not contain matter superfields, and $S_{\mbox{\scriptsize matter}}$ is a sum of all terms in $S_{\mbox{\scriptsize reg}}$ containing them.
Note that the last term in Eq. (\ref{Singularities_Parts}) has a structure different from the others, because it comes from $\mbox{Det}(PV,M)^c$, which cannot in general be presented as a functional integral (unlike $\mbox{Det}(PV,M)^{-1}$ and $\mbox{Det}(PV,M_\varphi)^{-1}$). We will always assume that in expressions like (\ref{Singularities_Parts}) sources are expressed in terms of (super)fields with the help of the standard equations, and the condition ``sources=0'' is equivalent to ``quantum fields=0''. However, in the expression  (\ref{Singularities_Parts}) we so far do not set $\bm{V}$ to 0 and $\bm{g}$ to 1. The derivatives of the actions $S_{\phi_0}$ and $S_{c_0}$ with respect to the background gauge superfield are not included into Eq. (\ref{Singularities_Parts}) because these terms vanish when the auxiliary sources $\phi_0$, $c_0$, and $\bar c_0$ are set to 0.

According to Eqs. (\ref{Beta_From_Singularities}) and (\ref{Singularities_Parts}) various higher order corrections to the $\beta$-function can be associated with various parts of the total action. Certainly, the corresponding contributions to the $\beta$-function are given by integrals of total derivatives (which originate from $\delta_a$). That is why only the ones produced by the massless superfields are nontrivial due to Eq. (\ref{Identity_With_Delta}), while the massive Pauli--Villars superfields cannot produce singularities. (However, closed loops of these superfields can be present in singular contributions produced by massless superfields.) Moreover, the Nielsen--Kallosh ghosts are essential only in the one-loop approximation, so that their contribution to the expression (\ref{Beta_From_Singularities}) also vanishes. This can be easily verified taking into account that their action is quadratic in quantum superfields and does not depend on the parameter $\bm{g}$. Thus, we obtain

\begin{eqnarray}\label{Sum_Of_Singularities}
&& \frac{\partial^2}{\partial g\, \partial g^*}\Big(\frac{\beta(\rho\alpha_0,\, \rho\lambda_0 \lambda_0^*, \, Y_0)}{\rho^2 \alpha_0^2}\Big) = \frac{2\pi}{r {\cal V}_4} \cdot \frac{\partial}{\partial a_\mu^A} \delta_a \int d^8x\, d^6z_1\, d^6\bar z_2\, \big(\theta^2\big)_{z_1} \big(v^B\big)^2_{z_1}\, \big(\bar\theta^2\big)_{z_2}\, \nonumber\\
&& \times  \big(\bar\theta^{\dot a} (\gamma_\mu)_{\dot a}{}^b \theta_b\big)_x\, \frac{d}{d\ln\Lambda} \frac{\delta^2}{\delta\mbox{\sl g}_{z_1} \delta{\mbox{\sl g}}{}^*_{z_2}} \Big\langle \frac{\delta S_{\mbox{\scriptsize gauge}}}{\delta\bm{V}_x^A} + \frac{\delta S_{\mbox{\scriptsize gf}}}{\delta\bm{V}_x^A} + \frac{\delta S_{\mbox{\scriptsize matter}}}{\delta\bm{V}_x^A} + \frac{\delta S_{\mbox{\scriptsize FP}}}{\delta\bm{V}_x^A} \Big\rangle\bigg|_{\mbox{\scriptsize fields} = 0;\, \mbox{\scriptsize \sl g}= 0}.\qquad\quad
\end{eqnarray}

\noindent
Therefore, we need to calculate only the contributions coming from $S_{\mbox{\scriptsize matter}}$, $S_{\mbox{\scriptsize FP}}$, and $S_{\mbox{\scriptsize gauge}} + S_{\mbox{\scriptsize gf}}$. Below we consider only the ones corresponding to cuts of matter and ghost propagators, which come from $S_{\mbox{\scriptsize matter}}$ and $S_{\mbox{\scriptsize FP}}$. Below we will demonstrate that their sums give the terms in the right hand side of Eq. (\ref{NSVZ_For_Green_Functions}) containing $(\ln G_\phi)_j{}^i$ and $\ln G_c$, respectively, or, equivalently, the terms containing $(\gamma_\phi)_j{}^i$ and $\gamma_c$ in Eq. (\ref{NSVZ_New_Equation}).

\subsection{The matter superfields}
\hspace*{\parindent}\label{Subsection_Matter}

We will start with calculating singular contributions generated by the matter superfields. The massive Pauli--Villars superfields do not produce singularities, so that (as we have already mentioned) it is sufficient to consider a part of the action containing only the usual chiral matter superfields $\phi_i$,

\begin{equation}\label{Action_For_Chiral_Matter}
S_{\mbox{\scriptsize matter}} = \frac{1}{4} \int d^8x\, \phi^{*i} \Big(F\Big(-\frac{\bar\nabla^2 \nabla^2}{16\Lambda^2}\Big) e^{2{\cal F}(V)}e^{2\bm{V}}\Big)_i{}^j \phi_j
+ \Big(\frac{1}{6} \lambda_0^{ijk} \int d^6x\,\bm{g}\, \phi_i \phi_j \phi_k + \mbox{c.c.}\Big),
\end{equation}

\noindent
where the covariant derivatives are given by Eq. (\ref{Covariant_Derivative_Definition}) and $\bm{V} = \bm{V}^A T^A$. The corresponding contribution to the left hand side of Eq. (\ref{Sum_Of_Singularities}) is given by the expression

\begin{eqnarray}\label{Matter_Singularities_Delta}
&& \Delta_{\mbox{\scriptsize matter}} =  \frac{2\pi}{r {\cal V}_4} \cdot \frac{\partial}{\partial a_\mu^A}\, \delta_a \int d^8x\, d^6z_1\, d^6\bar z_2\, \big(\theta^2\big)_{z_1} \big(v^B\big)^2_{z_1}\, \big(\bar\theta^2\big)_{z_2}\, \nonumber\\
&&\qquad\qquad\qquad\qquad\qquad\qquad \times  \big(\bar\theta^{\dot a} (\gamma_\mu)_{\dot a}{}^b \theta_b\big)_x \frac{d}{d\ln\Lambda} \frac{\delta^2}{\delta\mbox{\sl g}_{z_1} \delta{\mbox{\sl g}}{}^*_{z_2}} \Big\langle  \frac{\delta S_{\mbox{\scriptsize matter}}}{\delta\bm{V}_x^A} \Big\rangle\bigg|_{\mbox{\scriptsize fields} = 0;\, \mbox{\scriptsize \sl g}= 0}.\qquad\quad
\end{eqnarray}

The derivative of $S_{\mbox{\scriptsize matter}}$ with respect to $\bm{V}^A$ entering this expression is calculated in Appendix \ref{Appendix_X_Matter}. After some transformations involving Eq. (\ref{Yukawa_Gauge_Invariance}) and the identity

\begin{equation}\label{Chiral_Identity}
\bar D^2 D^2 \phi = -16\partial^2 \phi
\end{equation}

\noindent
valid for an arbitrary chiral superfield $\phi$, the result can be written as

\begin{eqnarray}\label{S_Matter_V_Derivative_Brief}
&& \frac{\delta S_{\mbox{\scriptsize matter}} }{\delta \bm{V}^A} =  \frac{\delta S_{\mbox{\scriptsize matter}}}{\delta\phi_i} \big(T^A\big)_i{}^j \frac{D^2}{4\partial^2} \phi_j - \bar D^{\dot a}\Big(\big(\bar X^A_{\dot a}\big)_{\mbox{\scriptsize WZ}} + \big(\bar X^A_{\dot a}\big)_{\mbox{\scriptsize HD}} + \big(\bar X^A_{\dot a}\big)_{\mbox{\scriptsize Yukawa}} \Big) \qquad\nonumber\\
&& +\mbox{terms containing $[\bm{V}, T^A]$}.\vphantom{\Big(}
\end{eqnarray}

\noindent
In this equation we use the notations

\begin{eqnarray}\label{X_WZ}
&&\hspace*{-7mm} \big(\bar X_{\dot a}^A\big)_{\mbox{\scriptsize WZ}} = -\bar D_{\dot a}\Big[F\Big(-\frac{\nabla^2 \bar\nabla^2}{16\Lambda^2}\Big) \phi^{+}\cdot e^{2{\cal F}(V)} e^{2\bm{V}}\Big]^i \big(T^A\big)_i{}^j \frac{D^2}{32\partial^2} \phi_j +  \Big[F\Big(-\frac{\nabla^2 \bar\nabla^2}{16\Lambda^2}\Big) \phi^{+}\cdot e^{2{\cal F}(V)} \nonumber\\
&&\hspace*{-7mm} \times  e^{2\bm{V}}\Big]^i \big(T^A\big)_i{}^j \frac{\bar D_{\dot a} D^2}{32\partial^2} \phi_j;\\
\label{X_HD}
&&\hspace*{-7mm} \big(\bar X_{\dot a}^A\big)_{\mbox{\scriptsize HD}} = \sum\limits_{\alpha=1}^\infty \sum\limits_{\beta=0}^{\alpha-1} f_\alpha \bigg\{ \frac{1}{2} \Big[\Big(-\frac{\nabla^2 \bar\nabla^2}{16\Lambda^2}\Big)^\beta  \phi^{+}\Big]^i \Big(e^{2{\cal F}(V)} T^A e^{-2{\cal F}(V)}\Big)_i{}^j \Big[\frac{\bar\nabla_{\dot a}\nabla^2}{16\Lambda^2}\Big(-\frac{\bar\nabla^2 \nabla^2}{16\Lambda^2}\Big)^{\alpha-\beta-1} \nonumber\\
&&\hspace*{-7mm} \times e^{2{\cal F}(V)} e^{2\bm{V}}\phi\Big]_j
- \frac{1}{2} \Big[\bar\nabla_{\dot a}\Big(-\frac{\nabla^2 \bar\nabla^2}{16\Lambda^2}\Big)^\beta \phi^{+}\Big]^i  \Big(e^{2{\cal F}(V)} T^A e^{-2{\cal F}(V)}\Big)_i{}^j \Big[\frac{\nabla^2}{16\Lambda^2}\Big(-\frac{\bar\nabla^2 \nabla^2}{16\Lambda^2}\Big)^{\alpha-\beta-1}\nonumber\\
&&\hspace*{-7mm} \times e^{2{\cal F}(V)} e^{2\bm{V}}\phi\Big]_j\bigg\};\\
\label{X_Yukawa}
&&\hspace*{-7mm} \big(\bar X_{\dot a}^A\big)_{\mbox{\scriptsize Yukawa}} = - \frac{2}{3}\, \bm{g}\, \lambda_0^{ijk} \big(T^A\big)_k{}^m \Big(\frac{D^2}{8\partial^2} \phi_m\, \frac{\bar D_{\dot a} D^2}{16\partial^2} \phi_i\, \phi_j - \frac{\bar D_{\dot a} D^2}{8\partial^2} \phi_m\, \frac{D^2}{16\partial^2} \phi_i\, \phi_j \Big),
\end{eqnarray}

\noindent
where the coefficients $f_\alpha$ are defined as $F(x) = 1 + f_1 x + f_2 x^2 +\ldots$

To calculate $\Delta_{\mbox{\scriptsize matter}}$, it is necessary to present the expression (\ref{Matter_Singularities_Delta}) in the form of an integral of a total derivative and find its singular part. As a starting point we consider

\begin{eqnarray}\label{Variation}
&& \frac{\partial}{\partial a_\mu^A} \delta_a \int d^8x\, \big(\bar\theta^{\dot a} (\gamma_\mu)_{\dot a}{}^b \theta_b\big)_x \left.\Big\langle\frac{\delta S_{\mbox{\scriptsize matter}}}{\delta \bm{V}^A_x}\Big\rangle\right|_{\mbox{\scriptsize fields}=0} = \frac{\partial}{\partial a_\mu^A} \delta_a \int d^8x\, \big(\bar\theta^{\dot a} (\gamma_\mu)_{\dot a}{}^b \theta_b\big)_x\qquad\qquad\nonumber\\
&&\qquad\quad \times \left.\Big\langle \frac{\delta S_{\mbox{\scriptsize matter}}}{\delta\phi_i} (T^A)_i{}^j \frac{D^2}{4\partial^2} \phi_j - \bar D^{\dot c}\Big(\big(\bar X^A_{\dot c}\big)_{\mbox{\scriptsize WZ}} + \big(\bar X^A_{\dot c}\big)_{\mbox{\scriptsize HD}} + \big(\bar X^A_{\dot c}\big)_{\mbox{\scriptsize Yukawa}}  \Big)\Big\rangle\right|_{\mbox{\scriptsize fields}=0},\qquad
\end{eqnarray}

\noindent
where the variation $\delta_a$ should be calculated before the fields are set to 0. The terms containing $[\bm{V}, T^A]$ vanish because due to Eq. (\ref{Delta_A_V})

\begin{equation}\label{Vanishing_Commutators}
\frac{\partial}{\partial a_\mu^A} \delta_a [\bm{V}, T^A]\Big|_{\bm{V}=0} =\bar\theta^{\dot a} (\gamma^\mu)_{\dot a}{}^b \theta_b\, [T^A,T^A] = 0,
\end{equation}

\noindent
and $\bm{V}$ should be set to 0 in the resulting expression due to the condition ``fields=0''.  It is possible to demonstrate that the first term in Eq. (\ref{Variation}) also vanishes. For this purpose we use the identity

\begin{equation}\label{Auxiliary_A_Identity}
\Big\langle  B\, \frac{\delta S_{\mbox{\scriptsize matter}}}{\delta\phi_i} \Big\rangle\bigg|_{\mbox{\scriptsize fields=0}} = i \Big\langle \frac{\delta B}{\delta\phi_i} \Big\rangle\bigg|_{\mbox{\scriptsize fields=0}},
\end{equation}

\noindent
where $B$ is a certain function(al) depending on the superfields of the theory. For proving this identity one should make the change of variables $\phi_i \to \phi_i + a_i$, where $a_i$ are chiral superfields, in the functional integral $Z\langle B \rangle$, see Eq. (\ref{Angular_Brackets}). Using Eq. (\ref{Auxiliary_A_Identity}) we see that the first term in the right hand side of Eq. (\ref{Variation}) vanishes, because it turns out to be proportional to $\mbox{tr}(T^A) = 0$.

In the remaining terms in the right hand side of Eq. (\ref{Variation}) we integrate by parts with respect to the derivative $\bar D^{\dot c}$. This allows rewriting the expression (\ref{Variation}) in the form

\begin{eqnarray}\label{Matter_Derivative_Variation}
\frac{\partial}{\partial a_\mu^A} \int d^8x\, (\gamma_\mu)^{\dot a b} \theta_b\, \delta_a \Big\langle \big(\bar X_{\dot a}^A\big)_{\mbox{\scriptsize WZ}} + \big(\bar X_{\dot a}^A\big)_{\mbox{\scriptsize HD}} + \big(\bar X_{\dot a}^A\big)_{\mbox{\scriptsize Yukawa}}  \Big)\Big\rangle\bigg|_{\mbox{\scriptsize fields}=0}.
\end{eqnarray}

\noindent
This expression is formally calculated in Appendix \ref{Appendix_X_Matter_Variation}. This formal calculation gives the vanishing result in agreement with the general argumentation of Ref. \cite{Stepanyantz:2019ihw}. However, in the momentum representation the result can be written as an integral of a total derivative which contains singular contributions. In this section we consider only singularities corresponding to the cuts of the matter superfield propagators. They are analysed in Appendix \ref{Appendix_X_Matter_Total_Derivatives}. It turns out that such singular contributions can appear from the terms containing $\big(\bar X_{\dot a}^A\big)_{\mbox{\scriptsize WZ}}$ and $\big(\bar X_{\dot a}^A\big)_{\mbox{\scriptsize Yukawa}}$, while the term containing $\big(\bar X_{\dot a}^A\big)_{\mbox{\scriptsize HD}}$ does not produce them.

Taking into account that, as we discussed in Sect. \ref{Subsection_Overview}, the variation $\delta_a$ is analogous to the integral $I$ (of a usual total derivative $\partial/\partial q_\mu$) defined by Eq. (\ref{I_Definition}), the result for the sum of singularities in Eq. (\ref{Matter_Derivative_Variation}) corresponding to the cuts of matter propagators obtained in Appendix \ref{Appendix_X_Matter_Total_Derivatives} can be written as

\begin{eqnarray}\label{Singularities_Original}
&&\hspace*{-4mm} \int \frac{d^4q}{(2\pi)^4} \frac{\partial}{\partial q^\mu} \int d^8x\, d^8y\, (\gamma^\mu)^{\dot a b} \theta_b\, \Big\langle \bigg\{ \Big[F\Big(-\frac{\nabla^2 \bar\nabla^2}{16\Lambda^2}\Big) \phi^{+}\cdot  e^{2{\cal F}(V)} e^{2\bm{V}}\Big]^i + \bm{g}\, \lambda_0^{imn} \phi_m \frac{D^2}{4\partial^2} \phi_n \bigg\}_x \quad\nonumber\\
&&\hspace*{-4mm}\times\, \delta^8_{y x}(q)\, C(R)_i{}^j \Big(\frac{\bar D_{\dot a} D^2}{16\partial^2} \phi_j\Big)_y\Big\rangle \bigg|_{\mbox{\scriptsize fields=0}},\qquad
\end{eqnarray}

\noindent
where the nonsingular higher derivative terms coming from $\big(\bar X_{\dot a}^A\big)_{\mbox{\scriptsize HD}}$ were omitted, and we introduced the notation

\begin{equation}\label{Delta_Function}
\delta^8_{xy}(q) \equiv \delta^4(\theta_x-\theta_y) e^{i q_\alpha (x^\alpha - y^\alpha)}.
\end{equation}

It is convenient to present the expression (\ref{Singularities_Original}) in a different form. For this purpose we introduce the scalar superfields $\phi_{0i}$ {\it which do not satisfy the chirality condition} in the same representation of the gauge group as $\phi_i$. Their action is written as

\begin{eqnarray}\label{S_Phi0}
&& S_{\phi_0} =  \int d^8x\, \Big\{\, \frac{1}{4} \Big[F\Big(-\frac{\nabla^2 \bar\nabla^2}{16\Lambda^2}\Big) \phi^{+} \cdot e^{2{\cal F}(V)}e^{2\bm{V}}\Big]^i \phi_{0i}
\nonumber\\
&&\qquad\qquad\qquad\qquad\qquad\qquad
+ \bm{g}\, \lambda_0^{imn} \phi_{0i}\, \phi_m \Big[e^{-2\bm{V}}\frac{1}{16(\bm{\nabla}_\mu)^2} \bm{\nabla}^2 (e^{2\bm{V}}\phi)\Big]_n + \mbox{c.c.}\Big\},\qquad
\end{eqnarray}

\noindent
where the background derivative $\bm{\nabla}_\mu$ is defined by the equation $\{\bm{\nabla}_{a},\bm{\bar\nabla}_{\dot b}\} = 2i(\gamma^\mu)_{a\dot b}\bm{\nabla}_\mu$ and in the lowest order in $\bm{V}$ coincides with the usual derivative $\partial_\mu$. (Due to the presence of the covariant derivatives the background gauge invariance remains unbroken.) Including $S_{\phi_0}$ into the generating functional (\ref{Z_Generating_Functional}) we can consider the superfields $\phi_{0i}$ as auxiliary sources. For $\bm{V}=0$ the derivative of the action (and, therefore, of the effective action) with respect to $\phi_{0i}$ is related to the derivative with respect to the chiral superfields $\phi_i$ by the identity

\begin{equation}\label{Phi0_Equation}
-\frac{1}{2} \bar D^2 \frac{\delta \Gamma}{\delta\phi_{0i}}\bigg|_{\phi_0,\bm{V}=0} = -\frac{1}{2} \bar D^2 \Big\langle\frac{\delta S_{\phi_0}}{\delta\phi_{0i}}\Big\rangle\bigg|_{\phi_0,\bm{V}=0} = \Big\langle\frac{\delta S}{\delta\phi_i}\Big\rangle\bigg|_{\phi_0,\bm{V}=0} = \frac{\delta \Gamma}{\delta\phi_i}\bigg|_{\phi_0,\bm{V}=0},
\end{equation}

\noindent
where the other (super)fields are not set to 0. Actually, Eq. (\ref{Phi0_Equation}) can be considered as the Schwinger--Dyson equation for the matter superfields. It can be derived with the help of the integration variable change $\phi_i \to \phi_i + a_i$  in the generating functional, where $a_i$ are chiral superfields.

With the help of the sources $\phi_{0i}$ it is possible to rewrite Eq. (\ref{Singularities_Original}) in the equivalent form

\begin{equation}
- \int \frac{d^4q}{(2\pi)^4} \frac{\partial}{\partial q^\mu} \int d^8x\, d^8y\, \delta^8_{yx}(q)\, (\gamma^\mu)_{\dot a}{}^b \theta_b\, \Big\langle\frac{\delta S_{\phi_0}}{\delta \phi_{0i,x}} C(R)_i{}^j \frac{\bar D^{\dot a} D^2}{4\partial^2} \phi_{j,y}\Big\rangle\bigg|_{\mbox{\scriptsize fields=0}},
\end{equation}

\noindent
where $\phi_{i,x} \equiv \phi_i(x^\mu,\theta_x)$. Thus, we obtain

\begin{eqnarray}\label{Matter_Singularities}
&& \Delta_{\mbox{\scriptsize matter}} = -\frac{2\pi}{r {\cal V}_4}\cdot \frac{d}{d\ln\Lambda} \int d^6z_1\, d^6\bar z_2\, \big(\theta^2\big)_{z_1} \big(v^B\big)_{z_1}^2\, \big(\bar\theta^2\big)_{z_2}\, \frac{\delta^2}{\delta\mbox{\sl g}_{z_1} \delta\mbox{\sl g}^*_{z_2}} \qquad\nonumber\\
&&\qquad\ \times  \int \frac{d^4q}{(2\pi)^4} \frac{\partial}{\partial q_\mu} \int d^8x\, d^8y\, \delta^8_{yx}(q)\,  (\gamma^\mu)_{\dot a}{}^b \theta_b\, \Big\langle \frac{\delta S_{\phi_0}}{\delta \phi_{0i,x}} C(R)_i{}^j \frac{\bar D^{\dot a} D^2}{4\partial^2} \phi_{j,y} \Big\rangle\bigg|_{\mbox{\scriptsize fields}=0;\ \mbox{\scriptsize {\sl g}} = 0}.\qquad
\end{eqnarray}

\noindent
This expression is quadratic in explicitly written $\bar\theta$-s and encodes a sum of certain (at least, connected) superdiagrams. Calculating superdiagrams it is impossible to obtain explicitly written $\theta$-s or $\bar\theta$-s (although  $\theta$-s and $\bar\theta$-s are also present inside the supersymmetric covariant derivatives). It is well known (see, e.g., \cite{West:1990tg}) that with the help of the $D$-algebra any supergraph can be reduced to the integral over $d^4\theta$. Therefore, terms linear in explicit $\bar\theta$ and terms which do not contain explicit $\bar\theta$ vanish. This implies that $\bar\theta^2$ in Eq. (\ref{Matter_Singularities}) can be shifted in an arbitrary point. Really, if $\bar\theta^2$ is shifted with the help of identities like

\begin{equation}
\big(\bar\theta^{\dot a}\big)_1 \frac{D_1^2 \bar D_1^2}{4\partial^2} \delta^8_{12} = \frac{D_1^2 \bar D_1^2}{4\partial^2} \Big(\big(\bar\theta^{\dot a}\big)_1\delta^8_{12}\Big) + O(1) = \big(\bar\theta^{\dot a}\big)_2 \frac{D_1^2 \bar D_1^2}{4\partial^2} \delta^8_{12} + O(1)
\end{equation}

\noindent
(where $O(1)$ denotes terms which do not explicitly depend on $\bar\theta$) we obtain only terms less than quadratic in $\bar\theta$, which are removed by the final integration over $d^4\theta$.

Using this possibility we can replace $\bar\theta^2$ to the point $x$. Moreover, we take into account the chirality of the superfield $\phi_j$ and use the identity

\begin{equation}
[\bar D^{\dot a}, D^2] = - 4i (\gamma^\nu)_c{}^{\dot a} D^c \partial_\nu.
\end{equation}

\noindent
Then it is possible to rewrite the expression (\ref{Matter_Singularities}) in the form

\begin{eqnarray}\label{Matter_Singularities_Correlator}
&& \Delta_{\mbox{\scriptsize matter}} = \frac{2\pi}{r {\cal V}_4} \cdot \frac{d}{d\ln\Lambda} \int d^6z_1\, d^6\bar z_2\, \big(\theta^2\big)_{z_1} \big(v^B\big)_{z_1}^2\, \frac{\delta^2}{\delta\mbox{\sl g}_{z_1} \delta\mbox{\sl g}^*_{z_2}} \int \frac{d^4q}{(2\pi)^4} \frac{\partial}{\partial q^\mu} \frac{q_\nu}{q^2}  \qquad\nonumber\\
&&\qquad\qquad\qquad\ \ \times \int d^8x\, d^8y\, \delta^8_{yx}(q)\, \big(\bar\theta^2\theta^b\big)_x (\gamma^\mu\gamma^\nu)_b{}^c\, \Big\langle \frac{\delta S_{\phi_0}}{\delta \phi_{0i,x}} C(R)_i{}^j D_c \phi_{j,y} \Big\rangle\bigg|_{\mbox{\scriptsize fields}=0;\ \mbox{\scriptsize {\sl g}} = 0},  \qquad\ \
\end{eqnarray}

\noindent
where the Minkowski momentum $q_\nu$ is obtained when the derivatives act on $\delta^8_{yx}(q)$ after integrating by parts. (Certainly, the momentum integrals should be calculated in the Euclidean space after the Wick rotation. This will be done later.)

Using the second identity in Eq. (\ref{Relation_Between_Derivatives})\footnote{The first one cannot be applied due to the presence of $(\theta^2)_{z_1}$ in Eq. (\ref{Matter_Singularities_Correlator}).} we can present the expression under consideration as

\begin{eqnarray}\label{Matter_Singularities_Correlator_New}
&& \Delta_{\mbox{\scriptsize matter}} = \frac{2\pi}{r {\cal V}_4}\cdot \frac{d}{d\ln\Lambda} \frac{\partial}{\partial g^*} \int d^6z_1\, \big(\theta^2\big)_{z_1} \big(v^B\big)_{z_1}^2\, \frac{\delta}{\delta\mbox{\sl g}_{z_1}} \int \frac{d^4q}{(2\pi)^4} \frac{\partial}{\partial q^\mu} \frac{q_\nu}{q^2}  \qquad\nonumber\\
&&\qquad\qquad\qquad\ \ \times \int d^8x\, d^8y\, \delta^8_{yx}(q)\, \big(\bar\theta^2\theta^b\big)_x (\gamma^\mu\gamma^\nu)_b{}^c\, \Big\langle \frac{\delta S_{\phi_0}}{\delta \phi_{0i,x}} C(R)_i{}^j D_c \phi_{j,y} \Big\rangle\bigg|_{\mbox{\scriptsize fields}=0;\ \mbox{\scriptsize {\sl g}} = 0}. \qquad\ \
\end{eqnarray}

\noindent
It is convenient to rewrite it in terms of the effective action. For this purpose we use the identity

\begin{equation}
\Big\langle \frac{\delta S_{\phi_0}}{\delta \phi_{0i,x}} \phi_{j,y}\Big\rangle\bigg|_{\mbox{\scriptsize fields}=0} = \frac{1}{i} \frac{\delta}{\delta j^j_{\,,y}} \Big\langle \frac{\delta S_{\phi_0}}{\delta \phi_{0i,x}} \Big\rangle\bigg|_{\mbox{\scriptsize fields}=0} = -i\, \frac{\delta^2\Gamma}{\delta j^j_{\,,y}\, \delta \phi_{0i,x}}\bigg|_{\mbox{\scriptsize fields}=0},
\end{equation}

\noindent
where the derivative with respect to the source should be expressed in terms of the variational derivatives with respect to the superfields of the theory,

\begin{equation}\label{Derivative_Wrt_Source}
\frac{\delta}{\delta j^j_{\,,y}} = -\int d^8w\,\bigg[\Big(\frac{\delta^2\Gamma}{\delta \phi_{j,y} \delta \phi^{*k}_{\,,w}}\Big)^{-1} \frac{\bar D^2}{8\partial^2} \frac{\delta}{\delta \phi^{*k}_{\,,w}} + \Big(\frac{\delta^2\Gamma}{\delta \phi_{j,y} \delta \phi_{k,w}}\Big)^{-1} \frac{D^2}{8\partial^2} \frac{\delta}{\delta \phi_{k,w}} + \ldots \bigg],
\end{equation}

\noindent
where dots denote similar terms with the derivatives with respect to the other superfields, which are present because the superfields in this equation are not so far set to 0. Therefore,

\begin{eqnarray}
&& \Delta_{\mbox{\scriptsize matter}} = - \frac{2\pi}{r {\cal V}_4}\cdot \frac{d}{d\ln\Lambda} \frac{\partial}{\partial g^*} \int d^6z_1\, \big(\theta^2\big)_{z_1} \big(v^B\big)_{z_1}^2\, \frac{\delta}{\delta\mbox{\sl g}_{z_1}} \int \frac{d^4q}{(2\pi)^4} \frac{\partial}{\partial q^\mu} \frac{i q_\nu}{q^2}  \qquad \nonumber\\
&&\qquad\qquad\qquad \times \int d^8x\, d^8y\, \delta^8_{yx}(q)\,  \big(\bar\theta^2\theta^b\big)_x  (\gamma^\mu\gamma^\nu)_b{}^c C(R)_i{}^j \big(D_c\big)_y \frac{\delta^2 \Gamma}{\delta j^j_{\,,y}\, \delta \phi_{0i,x}} \bigg|_{\mbox{\scriptsize fields}=0;\ \mbox{\scriptsize {\sl g}} = 0}. \qquad
\end{eqnarray}

\begin{figure}[ht]
\begin{picture}(0,3)
\put(3.5,0.3){\includegraphics[scale=0.2]{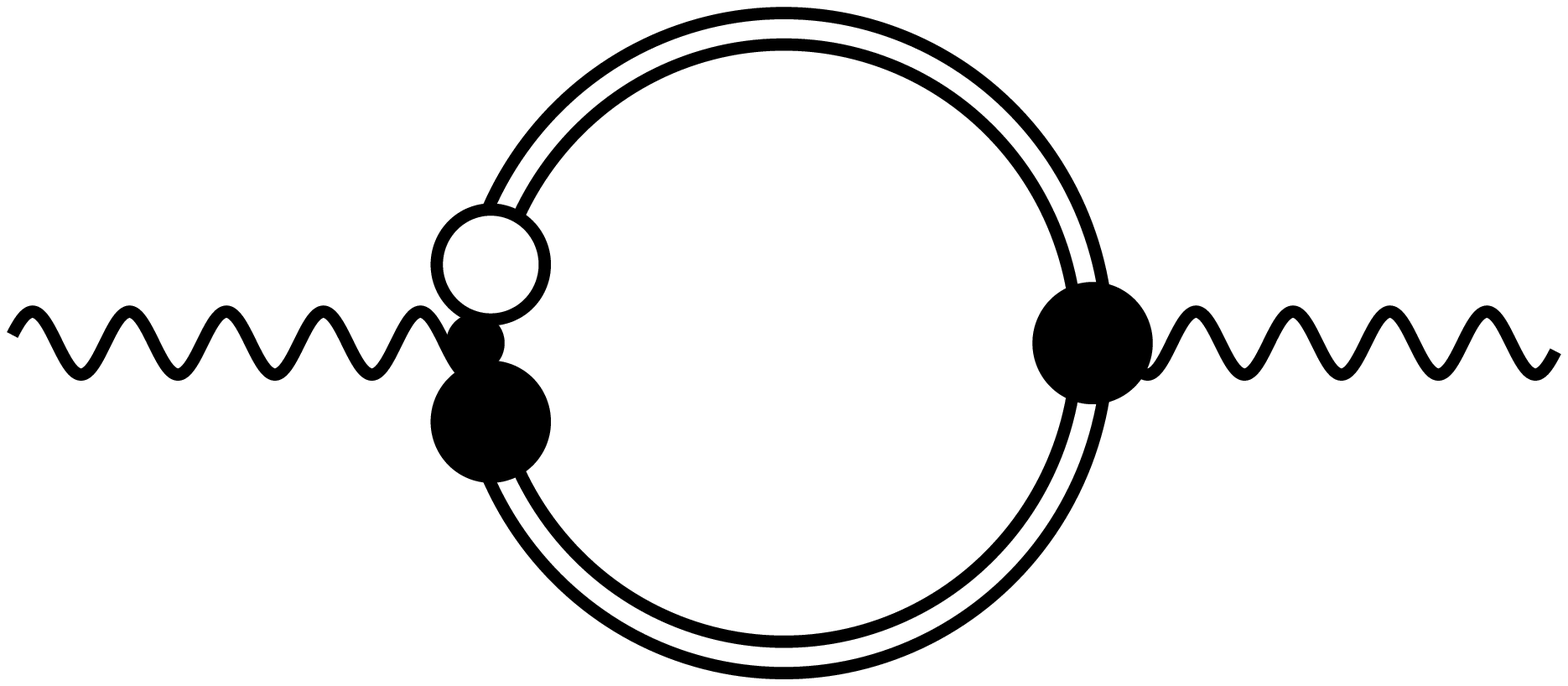}}
\put(7.1,1.35){$\theta^2$}
\put(3.3,1.35){$\bar\theta^2\theta^b$}
\put(3.0,2.2){(M1)}

\put(9.2,0){\includegraphics[scale=0.2]{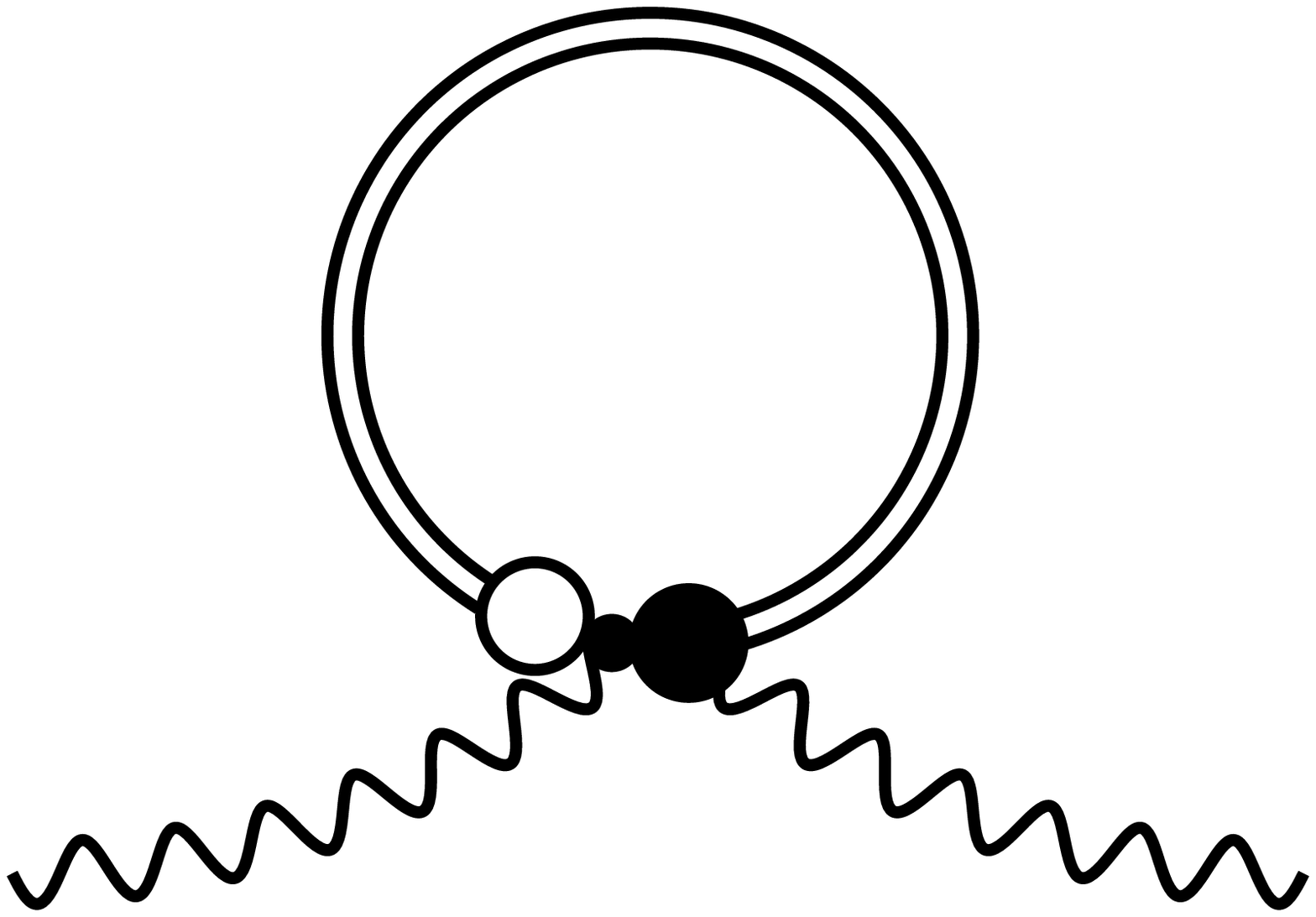}}
\put(12.3,0.3){$\theta^2$}
\put(8.7,0.3){$\bar\theta^2\theta^b$}
\put(8.9,2.2){(M2)}
\end{picture}
\caption{These two effective diagrams encode contributions to the $\beta$-function coming from the chiral matter superfields.}\label{Figure_Effective_Diagrams_For_Singularities}
\end{figure}

This expression can be presented as a sum of certain effective superdiagrams using the method similar to the one proposed in Ref. \cite{Stepanyantz:2004sg}. For this purpose it is necessary to commute the derivative with respect to the source $j^j_{\,,y}$ with the derivative with respect to $\mbox{\sl g}_{z_1}$. Note that these derivatives do not commute, because we consider the superfields ($V$, $\phi$, etc.) and the source $\mbox{\sl g}$ as independent variables. With the help of Eq. (\ref{Derivative_Wrt_Source}) after some transformations the result can be presented as

\begin{eqnarray}\label{Singularities_Effective_Diagrams}
&& \Delta_{\mbox{\scriptsize matter}} = -\frac{2\pi}{r {\cal V}_4}\cdot \frac{d}{d\ln\Lambda} \frac{\partial}{\partial g^*} \int d^6z_1\, \big(\theta^2\big)_{z_1} \big(v^B\big)_{z_1}^2  \int \frac{d^4q}{(2\pi)^4} \frac{\partial}{\partial q^\mu} \frac{i q_\nu}{q^2} \int d^8x\, d^8y\, \delta^8_{yx}(q)\,    \nonumber\\
&& \times \big(\bar\theta^2\theta^b\big)_x (\gamma^\mu\gamma^\nu)_b{}^c\, C(R)_i{}^j \big(D_c\big)_y  \bigg[\,
\frac{\delta}{\delta j^j_{\,,y}} \frac{\delta^2\Gamma}{\delta \mbox{\sl g}_{z_1}\, \delta \phi_{0i,x}} + \int d^8w\, \frac{\delta}{\delta j^j_{\,,y}} \frac{\delta^2\Gamma}{\delta \mbox{\sl g}_{z_1}\, \delta \phi_{k,w}}\cdot \Big(\frac{D^2}{8\partial^2}\Big)_w\,  \qquad\nonumber\\
&& \times \frac{\delta^2 \Gamma}{\delta j^k_{\,,w}\, \delta \phi_{0i,x}} + \int d^8w\, \frac{\delta}{\delta j^j_{\,,y}} \frac{\delta^2\Gamma}{\delta \mbox{\sl g}_{z_1}\, \delta \phi^{*k}_{\ ,w}}\cdot \Big(\frac{\bar D^2}{8\partial^2}\Big)_w\, \frac{\delta^2 \Gamma}{\delta j^*_{k,w}\, \delta \phi_{0i,x}} \bigg]  \bigg|_{\mbox{\scriptsize fields}=0;\ \mbox{\scriptsize {\sl g}} = 0},
\end{eqnarray}

\noindent
where, again, the derivatives with respect to the sources should be expressed in terms of the derivatives with respect to the fields. Eq. (\ref{Singularities_Effective_Diagrams}) admits a graphical interpretation as a sum of two effective diagrams, which are presented in Fig. \ref{Figure_Effective_Diagrams_For_Singularities}. In these diagrams external lines correspond to $(\theta^2)_{z_1} (v^B)_{z_1}^2$ and $(\bar\theta^2\theta^b)_x$. The double solid lines denote the exact propagators of the matter superfields and the large disks denote the effective vertices. The small disk encodes the derivative with respect to the sources $\phi_{0i}$, and the circle corresponds to the total derivative with respect to the loop momentum $q^\mu$.

Let us find the results for both these superdiagrams. We will start with calculating the diagram (M1) in Fig. \ref{Figure_Effective_Diagrams_For_Singularities} (which is given by the sum of the terms containing the integration over $w$). This can be done with the help of the identities \cite{Stepanyantz:2014ima}

\begin{equation}\label{Auxiliary Identity}
\frac{\delta^2\Gamma}{\delta j^k_{\,,w} \delta \phi_{0i,x}}\bigg|_{\mbox{\scriptsize fields}=0;\ \mbox{\scriptsize {\sl g}} = 0} = \delta_k^i\, \Big(\frac{\bar D^2 D^2}{16\partial^2}\Big)_w \delta^8_{xw};\qquad\qquad
\frac{\delta^2\Gamma}{\delta j^*_{k,w} \delta \phi_{0i,x}}\bigg|_{\mbox{\scriptsize fields}=0;\ \mbox{\scriptsize {\sl g}} = 0} = 0.
\end{equation}

\noindent
To obtain the first equation one should differentiate Eq. (\ref{Phi0_Equation}) with respect to $j^k_{\,,w}$ and take into account that from dimensional and chirality considerations the left hand side should be proportional to $\big(\bar D^2 D^2/\partial^2\big)_w$ in the massless case. The second identity is derived in a similar way by differentiating Eq. (\ref{Phi0_Equation}) with respect to $j^*_{k,w}$. Its left hand side in the massless case should be proportional to $\big(D^2 \bar D^2/\partial^2\big)_w$.

Using Eq. (\ref{Auxiliary Identity}) after some transformations the expression for the diagram (M1) in Fig. \ref{Figure_Effective_Diagrams_For_Singularities} can be presented as

\begin{eqnarray}\label{Second_Diagram}
&& \mbox{(M1)} = -\frac{2\pi}{r {\cal V}_4}\cdot \frac{d}{d\ln\Lambda} \frac{\partial}{\partial g^*} \int d^6z_1\, \big(\theta^2\big)_{z_1} \big(v^B\big)_{z_1}^2  \int \frac{d^4q}{(2\pi)^4} \frac{\partial}{\partial q^\mu} \frac{i q_\nu}{8 q^4} \int d^8x\, d^8y\, \delta^8_{yx}(q) C(R)_i{}^j   \qquad\nonumber\\
&& \times \big(\bar\theta^2\theta^b\big)_x\, (\gamma^\mu\gamma^\nu)_b{}^c  \big(D_c\big)_y \big(D^2\big)_x \frac{\delta}{\delta j^j_{\,,y}} \frac{\delta^2\Gamma}{\delta \mbox{\sl g}_{z_1}\, \delta \phi_{i,x}}  \bigg|_{\mbox{\scriptsize fields}=0;\ \mbox{\scriptsize {\sl g}} = 0}. \qquad
\end{eqnarray}

\noindent
Then we integrate by parts with respect to the derivative $(D_c)_y$ and use the identity $(D_c)_y \delta^8_{yx}(q) = - (D_c)_x \delta^8_{yx}(q)$. Next, we again integrate by parts with respect to the derivative $(D_c)_x$. As a result, the expression under consideration takes the form

\begin{eqnarray}\label{M1_Contribution}
&& \mbox{(M1)} = -\frac{2\pi}{r {\cal V}_4}\cdot \frac{d}{d\ln\Lambda} \frac{\partial}{\partial g^*} \int d^6z_1\, \big(\theta^2\big)_{z_1} \big(v^B\big)_{z_1}^2  \int \frac{d^4q}{(2\pi)^4} \frac{\partial}{\partial q^\mu} \frac{i q^\mu}{4 q^4} \int d^8x\, d^8y\, \big(\bar\theta^2\big)_x \delta^8_{yx}(q) \qquad\nonumber\\
&& \times\,  C(R)_i{}^j  \big(D^2\big)_x \frac{\delta}{\delta j^j_{\,,y}} \frac{\delta^2\Gamma}{\delta \mbox{\sl g}_{z_1}\, \delta \phi_{i,x}}  \bigg|_{\mbox{\scriptsize fields}=0;\ \mbox{\scriptsize {\sl g}} = 0}.  \qquad
\end{eqnarray}

Here we can shift $\theta^2$ from the point $z_1$ to the point $y$, because the expression under consideration is quadratic in the explicitly written (right components of) $\theta$.  As we discussed above, in this case various shifts of $\theta^2$ generate either linear terms or terms without $\theta_a$, which are removed by the final integration over $d^4\theta$. Also it is possible to replace $(v^B)^2$ from the point $z_1$ to the point $y$, because the difference is proportional to $1/(\Lambda^2 R^2) \to 0$, see Ref. \cite{Stepanyantz:2019ihw} for details. After these shifts one can use the first identity in Eq. (\ref{Relation_Between_Derivatives}) for rewriting the result in terms of the derivative with respect to the coordinate independent parameter $g$,

\begin{eqnarray}\label{M1_Contribution2}
&& \mbox{(M1)} = -\frac{2\pi}{r {\cal V}_4}\cdot \frac{d}{d\ln\Lambda} \frac{\partial}{\partial g^*} \int \frac{d^4q}{(2\pi)^4} \frac{\partial}{\partial q^\mu} \frac{i q^\mu}{4 q^4} \int d^8x\, d^8y\, \big(\theta^4\big)_y \big(v^B\big)_y^2\, \nonumber\\
&&\qquad\qquad\qquad\qquad\qquad\qquad\qquad\quad
\times\,  \delta^8_{yx}(q)\,  C(R)_i{}^j \big(D^2\big)_x\, \frac{\delta}{\delta j^j_{\,,y}} \frac{\partial}{\partial g} \frac{\delta\Gamma}{\delta \phi_{i,x}}  \bigg|_{\mbox{\scriptsize fields}=0;\ \mbox{\scriptsize {\sl g}} = 0}. \qquad\quad
\end{eqnarray}

\noindent
Certainly, here the derivative with respect to the source $j^j$ should be expressed in terms of the derivatives with respect to superfields with the help of Eq. (\ref{Derivative_Wrt_Source}).

The expression (\ref{M1_Contribution2}) contains only two-point Green functions. The only one which survives in the massless limit is

\begin{equation}\label{Exact_Green_Function}
\frac{\delta^2\Gamma}{\delta \phi_{j,x} \delta \phi^{*i}_{\,,y}}\bigg|_{\mbox{\scriptsize fields}=0;\ \mbox{\scriptsize {\sl g}} = 0} = \big(G_\phi\big)_i{}^j\, \frac{\bar D_x^2 D_y^2}{16} \delta^8_{xy},
\end{equation}

\noindent
where $(G_\phi)_i{}^j$ defined by Eq. (\ref{Gamma_Phi_Part}) is a dimensionless function of $\alpha_0$, $\lambda_0$, $Y_0$, $g$, $\Lambda^2/\partial^2$, $a_\varphi = M_\varphi/\Lambda$, and $a = M/\Lambda$. The corresponding inverse Green function has the form (see, e.g., \cite{Stepanyantz:2004sg})

\begin{equation}\label{Exact_Inverse_Functions}
\Big(\frac{\delta^2\Gamma}{\delta \phi_{i,x} \delta\phi^{*j}_{\,,y}}\Big)^{-1}\bigg|_{\mbox{\scriptsize fields}=0;\ \mbox{\scriptsize {\sl g}} = 0} = - \big(G_\phi^{-1}\big)_i{}^j\, \frac{\bar D_x^2 D_y^2}{4\partial^2} \delta^8_{xy}.
\end{equation}

\noindent
The functions (\ref{Exact_Green_Function}) and (\ref{Exact_Inverse_Functions}) should be substituted into Eq. (\ref{M1_Contribution2}). After this it is necessary to perform the Wick rotation $q^\mu \to Q^\mu$, where $Q^\mu$ is the Euclidean momentum, $q^0 = i Q^4$,\ $q^i = Q^i$, so that $q^\mu q_\mu = - Q^\mu Q_\mu$ and $d^4q = i d^4Q$. After some calculations the expression for the diagram (M1) can be rewritten as

\begin{equation}
\mbox{(M1)} = \frac{2\pi}{r {\cal V}_4}\cdot \frac{d}{d\ln\Lambda} \frac{\partial^2}{\partial g\,\partial g^*} \int d^8x\, \big(\theta^4\big)_x \big(v^B\big)_x^2\, C(R)_i{}^j \int \frac{d^4Q}{(2\pi)^4}\,\frac{\partial}{\partial Q^\mu} \Big(\frac{Q^\mu}{2Q^4}\, \big(\ln G_\phi\big)_j{}^i\Big).
\end{equation}

\noindent
The integral in this expression does not vanish due to the singular contribution which comes from the integration over the sphere $S^3_\varepsilon$ of the infinitely small radius $\varepsilon$ surrounding the point $Q_\mu=0$,

\begin{equation}\label{Matter_Singularities_Momentum_Integral}
\mbox{(M1)} = - \frac{2\pi}{r {\cal V}_4}\cdot \frac{d}{d\ln\Lambda} \frac{\partial^2}{\partial g\,\partial g^*} \int d^8x\, \big(\theta^4\big)_x \big(v^B\big)_x^2\, C(R)_i{}^j \oint\limits_{S^3_\varepsilon} \frac{dS}{(2\pi)^4}\, \frac{1}{2Q^3}\, \big(\ln G_\phi\big)_j{}^i.
\end{equation}

\noindent
(The integral over the infinitely large sphere $S^3_\infty$ in the momentum space vanishes due to the higher covariant derivative regularization.) The integral in Eq. (\ref{Matter_Singularities_Momentum_Integral}) can be easily taken and the contribution of the effective superdiagram (M1) in Fig. \ref{Figure_Effective_Diagrams_For_Singularities} appears to be

\begin{equation}
\mbox{(M1)} = - \frac{1}{2\pi r} C(R)_i{}^j \frac{d}{d\ln\Lambda} \frac{\partial^2}{\partial g\,\partial g^*}\,\big(\ln G_\phi\big)_j{}^i\Big|_{Q=0}.
\end{equation}

Now, let us consider the remaining contribution corresponding to the effective diagram (M2) in Fig. \ref{Figure_Effective_Diagrams_For_Singularities}, which is given by the expression

\begin{eqnarray}
&& \mbox{(M2)} = -\frac{2\pi}{r {\cal V}_4}\cdot \frac{d}{d\ln\Lambda} \frac{\partial}{\partial g^*} \int d^6z_1\, \big(\theta^2\big)_{z_1} \big(v^B\big)_{z_1}^2  \int \frac{d^4q}{(2\pi)^4} \frac{\partial}{\partial q_\mu} \frac{i q_\nu}{q^2} \int d^8x\, d^8y\, \delta^8_{yx}(q) C(R)_i{}^j \qquad\nonumber\\
&& \times\, \big(\bar\theta^2 \theta^b\big)_x (\gamma^\mu\gamma^\nu)_b{}^c \big(D_c\big)_y\, \frac{\delta}{\delta j^j_{\,,y}} \frac{\delta^2\Gamma}{\delta \mbox{\sl g}_{z_1}\, \delta \phi_{0i,x}} \bigg|_{\mbox{\scriptsize fields}=0;\ \mbox{\scriptsize {\sl g}} = 0}.\qquad\quad
\end{eqnarray}

\noindent
As for the previous superdiagram, the integral can be non-vanishing due to the singularity. After the Wick rotation the singular contribution appears from the integral over the sphere $S^3_\varepsilon$. The integral over $S^3_\infty$ vanishes, because the Green functions rapidly decrease at infinity ($Q\to \infty$) due to the higher covariant derivative regularization.

After expressing the derivative with respect to $j^j_{\,,y}$ in terms of the derivatives with respect to the superfields with the help of Eqs. (\ref{Derivative_Wrt_Source}) and using Eq. (\ref{Exact_Inverse_Functions}), the above expression can be rewritten as

\begin{eqnarray}\label{M2_Diagram}
&& \mbox{(M2)} = -\frac{2\pi}{r {\cal V}_4}\cdot \frac{d}{d\ln\Lambda} \frac{\partial}{\partial g^*} \int \frac{d^4q}{(2\pi)^4} \frac{\partial}{\partial q_\mu} \frac{iq_\nu}{2q^4} \int d^8x\, d^8y\, d^6z_1\, (\theta^2)_{z_1} \big(v^B\big)_{z_1}^2\, \delta^8_{yx}(q)\, \big(\bar\theta^2\theta^b\big)_x   \qquad\nonumber\\
&& \times  C(R)_i{}^j \big(G^{-1}_\phi(q)\big)_j{}^k\, (\gamma^\mu\gamma^\nu)_b{}^c \big(D_c \bar D^2\big)_y\, \frac{\delta^3\Gamma}{\delta \phi^{*k}_{\,,y}\, \delta \mbox{\sl g}_{z_1} \delta \phi_{0i,x}}\bigg|_{\mbox{\scriptsize fields}=0;\ \mbox{\scriptsize {\sl g}} = 0}.\qquad\quad
\end{eqnarray}

\noindent
From chirality considerations, we see that the Green function entering this expression can be presented in the form

\begin{eqnarray}\label{Explicit_Phi0_Functions}
&& \frac{\delta^3\Gamma}{\delta \phi^{*k}_{\,,y}\, \delta \mbox{\sl g}_{z_1} \delta \phi_{0i,x}}\bigg|_{\mbox{\scriptsize fields}=0;\ \mbox{\scriptsize {\sl g}} = 0} = \big(D^2\big)_y \big(\bar D^2\big)_{z_1} \int \frac{d^4p}{(2\pi)^4} \frac{d^4k}{(2\pi)^4} \Big(a(k,p)_k{}^i + b_\mu(k,p)_k{}^i\, (\gamma^\mu)^{a\dot b} \qquad \nonumber\\
&& \times  \big(D_a\big)_{z_1} \big(\bar D_{\dot b}\big)_y  + d(k,p)_k{}^i\, \big(D^2\big)_{z_1} \big(\bar D^2\big)_y  \Big) \Big(\delta^8_{xy}(-k-p)\, \delta^8_{xz_1}(p)\Big),\vphantom{\frac{1}{2}}
\end{eqnarray}

\noindent
where $a(k,p)_k{}^i$, $b_\mu(k,p)_k{}^i$, and $d(k,p)_k{}^i$ are functions of the external momenta $p$ and $k$, which correspond to the external $\mbox{\sl g}$-leg and the external $\phi_0$-leg, respectively. Substituting Eq. (\ref{Explicit_Phi0_Functions}) into Eq. (\ref{M2_Diagram}), after some transformations involving the algebra of the supersymmetric covariant derivatives we find the expression for the diagram (M2) in Fig. \ref{Figure_Effective_Diagrams_For_Singularities},

\begin{equation}\label{M2_Diagram_Final}
\mbox{(M2)} = - \frac{128 i\pi}{r} C(R)_i{}^j \frac{d}{d\ln\Lambda} \frac{\partial}{\partial g^*} \int \frac{d^4q}{(2\pi)^4} \frac{\partial}{\partial q_\mu}\Big[\,\frac{1}{q^2} \big(G_\phi^{-1}(q)\big)_j{}^k b_\mu(q,0)_k{}^i\Big] = 0.
\end{equation}

\noindent
The last equality follows from the fact that the surface integral over the sphere $S^3_\varepsilon$ evidently vanishes. (The integrals giving the functions $G_\phi$ and $b_\mu$ do not contain propagators with the momentum $q$.)

Thus, we obtain that the total contribution to the left hand side of Eq. (\ref{Sum_Of_Singularities}) coming from the matter singularities has the form

\begin{eqnarray}
&& \Delta_{\mbox{\scriptsize matter}} = \mbox{(M1)} + \mbox{(M2)} = - \frac{1}{2\pi r} C(R)_i{}^j \frac{d}{d\ln\Lambda} \frac{\partial^2}{\partial g\,\partial g^*}\,\big(\ln G_\phi\big)_j{}^i\Big|_{Q=0}\nonumber\\
&&\qquad\qquad\qquad\qquad\qquad\qquad\qquad = - \frac{1}{2\pi r} C(R)_i{}^j \frac{\partial^2}{\partial g\,\partial g^*}\, \big(\gamma_\phi\big)_j{}^i(\rho\alpha_0,\rho\lambda_0\lambda_0^*, Y_0) \qquad
\end{eqnarray}

\noindent
and coincides with the last term in the right hand side of Eq. (\ref{NSVZ_For_Green_Functions}). Evidently, after the integration (\ref{Rho_Integral}) we obtain the last term in the new form of the NSVZ equation (\ref{NSVZ_New_Equation}).

\subsection{The Faddeev--Popov ghosts}
\hspace*{\parindent}\label{Subsection_Ghosts}

Let us proceed to calculating the Faddeev--Popov ghost contribution to the sum of singularities (\ref{Sum_Of_Singularities}). The relevant part of the total action is given by Eq. (\ref{Ghosts_Faddeev-Popov}). First, it is necessary to calculate the derivative of $S_{\mbox{\scriptsize FP}}$ with respect to the background gauge superfield. The explicit expression for this derivative (given by Eq. (\ref{S_FP_V_Derivative_Original})) is presented in Appendix \ref{Appendix_X_FP}, where we demonstrate that it can be written in the form

\begin{eqnarray}\label{S_FP_V_Derivative_Brief}
&& \frac{\delta S_{\mbox{\scriptsize FP}} }{\delta \bm{V}^A} = - \frac{\delta S_{\mbox{\scriptsize FP}}}{\delta \bar c^B}\, \big(T^A_{Adj}\big)_{BD}\, \frac{D^2}{4\partial^2} \bar c^D - \frac{\delta S_{\mbox{\scriptsize FP}}}{\delta c^B}\, \big(T^A_{Adj}\big)_{BD}\, \frac{D^2}{4\partial^2} c^D \qquad\nonumber\\
&&\qquad\qquad\qquad\qquad\qquad\qquad - \bar D^{\dot a} \big(\bar X^A_{\dot a}\big)_{\mbox{\scriptsize FP}} +\mbox{terms containing $[\bm{V}, T^A]_{Adj}$}. \vphantom{\frac{1}{2}}\qquad
\end{eqnarray}

\noindent
Note that the minus signs in the first two terms originate from the anticommutation of the ghost superfields. The (rather large) expression for $\big(\bar X^A_{\dot a}\big)_{\mbox{\scriptsize FP}}$ is also presented in Appendix \ref{Appendix_X_FP}, see Eq. (\ref{X_FP}).

The Faddeev--Popov ghost contribution to the left hand side of Eq. (\ref{Sum_Of_Singularities}) is written as

\begin{eqnarray}
&&\Delta_{\mbox{\scriptsize ghost}} = \frac{2\pi}{r {\cal V}_4} \cdot \frac{\partial}{\partial a_\mu^A} \delta_a \int d^8x\, d^6z_1\, d^6\bar z_2\, \big(\theta^2\big)_{z_1} \big(v^B\big)_{z_1}^2\, \nonumber\\
&&\qquad\qquad\qquad\qquad\qquad \times\, \big(\bar\theta^2\big)_{z_2}\, \big(\bar\theta^{\dot a} (\gamma^\mu)_{\dot a}{}^b \theta_b\big)_x \frac{d}{d\ln\Lambda} \frac{\delta^2}{\delta\mbox{\sl g}_{z_1} \delta\mbox{\sl g}^*_{z_2}} \Big\langle \frac{\delta S_{\mbox{\scriptsize FP}}}{\delta \bm{V}^A_x}\Big\rangle \bigg|_{\mbox{\scriptsize fields}=0;\ \mbox{\scriptsize {\sl g}} = 0},\qquad\quad
\end{eqnarray}

\noindent
where $\delta_a$ denotes the variation of various superfields under the transformations (\ref{Background_Gauge_Transformations_Original}) with the parameter (\ref{Parameters}). Using Eq. (\ref{S_FP_V_Derivative_Brief}) a part of this expression can be presented in the form

\begin{eqnarray}\label{Variation_FP}
&& \frac{\partial}{\partial a_\mu^A} \delta_a \int d^8x\, \big(\bar\theta^{\dot a} (\gamma^\mu)_{\dot a}{}^b \theta_b\big)_x \left.\Big\langle\frac{\delta S_{\mbox{\scriptsize FP}}}{\delta \bm{V}^A_x}\Big\rangle\right|_{\mbox{\scriptsize fields}=0} = \frac{\partial}{\partial a_\mu^A} \delta_a \int d^8x\, \big(\bar\theta^{\dot a} (\gamma^\mu)_{\dot a}{}^b \theta_b\big)_x\qquad\nonumber\\
&& \times \left.\Big\langle- \frac{\delta S_{\mbox{\scriptsize FP}}}{\delta \bar c^B}\, \big(T^A_{Adj}\big)_{BD}\, \frac{D^2}{4\partial^2} \bar c^D - \frac{\delta S_{\mbox{\scriptsize FP}}}{\delta c^B}\, \big(T^A_{Adj}\big)_{BD}\, \frac{D^2}{4\partial^2} c^D - \bar D^{\dot c} \big(\bar X^A_{\dot c}\big)_{\mbox{\scriptsize FP}}\Big\rangle\right|_{\mbox{\scriptsize fields}=0}.\qquad
\end{eqnarray}

\noindent
Note that the terms containing $[\bm{V}, T^A]_{Adj}$ vanish due to Eq. (\ref{Vanishing_Commutators}) and the condition ``fields=0'' exactly as for the matter contribution considered in Sect. \ref{Subsection_Matter}.

Setting the (super)fields and the auxiliary sources $c_0$ to 0 and making in the functional integral for $Z\langle B\rangle$ the changes of the integration variables $\bar c^A \to \bar c^A + a^A$ or $c^A \to c^A + a^A$, where $a^A$ are anticommuting chiral superfields, we obtain the identities

\begin{equation}\label{Auxiliary_A_FP_Identities}
\Big\langle  \frac{\delta S_{\mbox{\scriptsize FP}}}{\delta \bar c^A} B \Big\rangle\bigg|_{\mbox{\scriptsize fields=0}} = i \Big\langle \frac{\delta B}{\delta \bar c^A} \Big\rangle\bigg|_{\mbox{\scriptsize fields=0}};\qquad\quad \Big\langle  \frac{\delta S_{\mbox{\scriptsize FP}}}{\delta c^A} B \Big\rangle\bigg|_{\mbox{\scriptsize fields=0}} = i \Big\langle \frac{\delta B}{\delta c^A} \Big\rangle\bigg|_{\mbox{\scriptsize fields=0}}.
\end{equation}

\noindent
Using these identities and taking into account that $\big(T^A_{Adj}\big)_{BB} = -i f^{ABB} = 0$, we conclude that the first two terms in Eq. (\ref{Variation_FP}) vanish. Therefore, after integrating by parts with respect to the derivative $\bar D^{\dot c}$, it is possible to present the expression (\ref{Variation_FP}) in the form

\begin{eqnarray}\label{FP_Derivative_Variation}
\frac{\partial}{\partial a_\mu^A} \int d^8x\, (\gamma^\mu)^{\dot a b} \theta_b\, \delta_a \Big\langle  (\bar X_{\dot a}^A)_{\mbox{\scriptsize FP}} \Big\rangle\bigg|_{\mbox{\scriptsize fields}=0}.
\end{eqnarray}

\noindent
Next, we introduce the anticommuting superfields $c_0$ and $\bar c_0$ in the adjoint representation, which are considered as auxiliary sources and do not satisfy the chirality condition, and add the term

\begin{eqnarray}\label{S_C0}
&&\hspace*{-8mm}\ S_{c_0} = \frac{1}{2} \int
d^8x\, \frac{\partial {\cal F}^{-1}(\widetilde V)^A}{\partial {\widetilde V}^B}\left.\vphantom{\frac{1}{2}}\right|_{\widetilde V = {\cal F}(V)} \Bigg[\left(\big(e^{2\bm{V}}\big)_{Adj} \bar c_0 \right)^A \left\{\vphantom{\frac{1}{2}}\smash{
\Big(\frac{{\cal F}(V)}{1-e^{2{\cal F}(V)}}\Big)_{Adj} c^+
+ \Big(\frac{{\cal F}(V)}{1-e^{-2{\cal F}(V)}}\Big)_{Adj}}\right.
\nonumber\\
&&\hspace*{-8mm} \left.\vphantom{\frac{1}{2}}\times \left(e^{2\bm{V}}\right)_{Adj} c\, \right\}^B
+ \left(e^{2\bm{V}}\bar c\, e^{-2\bm{V}} +
\bar c^+ \right)^A \left\{\vphantom{\frac{1}{2}}\smash{\Big(\frac{{\cal F}(V)}{1-e^{-2{\cal F}(V)}} e^{2\bm{V}} \Big)_{Adj} c_0}\right\}^B\Bigg] + \mbox{c.c.}
\end{eqnarray}

\noindent
to the action. Then, in analogy to Eq. (\ref{Phi0_Equation}), it is possible to derive the identities

\begin{equation}\label{C0_Equations}
\left.-\frac{1}{2} \bar D^2 \frac{\delta \Gamma}{\delta \bar c_0^B}\right|_{c_0,\bar c_0 = 0} = \left.\frac{\delta \Gamma}{\delta \bar c^B}\right|_{c_0,\bar c_0 = 0};\qquad \left.-\frac{1}{2} \bar D^2 \frac{\delta \Gamma}{\delta c_0^B}\right|_{c_0,\bar c_0 = 0} = \left.\frac{\delta \Gamma}{\delta c^B}\right|_{c_0,\bar c_0 = 0},
\end{equation}

\noindent
which relate the derivatives of the effective action with respect to $\bar c_0$ and $c_0$ to the derivatives with respect to $\bar c$ and $c$, respectively.

Similar to the case of the matter singularities considered in the previous section, the sum of singularities corresponding to the cuts of ghost propagators in the expression (\ref{FP_Derivative_Variation}) can be written as

\begin{equation}
- C_2 \int \frac{d^4q}{(2\pi)^4} \frac{\partial}{\partial q^\mu} \int d^8x\, d^8y\, \delta^8_{yx}(q)\, (\gamma^\mu)_{\dot a}{}^b \theta_b\, \Big\langle \frac{\delta S_{c_0}}{\delta \bar c_{0,x}^A}\, \frac{\bar D^{\dot a} D^2}{4\partial^2} \bar c^A_y + \frac{\delta S_{c_0}}{\delta c_{0,x}^A}\, \frac{\bar D^{\dot a} D^2}{4\partial^2} c^A_y \Big\rangle\bigg|_{\mbox{\scriptsize fields = 0}}.
\end{equation}

\noindent
(Singularities corresponding to the cuts of the quantum gauge superfield propagators are not considered in this paper.) This implies that total contribution produced by the Faddeev--Popov ghosts is

\begin{eqnarray}\label{FP_Singularities_Correlator}
&& \Delta_{\mbox{\scriptsize{ghost}}} = \frac{2\pi}{r {\cal V}_4}\cdot C_2 \frac{d}{d\ln\Lambda} \int d^6z_1\, d^6\bar z_2\, \big(\theta^2\big)_{z_1} \big(v^B\big)_{z_1}^2\, \big(\bar\theta^2\big)_{z_2}\,\frac{\delta^2}{\delta\mbox{\sl g}_{z_1} \delta\mbox{\sl g}^*_{z_2}} \int \frac{d^4q}{(2\pi)^4} \frac{\partial}{\partial q^\mu} \frac{q_\nu}{q^2} \qquad\nonumber\\
&& \times \int d^8x\, d^8y\, \delta^8_{yx}(q)\, \big(\theta^b\big)_x (\gamma^\mu\gamma^\nu)_b{}^d \Big\langle \frac{\delta S_{c_0}}{\delta \bar c_{0,x}^A} D_d \bar c^A_y +  \frac{\delta S_{c_0}}{\delta c_{0,x}^A} D_d c^A_y \Big\rangle\bigg|_{\mbox{\scriptsize fields}=0;\ \mbox{\scriptsize {\sl g}} = 0}. \qquad
\end{eqnarray}

\noindent
Because this expression is quadratic in $\bar\theta$, it is possible to shift $\bar\theta^2$ from the point $z_2$ to the point $x$. After this, one can convert the integral of the derivative with respect to $\mbox{\sl{g}}^*_{z_2}$ to the derivative with respect to $g^*$ with the help of Eq. (\ref{Relation_Between_Derivatives}). At the next step we rewrite the resulting expression in terms of the derivatives with respect to the sources $\bar j_c$ and $j_c$,

\begin{eqnarray}
&&\hspace*{-4mm} \Delta_{\mbox{\scriptsize{ghost}}} = -\frac{2\pi}{r {\cal V}_4}\cdot C_2\, \frac{d}{d\ln\Lambda}\, \frac{\partial}{\partial g^*}\int d^6z_1\, \big(\theta^2\big)_{z_1} \big(v^B\big)_{z_1}^2\, \frac{\delta}{\delta\mbox{\sl g}_{z_1}} \int \frac{d^4q}{(2\pi)^4} \frac{\partial}{\partial q^\mu} \frac{i q_\nu}{q^2} \int d^8x\, d^8y\, \delta^8_{yx}(q)\, \qquad\nonumber\\
&&\hspace*{-4mm} \times\,  \big(\bar\theta^2 \theta^b\big)_x (\gamma^\mu\gamma^\nu)_b{}^d \Big(\big(D_d\big)_y\, \frac{\delta^2 \Gamma}{\delta \bar j_{c,y}^A \, \delta \bar c_{0,x}^A} + \big(D_d\big)_y\, \frac{\delta^2 \Gamma}{\delta j_{c,y}^A\, \delta c_{0,x}^A}\Big)\bigg|_{\mbox{\scriptsize fields}=0;\ \mbox{\scriptsize {\sl g}} = 0}, \quad
\end{eqnarray}

\noindent
where, for example,

\begin{equation}
\frac{\delta}{\delta \bar j_{c,y}^A} = \int d^8w \bigg[\Big(\frac{\delta^2\Gamma}{\delta \bar c^A_y\,\delta c^{*B}_w}\Big)^{-1} \Big(\frac{\bar D^2}{8\partial^2}\Big)_w \frac{\delta}{\delta c_w^{*B}} +\ldots\bigg].
\end{equation}

\noindent
The dots again denote the analogous terms containing derivatives with respect to the other superfields of the theory, but only the explicitly written term survives under the condition ``fields=0;\ {\sl g}=0''.

\begin{figure}[ht]
\begin{picture}(0,3)
\put(3.5,0.3){\includegraphics[scale=0.2]{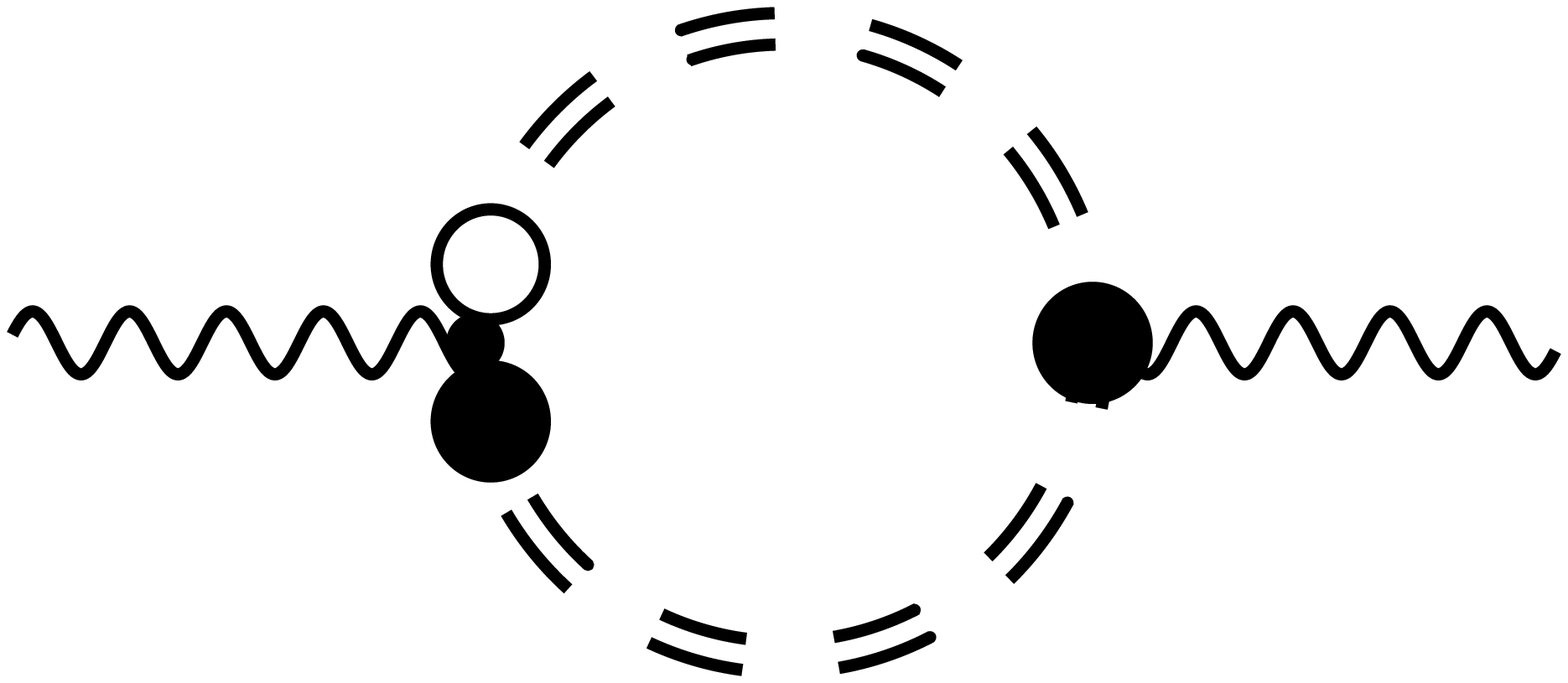}}
\put(7.1,1.35){$\theta^2$}
\put(3.3,1.35){$\bar\theta^2\theta^b$}
\put(3.0,2.2){(G1)}

\put(9.2,0){\includegraphics[scale=0.2]{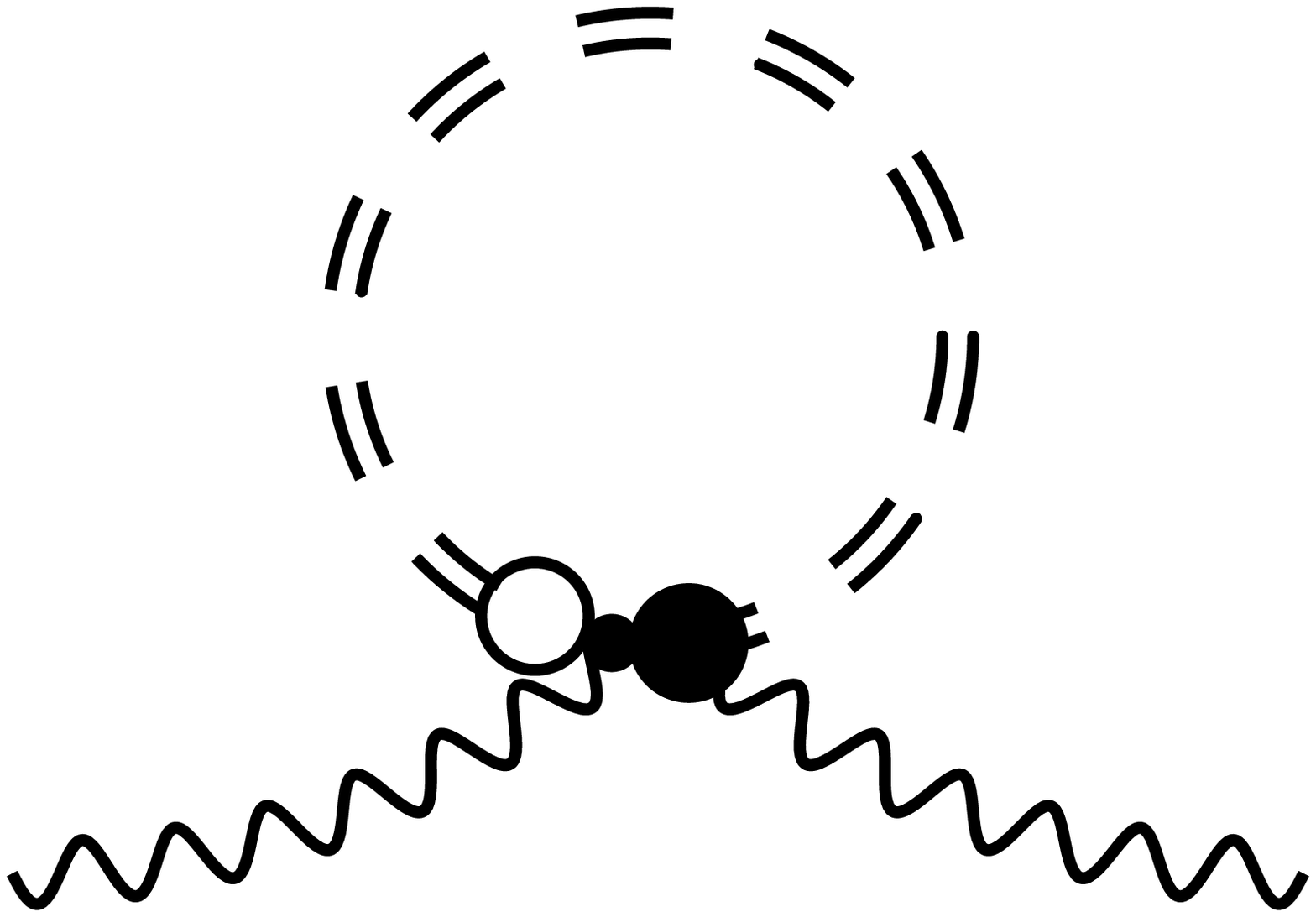}}
\put(12.3,0.3){$\theta^2$}
\put(8.8,0.3){$\bar\theta^2\theta^b$}
\put(8.9,2.2){(G2)}
\end{picture}
\caption{Effective superdiagrams giving the contribution of the Faddeev--Popov ghosts to the $\beta$-function.}\label{Figure_FP_Effective_Diagrams_For_Singularities}
\end{figure}

Commuting the derivatives with respect to the sources and the derivative with respect to $\mbox{\sl g}_{z_1}$ we present the result as a sum of two effective diagrams similar to the ones presented in Fig. \ref{Figure_Effective_Diagrams_For_Singularities}. However, now the loop consists of the effective ghost propagators. These diagrams are presented in Fig. \ref{Figure_FP_Effective_Diagrams_For_Singularities}. The analytic expression for them is given by the expression

\begin{eqnarray}\label{FP_Singularities_Effective_Diagrams}
&&\hspace*{-7mm} \Delta_{\mbox{\scriptsize{ghost}}} = - \frac{2\pi}{r {\cal V}_4}\cdot C_2\, \frac{d}{d\ln\Lambda} \frac{\partial}{\partial g^*} \int d^6z_1\, \big(\theta^2\big)_{z_1} \big(v^B\big)_{z_1}^2  \int \frac{d^4q}{(2\pi)^4} \frac{\partial}{\partial q^\mu} \frac{i q_\nu}{q^2} \int d^8x\, d^8y\, \delta^8_{yx}(q)\, \big(\bar\theta^2\theta^b\big)_x   \nonumber\\
&&\hspace*{-7mm}  \times (\gamma^\mu\gamma^\nu)_b{}^d \big(D_d\big)_y  \bigg[
\frac{\delta}{\delta \bar j_{c,y}^A} \frac{\delta^2\Gamma}{\delta \mbox{\sl g}_{z_1}\, \delta \bar c_{0,x}^A} - \int d^8w\, \frac{\delta}{\delta \bar j_{c,y}^A} \frac{\delta^2\Gamma}{\delta \mbox{\sl g}_{z_1}\, \delta \bar c_{0,w}^D} \Big(\frac{D^2}{8\partial^2}\Big)_w \frac{\delta^2 \Gamma}{\delta \bar j_{c,w}^D\, \delta \bar c_{0,x}^A} + \frac{\delta}{\delta j_{c,y}^A} \frac{\delta^2\Gamma}{\delta \mbox{\sl g}_{z_1}\, \delta c_{0,x}^A} \nonumber\\
&& \hspace*{-7mm}  - \int d^8w\, \frac{\delta}{\delta j_{c,y}^A} \frac{\delta^2\Gamma}{\delta \mbox{\sl g}_{z_1}\, \delta c_{0,w}^D} \Big(\frac{D^2}{8\partial^2}\Big)_w \frac{\delta^2 \Gamma}{\delta j_{c,w}^D\, \delta c_{0,x}^A} \bigg]  \bigg|_{\mbox{\scriptsize fields}=0;\ \mbox{\scriptsize {\sl g}} = 0}.\vphantom{\frac{1}{2}}
\end{eqnarray}

The superdiagram (G1) in Fig. \ref{Figure_FP_Effective_Diagrams_For_Singularities} corresponds to the sum of the terms containing the integrals over $w$. For calculating them we use the identities

\begin{equation}\label{FP_Auxiliary Identity}
\frac{\delta^2\Gamma}{\delta \bar j_{c,w}^D\, \delta \bar c_{0,x}^A}\bigg|_{\mbox{\scriptsize fields}=0;\ \mbox{\scriptsize {\sl g}} = 0} =
\frac{\delta^2\Gamma}{\delta j_{c,w}^D\, \delta c_{0,x}^A}\bigg|_{\mbox{\scriptsize fields}=0;\ \mbox{\scriptsize {\sl g}} = 0} = - \delta_{AD}\, \Big(\frac{\bar D^2 D^2}{16\partial^2}\Big)_w \delta^8_{xw}.
\end{equation}

\noindent
To obtain them, it is necessary to differentiate Eq. (\ref{C0_Equations}) and take into account dimensional and chirality considerations. Making the transformations similar to the ones described in Sect. \ref{Subsection_Matter} it is possible to present the expression for the considered superdiagram (G1) in the form

\begin{eqnarray}\label{FP_Second_Contribution2}
&& \mbox{(G1)} = - \frac{2\pi}{r {\cal V}_4}\cdot C_2\, \frac{d}{d\ln\Lambda} \frac{\partial}{\partial g^*} \int \frac{d^4q}{(2\pi)^4} \frac{\partial}{\partial q^\mu} \frac{i q_\mu}{4 q^4} \int d^8x\, d^8y\, \big(\theta^4\big)_y \big(v^B\big)_y^2\, \nonumber\\
&&\qquad\qquad\qquad\qquad\qquad\qquad \times \delta^8_{yx}(q)\, \big(D^2\big)_x \bigg[\,\frac{\delta}{\delta \bar j_{c,y}^A} \frac{\partial}{\partial g} \frac{\delta\Gamma}{\delta \bar c^A_x} + \frac{\delta}{\delta j_{c,y}^A} \frac{\partial}{\partial g} \frac{\delta\Gamma}{\delta c^A_x}\, \bigg]\bigg|_{\mbox{\scriptsize fields}=0;\ \mbox{\scriptsize {\sl g}} = 0}. \qquad\quad
\end{eqnarray}

\noindent
The two-point Green functions of the Faddeev--Popov ghosts entering this equation are obtained by differentiating Eq. (\ref{Gamma_C_Part}) with respect to the (anti)ghost superfields. These Green functions and the corresponding inverse functions are written as

\begin{eqnarray}\label{FP_Exact_Green_Function}
&& \frac{\delta^2\Gamma}{\delta \bar c^A_x\, \delta c^{*B}_y}\bigg|_{\mbox{\scriptsize fields}=0;\ \mbox{\scriptsize {\sl g}} = 0} = \frac{\delta^2\Gamma}{\delta c^A_x\, \delta \bar c^{*B}_y}\bigg|_{\mbox{\scriptsize fields}=0;\ \mbox{\scriptsize {\sl g}} = 0} = \delta_{AB} G_c\, \frac{\bar D_x^2 D_y^2}{16} \delta^8_{xy};\\
\label{FP_Exact_Inverse_Functions}
&& \Big(\frac{\delta^2\Gamma}{\delta \bar c^A_x\, \delta c^{*B}_y}\Big)^{-1}\bigg|_{\mbox{\scriptsize fields}=0;\ \mbox{\scriptsize {\sl g}} = 0} = \Big(\frac{\delta^2\Gamma}{\delta c^A_x\, \delta \bar c^{*B}_y}\Big)^{-1}\bigg|_{\mbox{\scriptsize fields}=0;\ \mbox{\scriptsize {\sl g}} = 0} = \delta_{AB} G_c^{-1}\, \frac{\bar D_x^2 D_y^2}{4\partial^2} \delta^8_{xy}.\qquad
\end{eqnarray}

\noindent
Substituting them into Eq. (\ref{FP_Second_Contribution2}) and making calculations analogous to those described in Sect. \ref{Subsection_Matter}, we find that the singular contribution encoded in the effective superdiagram (G1) in Fig. \ref{Figure_FP_Effective_Diagrams_For_Singularities} is given by the expression

\begin{equation}\label{FP_Singularities_Result}
\mbox{(G1)} = \frac{C_2}{\pi}\, \frac{d}{d\ln\Lambda} \frac{\partial^2}{\partial g\,\partial g^*}\, \ln G_c\Big|_{Q=0}.
\end{equation}

The superdiagram (G2) in Fig. \ref{Figure_FP_Effective_Diagrams_For_Singularities} vanishes similarly to the case of the matter contribution. Really, it corresponds to the terms in Eq. (\ref{FP_Singularities_Effective_Diagrams}) which do not contain the integral over $w$,

\begin{eqnarray}\label{FP_First_Diagram}
&&\hspace*{-5mm} \mbox{(G2)} = - \frac{2\pi}{r {\cal V}_4}\cdot C_2\, \frac{d}{d\ln\Lambda} \frac{\partial}{\partial g^*} \int d^6z_1\, \big(\theta^2\big)_{z_1} \big(v^B\big)_{z_1}^2  \int \frac{d^4q}{(2\pi)^4} \frac{\partial}{\partial q^\mu} \frac{i q_\nu}{q^2} \int d^8x\, d^8y\, \delta^8_{yx}(q)
\big(\bar\theta^2\theta^b\big)_x \qquad\ \nonumber\\
&&\hspace*{-5mm} \times (\gamma^\mu\gamma^\nu)_b{}^d \big(D_d\big)_y
\Big(\frac{\delta}{\delta \bar j_{c,y}^A} \frac{\delta^2\Gamma}{\delta \mbox{\sl g}_{z_1}\, \delta \bar c_{0,x}^A} + \frac{\delta}{\delta j_{c,y}^A} \frac{\delta^2\Gamma}{\delta \mbox{\sl g}_{z_1}\, \delta c_{0,x}^A} \Big)\bigg|_{\mbox{\scriptsize fields}=0;\ \mbox{\scriptsize {\sl g}} = 0}.\qquad\quad
\end{eqnarray}

\noindent
Substituting in this equation the expressions (\ref{FP_Exact_Inverse_Functions}) for the inverse Green functions (which enter the derivatives with respect to the sources) and the Green functions

\begin{eqnarray}\label{Explicit_C0_Function1}
&&\hspace*{-3mm} \frac{\delta^3\Gamma}{\delta c^{*A}_y \delta \mbox{\sl g}_{z_1} \delta\bar c_{0,x}^B}\bigg|_{\mbox{\scriptsize fields}=0;\ \mbox{\scriptsize {\sl g}} = 0} = \delta_{AB}\, \big(D^2\big)_y \big(\bar D^2\big)_{z_1} \int \frac{d^4p}{(2\pi)^4} \frac{d^4k}{(2\pi)^4} \Big(a_c(k,p) + (b_c)_\mu(k,p)\, (\gamma^\mu)^{a\dot b} \qquad\nonumber\\
&&\hspace*{-3mm} \times  \big(D_a\big)_{z_1} \big(\bar D_{\dot b}\big)_y + d_c(k,p)\, \big(D^2\big)_{z_1} \big(\bar D^2\big)_y  \Big) \Big(\delta^8_{xy}(-k-p)\, \delta^8_{xz_1}(p)\Big);\vphantom{\frac{1}{2}}\qquad\\
\label{Explicit_C0_Function2}
&&\hspace*{-3mm} \frac{\delta^3\Gamma}{\delta \bar c^{*A}_y \delta \mbox{\sl g}_{z_1} \delta c_{0,x}^B}\bigg|_{\mbox{\scriptsize fields}=0;\ \mbox{\scriptsize {\sl g}} = 0} = \delta_{AB}\, \big(D^2\big)_y \big(\bar D^2\big)_{z_1} \int \frac{d^4p}{(2\pi)^4} \frac{d^4k}{(2\pi)^4} \Big(\bar a_c(k,p) + (\bar b_c)_\mu(k,p)\, (\gamma^\mu)^{a\dot b} \qquad\nonumber\\
&&\hspace*{-3mm} \times \big(D_a\big)_{z_1} \big(\bar D_{\dot b}\big)_y + \bar d_c(k,p)\, \big(D^2\big)_{z_1} \big(\bar D^2\big)_y  \Big) \Big(\delta^8_{xy}(-k-p)\, \delta^8_{xz_1}(p)\Big),\vphantom{\frac{1}{2}}\qquad
\end{eqnarray}

\noindent
written from chirality considerations, we obtain

\begin{equation}\label{FP_First_Diagram_Final}
\mbox{(G2)} = - 128 i\pi C_2 \frac{d}{d\ln\Lambda} \frac{\partial}{\partial g^*} \int \frac{d^4q}{(2\pi)^4} \frac{\partial}{\partial q_\mu}\Big[\,\frac{1}{q^2} G_c^{-1}(q) \Big(b_c(q,0)+ \bar b_c(q,0)\Big)\Big] = 0.
\end{equation}

\noindent
Note that the result is equal to 0 because the denominator contains only $q^{-2}$, so that the Euclidean integral over the sphere $S^3_\varepsilon$ surrounding the singularity at $Q_\mu=0$ vanishes.

Therefore, the sum of singular contributions produced by the Faddeev--Popov ghosts is equal to the term in the right hand side of Eq. (\ref{NSVZ_For_Green_Functions}) containing the function $G_c$,

\begin{equation}
\Delta_{\mbox{\scriptsize ghost}} = \mbox{(G1)} + \mbox{(G2)} = \frac{C_2}{\pi}\, \frac{d}{d\ln\Lambda} \frac{\partial^2}{\partial g\, \partial g^*}\, \ln G_c \Big|_{Q=0} = \frac{C_2}{\pi}\,\frac{\partial^2}{\partial g\, \partial g^*}\, \gamma_c(\rho\alpha_0, \rho\lambda_0 \lambda_0^*, Y_0).
\end{equation}

After the integration (\ref{Rho_Integral}) we obtain the term with $\gamma_c$ in the NSVZ equation (\ref{NSVZ_New_Equation}).

Thus, the sums of singular contributions produced by the matter superfields and by the Faddeev--Popov ghosts (coming from the corresponding terms in Eq. (\ref{Sum_Of_Singularities})) give the terms containing $(\gamma_\phi)_j{}^i$ and $\gamma_c$ in Eq. (\ref{NSVZ_New_Equation}), respectively. Therefore, for finishing the all-order perturbative derivation of the NSVZ equation it remains to obtain only the term containing $\gamma_V$ in Eq. (\ref{NSVZ_New_Equation}).

\section{Conclusion}
\hspace*{\parindent}

According to Ref. \cite{Stepanyantz:2019ihw}, all loop integrals giving the $\beta$-function of ${\cal N}=1$ supersymmetric gauge theories (defined in terms of the bare couplings) are integrals of double total derivatives with respect to the loop momenta if the higher covariant derivatives are used for regularization. These integrals do not vanish, because the integrands contain singularities. Earlier it was supposed \cite{Stepanyantz:2016gtk} that the sum of singular contributions produces the NSVZ $\beta$-function written in the form (\ref{NSVZ_New_Equation}).

In this paper we have calculated the sum of singular contributions produced by the matter and Faddeev--Popov ghost superfields exactly in all orders for the theory with a simple gauge group containing the Yukawa term in the action. The result can be written in the form

\begin{eqnarray}\label{Result}
&& \frac{\beta(\alpha_0,\lambda_0,Y_0)}{\alpha_0^2} = - \frac{1}{2\pi}\Big(3 C_2 - T(R) + \frac{1}{r} C(R)_i{}^j \big(\gamma_\phi\big)_j{}^i(\alpha_0,\lambda_0,Y_0) - 2C_2 \gamma_c(\alpha_0,\lambda_0,Y_0)\Big)\qquad\nonumber\\
&& +\ \mbox{singularities produced by the quantum gauge superfield}.\vphantom{\Big(}
\end{eqnarray}

\noindent
The first two terms correspond to the one-loop approximation, the matter superfields give the term containing the anomalous dimension $(\gamma_\phi)_j{}^i$, and the Faddeev--Popov ghosts give the term with the anomalous dimension $\gamma_c$. Certainly, both these anomalous dimensions are defined in terms of the bare couplings, so that Eq. (\ref{Result}) is valid for an arbitrary renormalization prescription which supplements the higher covariant derivative regularization. The result (\ref{Result}) agrees with Eq. (\ref{NSVZ_New_Equation}), but the term containing $\gamma_V$ is absent. Presumably, this term comes from the sum of singularities produced by the quantum gauge superfield. The calculation of this contribution will be described separately. Now this work is in preparation.

\section*{Acknowledgments}
\hspace*{\parindent}

The work was supported by the Foundation for the Advancement of Theoretical Physics and Mathematics `BASIS', grant No. 19-1-1-45-1.

The author is very grateful to S.~S.~Aleshin, A.~E.~Kazantsev, M.~D.~Kuzmichev, N.~P.~Meshcheriakov, S.~V.~Novgorodtsev, and I.~E.~Shirokov for numerous valuable discussions and useful comments on the manuscript.

\appendix

\section{Derivatives of $S_{\mbox{\scriptsize matter}}$ and $S_{\mbox{\scriptsize FP}}$ with respect to the background gauge superfield}
\label{Appendix_X}

\subsection{Action for the matter superfields}
\hspace*{\parindent}\label{Appendix_X_Matter}

Let us calculate the derivative of the action for the chiral matter superfields given by Eq. (\ref{Action_For_Chiral_Matter}) with respect to the background gauge superfield. Note that in this part of the action $\bm{V} = \bm{V}^A T^A$, so that

\begin{eqnarray}\label{Derivative_Of_Exponent}
&& \frac{\delta}{\delta \bm{V}_x^A} \big(e^{2\bm{V}}\big)_y = 2\,\delta^8_{xy}\, e^{2\bm{V}} T^A + \mbox{terms containing $[\bm{V}, T^A]$};\nonumber\\
&& \frac{\delta}{\delta \bm{V}_x^A} \big(e^{-2\bm{V}}\big)_y = - 2\,\delta^8_{xy}\, T^A e^{-2\bm{V}} + \mbox{terms containing $[\bm{V}, T^A]$}.\quad
\end{eqnarray}

\noindent
Such exponents are present, in particular, in the expressions for the left supersymmetric covariant derivatives (\ref{Covariant_Derivative_Definition}). That is why it is also expedient to calculate the derivative

\begin{equation}\label{Derivative_Of_Derivative}
\frac{\delta}{\delta\bm{V}_x^A} \big(e^{2\bm{V}} \bar D_{\dot a} e^{-2\bm{V}}\big)_y  = 2\, T^A \big(\bar D_{\dot a}\big)_x \delta^8_{xy} +\mbox{terms containing $[\bm{V}, T^A]$},
\end{equation}

\noindent
where $\bar D_{\dot a}$ in the right hand side acts only on the $\delta$-function (and does not act on the expression standing on the right). With the help of Eqs. (\ref{Derivative_Of_Exponent}) and (\ref{Derivative_Of_Derivative}) one can obtain the derivative of the action (\ref{Action_For_Chiral_Matter}). To write the result, we present the higher derivative regulator $F(x)$ as a series

\begin{equation}
F(x) \equiv 1 + f_1 x + f_2 x^2 + \ldots
\end{equation}

\noindent
Then the expression under consideration takes the form

\begin{eqnarray}\label{S_Matter_V_Derivative}
&&\hspace*{-5mm} \frac{\delta S_{\mbox{\scriptsize matter}} }{\delta \bm{V}^A} =  \frac{1}{2} \Big[F\Big(-\frac{\nabla^2 \bar\nabla^2}{16\Lambda^2}\Big) \phi^{+}\cdot e^{2{\cal F}(V)} e^{2\bm{V}} \Big]^i \big(T^A\big)_i{}^j \phi_j + \frac{1}{2} \sum\limits_{\alpha=1}^\infty \sum\limits_{\beta=0}^{\alpha-1} f_\alpha \bar D^{\dot a} \bigg\{ \Big[\bar\nabla_{\dot a}\Big(-\frac{\nabla^2 \bar\nabla^2}{16\Lambda^2}\Big)^\beta \phi^{+}\Big]^i \nonumber\\
&&\hspace*{-5mm} \times \Big(e^{2{\cal F}(V)} T^A e^{-2{\cal F}(V)}\Big)_i{}^j \Big[\frac{\nabla^2}{16\Lambda^2}\Big(-\frac{\bar\nabla^2 \nabla^2}{16\Lambda^2}\Big)^{\alpha-\beta-1} e^{2{\cal F}(V)} e^{2\bm{V}} \phi\Big]_j - \Big[\Big(-\frac{\nabla^2 \bar\nabla^2}{16\Lambda^2}\Big)^\beta \phi^{+}\Big]^i \Big(e^{2{\cal F}(V)} T^A \nonumber\\
&&\hspace*{-5mm} \times  e^{-2{\cal F}(V)}\Big)_i{}^j \Big[\frac{\bar\nabla_{\dot a}\nabla^2}{16\Lambda^2}\Big(-\frac{\bar\nabla^2 \nabla^2}{16\Lambda^2}\Big)^{\alpha-\beta-1}e^{2{\cal F}(V)} e^{2\bm{V}} \phi\Big]_j \bigg\}
+\mbox{terms containing $[\bm{V}, T^A]$}.\vphantom{\Big(}
\end{eqnarray}

\noindent
We see that the terms coming from the covariant derivatives contained inside the the regulator $F$ are proportional to $1/\Lambda^2$ and can be written as the left covariant derivative $\bar D^{\dot a}$ acting on a rather complicated expression. One can also try to construct a similar representation for the first term of Eq. (\ref{S_Matter_V_Derivative}). For this purpose we equivalently rewrite it in the form

\begin{eqnarray}\label{First_Term}
&&\hspace*{-5mm}  - \bar D^2 \Big[F\Big(-\frac{\nabla^2 \bar\nabla^2}{16\Lambda^2}\Big) \phi^{+} \cdot e^{2{\cal F}(V)} e^{2\bm{V}}\Big]^i \big(T^A\big)_i{}^j \frac{D^2}{32\partial^2} \phi_j + \bar D^{\dot a} \bigg\{ \bar D_{\dot a}\Big[F\Big(-\frac{\nabla^2 \bar\nabla^2}{16\Lambda^2}\Big) \phi^{+} \cdot e^{2{\cal F}(V)} e^{2\bm{V}}\Big]^i \nonumber\\
&&\hspace*{-5mm} \times \big(T^A\big)_i{}^j \frac{D^2}{32\partial^2} \phi_j -  \Big[F\Big(-\frac{\nabla^2 \bar\nabla^2}{16\Lambda^2}\Big) \phi^{+}\cdot e^{2{\cal F}(V)} e^{2\bm{V}}\Big]^i \big(T^A\big)_i{}^j \frac{\bar D_{\dot a} D^2}{32\partial^2} \phi_j \bigg\},
\end{eqnarray}

\noindent
where the equality can easily be verified with the help of the product rule for the supersymmetric covariant derivative $\bar D^{\dot a}$ and the identity (\ref{Chiral_Identity}). To present this expression in the most convenient form, we extract a term containing the derivative of the action (\ref{Action_For_Chiral_Matter}) with respect to the chiral matter superfield,

\begin{equation}
\frac{\delta S_{\mbox{\scriptsize matter}} }{\delta \phi_i} = - \frac{1}{8} \bar D^2\Big[ F\Big(-\frac{\nabla^2 \bar\nabla^2}{16\Lambda^2}\Big) \phi^{+}\cdot e^{2{\cal F}(V)} e^{2\bm{V}} \Big]^i
+ \frac{1}{2} \lambda_0^{imn} \bm{g}\, \phi_m \phi_n.
\end{equation}

\noindent
Then the first term of Eq. (\ref{First_Term}) can be rewritten as

\begin{eqnarray}
&&\hspace*{-7mm} \frac{\delta S_{\mbox{\scriptsize matter}} }{\delta \phi_i} \big(T^A\big)_i{}^j \frac{D^2}{4\partial^2} \phi_j - \lambda_0^{imn} \bm{g}\, \phi_m \phi_n \big(T^A\big)_i{}^j \frac{D^2}{8\partial^2} \phi_j\\
&&\hspace*{-7mm} = \frac{\delta S_{\mbox{\scriptsize matter}} }{\delta \phi_i} \big(T^A\big)_i{}^j \frac{D^2}{4\partial^2} \phi_j + \frac{2}{3} \lambda_0^{imn} \big(T^A\big)_i{}^j {\bar D}^{\dot a} \Big(\bm{g}\,\phi_m\, \frac{\bar D_{\dot a} D^2}{16\partial^2} \phi_n\,  \frac{D^2}{8\partial^2} \phi_j - \bm{g}\,\phi_m\, \frac{D^2}{16\partial^2} \phi_n\, \frac{\bar D_{\dot a} D^2}{8\partial^2} \phi_j   \Big).\nonumber
\end{eqnarray}

\noindent
Again, the last equality can be verified with the help of the product rule for the supersymmetric covariant derivative $\bar D^{\dot a}$ using Eqs. (\ref{Yukawa_Gauge_Invariance}) and (\ref{Chiral_Identity}).

Thus, the considered expression (\ref{S_Matter_V_Derivative}) takes the form

\begin{eqnarray}
&&\hspace*{-6mm} \frac{\delta S_{\mbox{\scriptsize matter}} }{\delta \bm{V}^A} = \frac{\delta S_{\mbox{\scriptsize matter}} }{\delta \phi_i} \big(T^A\big)_i{}^j \frac{D^2}{4\partial^2} \phi_j + \bar D^{\dot a}\bigg\{\bar D_{\dot a}\Big[F\Big(-\frac{\nabla^2 \bar\nabla^2}{16\Lambda^2}\Big) \phi^{+} \cdot e^{2{\cal F}(V)} e^{2\bm{V}}\Big]^i \big(T^A\big)_i{}^j \frac{D^2}{32\partial^2} \phi_j \nonumber\\
&&\hspace*{-6mm} -  \Big[F\Big(-\frac{\nabla^2 \bar\nabla^2}{16\Lambda^2}\Big) \phi^{+}\cdot e^{2{\cal F}(V)} e^{2\bm{V}}\Big]^i \big(T^A\big)_i{}^j \frac{\bar D_{\dot a} D^2}{32\partial^2} \phi_j
+ \frac{1}{2} \sum\limits_{\alpha=1}^\infty \sum\limits_{\beta=0}^{\alpha-1} f_\alpha \Big[\bar\nabla_{\dot a}\Big(-\frac{\nabla^2 \bar\nabla^2}{16\Lambda^2}\Big)^\beta  \phi^{+}\Big]^i \Big(e^{2{\cal F}(V)}\nonumber\\
&&\hspace*{-6mm} \times T^A e^{-2{\cal F}(V)}\Big)_i{}^j \Big[\frac{\nabla^2}{16\Lambda^2}\Big(-\frac{\bar\nabla^2 \nabla^2}{16\Lambda^2}\Big)^{\alpha-\beta-1}e^{2{\cal F}(V)} e^{2\bm{V}}\phi\Big]_j
- \frac{1}{2} \sum\limits_{\alpha=1}^\infty \sum\limits_{\beta=0}^{\alpha-1} f_\alpha \Big[\Big(-\frac{\nabla^2 \bar\nabla^2}{16\Lambda^2}\Big)^\beta \phi^{+}\Big]^i  \Big(e^{2{\cal F}(V)} \nonumber\\
&&\hspace*{-6mm} \times T^A e^{-2{\cal F}(V)}\Big)_i{}^j \Big[\frac{\bar\nabla_{\dot a}\nabla^2}{16\Lambda^2}\Big(-\frac{\bar\nabla^2 \nabla^2}{16\Lambda^2}\Big)^{\alpha-\beta-1}e^{2{\cal F}(V)} e^{2\bm{V}}\phi\Big]_j
+ \frac{2}{3} \lambda_0^{imn} \big(T^A\big)_i{}^j \Big(\bm{g}\,\phi_m\, \frac{\bar D_{\dot a} D^2}{16\partial^2} \phi_n\, \frac{D^2}{8\partial^2} \phi_j
\nonumber\\
&&\hspace*{-6mm} - \bm{g}\,\phi_m\, \frac{D^2}{16\partial^2} \phi_n\, \frac{\bar D_{\dot a} D^2}{8\partial^2} \phi_j   \Big)\bigg\}
+ \mbox{terms containing $[\bm{V}, T^A]$}.
\end{eqnarray}

\noindent
This equation can be equivalently presented in the form (\ref{S_Matter_V_Derivative_Brief}) in which $\big(\bar X^A_{\dot a}\big)_{\mbox{\scriptsize WZ}}$, $\big(\bar X^A_{\dot a}\big)_{\mbox{\scriptsize HD}}$, and $\big(\bar X^A_{\dot a}\big)_{\mbox{\scriptsize Yukawa}}$ are given by Eqs. (\ref{X_WZ}), (\ref{X_HD}), and (\ref{X_Yukawa}), respectively.

\subsection{The Faddeev--Popov ghost action}
\hspace*{\parindent}\label{Appendix_X_FP}

The derivative of the Faddeev--Popov ghost action (\ref{Ghosts_Faddeev-Popov}) with respect to the background gauge superfield can be written as

\begin{eqnarray}\label{S_FP_V_Derivative_Original}
&&\hspace*{-6mm} \ \frac{\delta S_{\mbox{\scriptsize FP}} }{\delta \bm{V}^A} =  \frac{\partial {\cal F}^{-1}(\widetilde V)^D}{\partial {\widetilde V}^B}\left.\vphantom{\frac{1}{2}}\right|_{\widetilde V = {\cal F}(V)} \left(\left(e^{2\bm{V}} T^A\right)_{Adj} \bar c\right)^D \left\{\vphantom{\frac{1}{2}}\smash{
\Big(\frac{{\cal F}(V)}{1-e^{2{\cal F}(V)}}\Big)_{Adj} c^+
+ \Big(\frac{{\cal F}(V)}{1-e^{-2{\cal F}(V)}}\Big)_{Adj}}\right.
\nonumber\\
&&\hspace*{-6mm} \left.\vphantom{\frac{1}{2}}\times \left(e^{2\bm{V}}\right)_{Adj} c \right\}^B
+ \frac{\partial {\cal F}^{-1}(\widetilde V)^D}{\partial {\widetilde V}^B}\left.\vphantom{\frac{1}{2}}\right|_{\widetilde V = {\cal F}(V)} \left(e^{2\bm{V}}\bar c\, e^{-2\bm{V}} +
\bar c^+ \right)^D \left\{\vphantom{\frac{1}{2}}\smash{
\Big(\frac{{\cal F}(V)}{1-e^{-2{\cal F}(V)}} e^{2\bm{V}} T^A \Big)_{Adj} c}\right\}^B\nonumber\\
&&\hspace*{-6mm}\ +\mbox{terms containing $[\bm{V}, T^A]_{Adj}$}.\vphantom{\frac{1}{2}}\qquad
\end{eqnarray}

\noindent
To present it in a more convenient form, we introduce

\begin{eqnarray}\label{X_FP}
&&\hspace*{-5mm} \big(\bar X_{\dot a}^A\big)_{\mbox{\scriptsize FP}} = -\frac{D^2}{16\partial^2} \bar c^E \, \bar D_{\dot a} \Bigg[\Big(\big(T^A e^{-2\bm{V}}\big)_{Adj} \Big)^{ED} \frac{\partial {\cal F}^{-1}(\widetilde V)^D}{\partial {\widetilde V}^B}\left.\vphantom{\frac{1}{2}}\right|_{\widetilde V = {\cal F}(V)} \left\{\vphantom{\frac{1}{2}}\smash{
\Big(\frac{{\cal F}(V)}{1-e^{2{\cal F}(V)}}\Big)_{Adj} c^+}
\right.\nonumber\\
\nonumber\\
&&\hspace*{-5mm} \left.+ \Big(\frac{{\cal F}(V)}{1-e^{-2{\cal F}(V)}}\Big)_{Adj} \left(e^{2\bm{V}}\right)_{Adj} c \right\}^B\Bigg] - \frac{\bar D_{\dot a} D^2}{16\partial^2} \bar c^E\, \Big(\big(T^A e^{-2\bm{V}}\big)_{Adj} \Big)^{ED}\, \frac{\partial {\cal F}^{-1}(\widetilde V)^D}{\partial {\widetilde V}^B}\left.\vphantom{\frac{1}{2}}\right|_{\widetilde V = {\cal F}(V)} \nonumber\\
&&\hspace*{-5mm} \times \left\{\vphantom{\frac{1}{2}}
\Big(\frac{{\cal F}(V)}{1-e^{2{\cal F}(V)}}\Big)_{Adj} c^+
+ \Big(\frac{{\cal F}(V)}{1-e^{-2{\cal F}(V)}}\Big)_{Adj} \left(e^{2\bm{V}}\right)_{Adj} c \right\}^B
\nonumber\\
&&\hspace*{-5mm} - \bar D_{\dot a}\Bigg[\frac{\partial {\cal F}^{-1}(\widetilde V)^D}{\partial {\widetilde V}^B}\left.\vphantom{\frac{1}{2}}\right|_{\widetilde V = {\cal F}(V)} \left(e^{2\bm{V}}\bar c\, e^{-2\bm{V}} +
\bar c^+ \right)^D \left\{\vphantom{\frac{1}{2}}\smash{
\Big(\frac{{\cal F}(V)}{1-e^{-2{\cal F}(V)}} e^{2\bm{V}} T^A \Big)_{Adj}}\right\}^{BE}\Bigg] \frac{D^2}{16\partial^2} c^E\qquad\nonumber\\
&&\hspace*{-5mm} - \frac{\partial {\cal F}^{-1}(\widetilde V)^D}{\partial {\widetilde V}^B}\left.\vphantom{\frac{1}{2}}\right|_{\widetilde V = {\cal F}(V)} \left(e^{2\bm{V}}\bar c\, e^{-2\bm{V}} +
\bar c^+ \right)^D \left\{\vphantom{\frac{1}{2}}\smash{
\Big(\frac{{\cal F}(V)}{1-e^{-2{\cal F}(V)}} e^{2\bm{V}} T^A \Big)_{Adj} \, \frac{\bar D_{\dot a} D^2}{16\partial^2} c}\right\}^B,
\end{eqnarray}

\noindent
where $f(X)_{Adj} \big(t^A Y^A\big) \equiv \big[f(X)_{Adj}\big]_{AB} Y^B t^A$. Applying the operator $\bar D^{\dot a}$ to the expression (\ref{X_FP}) after some transformations involving the product rule for the supersymmetric covariant derivative and the identity (\ref{Chiral_Identity}), we obtain
the equation

\begin{eqnarray}
&& \bar D^{\dot a} \big(\bar X^A_{\dot a}\big)_{\mbox{\scriptsize FP}} = -\frac{\delta S_{\mbox{\scriptsize FP}}}{\delta \bm{V}^A} - \frac{\delta S_{\mbox{\scriptsize FP}}}{\delta \bar c^B}\, \big(T^A_{Adj}\big)_{BD}\, \frac{D^2}{4\partial^2} \bar c^D - \frac{\delta S_{\mbox{\scriptsize FP}}}{\delta c^B}\, \big(T^A_{Adj}\big)_{BD}\, \frac{D^2}{4\partial^2} c^D \qquad\nonumber\\
&& +\mbox{terms containing $[\bm{V}, T^A]_{Adj}$}, \vphantom{\frac{1}{2}}\qquad
\end{eqnarray}

\noindent
which is equivalent to Eq. (\ref{S_FP_V_Derivative_Brief}).

\section{The formal calculation of the expression (\ref{Matter_Derivative_Variation})}
\hspace*{\parindent}\label{Appendix_X_Matter_Variation}

In this section we will formally calculate the expression (\ref{Matter_Derivative_Variation})

\begin{equation}\label{We_Variate}
\frac{\partial}{\partial a_\mu^A} \int d^8x\, (\gamma_\mu)^{\dot a b} \theta_b\, \delta_a \Big\langle (\bar X_{\dot a}^A)_{\mbox{\scriptsize WZ}} + (\bar X_{\dot a}^A)_{\mbox{\scriptsize HD}} + (\bar X_{\dot a}^A)_{\mbox{\scriptsize Yukawa}}  \Big)\Big\rangle\bigg|_{\mbox{\scriptsize fields=0}},
\end{equation}

\noindent
where $\delta_a$ denotes the variation under the background gauge transformations (\ref{Infinitesimal_Background_Gauge_Transformations}) with the parameter $A = i a_\mu^A T^A y^\mu = i a_\mu^A T^A (x^\mu + i\bar\theta^{\dot a}(\gamma^\mu)_{\dot a}{}^b \theta_b)$. The expressions $(\bar X^A_{\dot a})_{\mbox{\scriptsize WZ}}$, $(\bar X_{\dot a}^A)_{\mbox{\scriptsize HD}}$, and $(\bar X^A_{\dot a})_{\mbox{\scriptsize Yukawa}}$ entering Eq. (\ref{Matter_Derivative_Variation}) have been found in Appendix \ref{Appendix_X_Matter} and are given by Eqs. (\ref{X_WZ}), (\ref{X_HD}), and (\ref{X_Yukawa}), respectively. The fields (including the background gauge superfield $\bm{V}$) are set to 0 only after calculating the variation $\delta_a$. The word ``formally'' means that we will ignore possible singularities similar the one in Eq. (\ref{Identity_With_Delta}). According to the general argumentation of Ref. \cite{Stepanyantz:2019ihw}, the result of the formal calculation should be equal to 0. Nevertheless, here we will verify this by the explicit calculation.

The variations of various {\it quantum} superfields $\varphi_I$ given by Eq. (\ref{Infinitesimal_Background_Gauge_Transformations}) can schematically be written as $\delta_a \varphi_I = M_I{}^J \varphi_J$, where $M_I{}^J$ are certain functions of the superspace coordinates. The quantum superfields $\varphi_I$ are the arguments of the effective action related to the sources in the standard way, $\varphi_I = \delta W/\delta j^I$. The corresponding transformation of the sources $j^I$ is written as $\delta_a j^I = - M_J{}^I j^J$. This change of the sources can be compensated by the change of the integration variables in the generating functional $\delta_a\varphi_I = M_I{}^J \varphi_J$. This implies that for a certain function(al) $B(\varphi_I)$

\begin{equation}\label{VEV_Variation}
\delta_a \langle B \rangle =  \langle \delta_a B \rangle.
\end{equation}

\noindent
(Certainly, for deriving this equation it is necessary to take into account the background gauge invariance of the total action and of the Pauli--Villars determinants.)

Now us proceed to calculating the expression (\ref{We_Variate}) with the help of Eq. (\ref{VEV_Variation}). First we note that $(\bar X_{\dot a}^A)_{\mbox{\scriptsize HD}}$ includes only the supersymmetric covariant derivatives (\ref{Covariant_Derivative_Definition}), so that its variation under the considered transformation can easily be calculated. Although the result is rather lengthy, it contains

\begin{equation}\label{Variation_HD_Final}
\frac{\partial}{\partial a_\mu^A} [A^+,T^A] = \frac{\partial}{\partial a_\mu^A} \Big[-i a_\nu^B T^B (y^\nu)^*, T^A\Big] = -i (y^\mu)^* [T^A, T^A] = 0
\end{equation}

\noindent
and, therefore, vanishes,

\begin{equation}\label{Result_For_HD}
\frac{\partial}{\partial a_\mu^A} \int d^8x\, (\gamma_\mu)^{\dot a b} \theta_b\, \Big\langle \delta_a (\bar X_{\dot a}^A)_{\mbox{\scriptsize HD}}\Big\rangle\bigg|_{\mbox{\scriptsize fields = 0}} = 0.
\end{equation}

The remaining terms in Eq. (\ref{We_Variate}) include the expression $D^2/\partial^2$. That is why for calculating them one should use the (formal) identity

\begin{equation}\label{Auxiliary_Commutator_Identity}
\Big[\frac{D^2}{\partial^2},\, y^\mu \Big] = \Big\{\bar D^{\dot b},\, (\gamma^\nu\gamma^\mu)_{\dot b}{}^{\dot a} \bar\theta_{\dot a}\, \frac{D^2 \partial_\nu}{\partial^4}\Big\},
\end{equation}

\noindent
which can be verified with the help of the equations $\{\bar D_{\dot b},\,\bar\theta^{\dot a}\} = \delta^{\dot a}_{\dot b}$,\, $[D_a,\,y^\mu] = -2i (\gamma^\mu)_a{}^{\dot b}\bar\theta_{\dot b}$, and $\{\bar D_{\dot a},\, D_b\} = 2i(\gamma^\mu)_{\dot a b} \partial_\mu$.

We will start with calculating the variation of the first term in the expression (\ref{We_Variate}). After integrating by parts with respect to the supersymmetric covariant derivative $\bar D_{\dot a}$ it can be presented in the form

\begin{eqnarray}\label{WZ_Variation}
&&\hspace*{-7mm}  \frac{\partial}{\partial a_\mu^A}  \delta_a \int d^8x\, (\gamma_\mu)^{\dot a b} \theta_b \Big\langle (\bar X_{\dot a}^A)_{\mbox{\scriptsize WZ}} \Big\rangle\bigg|_{\mbox{\scriptsize fields = 0}} = \frac{\partial}{\partial a_\mu^A}\,  \delta_a \int d^8x\, (\gamma_\mu)^{\dot a b} \theta_b \nonumber\\
&&\hspace*{-7mm}\qquad\qquad\qquad\qquad\quad \times \Big\langle \Big[F\Big(-\frac{\nabla^2 \bar\nabla^2}{16\Lambda^2}\Big) \phi^{+} \cdot e^{2{\cal F}(V)} e^{2\bm{V}}\Big]^i  (T^A)_i{}^j \frac{\bar D_{\dot a} D^2}{16\partial^2} \phi_j \Big\rangle\bigg|_{\mbox{\scriptsize fields = 0}}.\qquad
\end{eqnarray}

\noindent
Then in this expression we perform the transformation (\ref{Infinitesimal_Background_Gauge_Transformations}) with the parameter (\ref{Parameters}) and apply the identity (\ref{Auxiliary_Commutator_Identity}). Using the well-known equation $\gamma^\mu \gamma^\nu \gamma_\mu = - 2\gamma^\nu$, after some transformations Eq. (\ref{WZ_Variation}) can be rewritten as

\begin{equation}\label{Varition_WZ_Final}
i \int d^8x\, \bar\theta \gamma^\mu \theta\, \Big\langle \bar D^2 \Big[F\Big(-\frac{\nabla^2 \bar\nabla^2}{16\Lambda^2}\Big) \phi^{+} \cdot e^{2{\cal F}(V)} e^{2\bm{V}}\Big]^i C(R)_i{}^j  \frac{D^2\partial_\mu}{16\partial^4} \phi_j \Big\rangle\bigg|_{\mbox{\scriptsize fields = 0}}.
\end{equation}

Next, let us proceed to calculating the variation of the third term in Eq. (\ref{We_Variate}) which contains $(\bar X_{\dot a}^A)_{\mbox{\scriptsize Yukawa}}$ given by Eq. (\ref{X_Yukawa}). After integrating by parts with respect to the derivative $\bar D_{\dot a}$ the considered expression takes the form

\begin{eqnarray}\label{Yukawa_Variation}
&& \frac{\partial}{\partial a_\mu^A}  \delta_a \int d^8x\, (\gamma_\mu)^{\dot a b} \theta_b\, \Big\langle (\bar X_{\dot a}^A)_{\mbox{\scriptsize Yukawa}} \Big\rangle\bigg|_{\mbox{\scriptsize fields = 0}}\nonumber\\
&&\qquad\quad = - \frac{4}{3}\, \lambda_0^{ijk} (T^A)_k{}^m \frac{\partial}{\partial a_\mu^A}\,  \delta_a \int d^8x\, \bm{g}\,(\gamma_\mu)^{\dot a b} \theta_b\, \Big\langle  \frac{D^2}{8\partial^2} \phi_m\, \frac{\bar D_{\dot a} D^2}{16\partial^2} \phi_i\, \phi_j \Big\rangle\bigg|_{\mbox{\scriptsize fields = 0}}. \qquad
\end{eqnarray}

\noindent
As earlier, we make the infinitesimal background gauge transformation (\ref{Infinitesimal_Background_Gauge_Transformations}) with the parameter (\ref{Parameters}) taking into account Eq. (\ref{Auxiliary_Commutator_Identity}) and the identity $\bar D_{\dot a} \bar D^{\dot c} = -\delta_{\dot a}^{\dot c}\bar D^2/2$. With the help of Eq. (\ref{Yukawa_Gauge_Invariance}) the result can be written as

\begin{eqnarray}
&& \bigg(\frac{2i}{3} \lambda_0^{ijk} (T^A)_k{}^m (T^A)_i{}^n  \int d^8x\, \bm{g}\,(\gamma_\mu)^{\dot a b} \theta_b \Big\langle  \frac{D^2}{8\partial^2} \phi_m\, \bar D^2 \Big((\gamma^\nu\gamma^\mu)_{\dot a}{}^{\dot d}\bar\theta_{\dot d}\frac{D^2 \partial_\nu}{16\partial^4} \phi_n\Big)\, \phi_j \Big\rangle\qquad\nonumber\\
&& - \frac{4i}{3} \lambda_0^{ijk} C(R)_k{}^m  \int d^8x\,  \bm{g}\, (\gamma_\mu)^{\dot a b} \theta_b \Big\langle \bar D^{\dot c}\Big((\gamma^\nu\gamma^\mu)_{\dot c}{}^{\dot d} \bar\theta_{\dot d} \frac{D^2\partial_\nu}{8\partial^4} \phi_m\Big)\, \frac{\bar D_{\dot a} D^2}{16\partial^2} \phi_i\, \phi_j \Big\rangle\bigg)\bigg|_{\mbox{\scriptsize fields=0}}.\qquad
\end{eqnarray}

\noindent
Integrating by parts with respect to the derivative $\bar D^2$ in the first term and with respect to the derivative $\bar D^{\dot c}$ in the second term, using the equation $\gamma_\mu \gamma^\nu \gamma^\mu = -2\gamma^\nu$ and the identity (\ref{Chiral_Identity}), we present this expression in the form

\begin{equation}\label{Variation_Yukawa}
\frac{4i}{3} \lambda_0^{ijk}  \int d^8x\,  \bm{g}\, \bar\theta\gamma^\mu \theta \Big\{ (T^A)_k{}^m (T^A)_i{}^n \Big\langle \phi_m\, \frac{D^2 \partial_\mu}{8\partial^4} \phi_n\, \phi_j \Big\rangle - C(R)_k{}^m \Big\langle \frac{D^2\partial_\mu}{8\partial^4} \phi_m\, \phi_i\, \phi_j \Big\rangle\Big\}\bigg|_{\mbox{\scriptsize fields = 0}}.
\end{equation}

\noindent
The product $\phi_m \phi_j$ in the first term is evidently symmetric with respect to the permutation of the indices $m$ and $j$. This implies that in this term it is possible to make the replacement

\begin{eqnarray}
&& \lambda_0^{ijk} (T^A)_k{}^m (T^A)_i{}^n \to \frac{1}{2} \Big(\lambda_0^{ijk} (T^A)_k{}^m + \lambda_0^{imk} (T^A)_k{}^j \Big) (T^A)_i{}^n\nonumber\\
&&\qquad\qquad\qquad\qquad\qquad\qquad = -\frac{1}{2} \lambda_0^{jmk} (T^A)_k{}^i (T^A)_i{}^n = -\frac{1}{2} \lambda_0^{jmk} C(R)_k{}^n,\qquad
\end{eqnarray}

\noindent
where we took Eq. (\ref{Yukawa_Gauge_Invariance}) into account. After this replacement Eq. (\ref{Variation_Yukawa}) takes the form

\begin{equation}\label{Variation_Yukawa_Final}
- i \lambda_0^{ijk} C(R)_k{}^m \int d^8x\, \bm{g}\, \bar\theta\gamma^\mu \theta\, \Big\langle \frac{D^2\partial_\mu}{4\partial^4} \phi_m\, \phi_i\, \phi_j \Big\rangle\bigg|_{\mbox{\scriptsize fields = 0}}.
\end{equation}

Summing up Eqs. (\ref{Result_For_HD}), (\ref{Varition_WZ_Final}) and (\ref{Variation_Yukawa_Final}) we obtain

\begin{eqnarray}\label{Variation_Matter_Total}
&& \frac{\partial}{\partial a_\mu^A} \int d^8x\, (\gamma_\mu)^{\dot a b} \theta_b \Big\langle  \delta_a (\bar X_{\dot a}^A)_{\mbox{\scriptsize WZ}} + \delta_a (\bar X_{\dot a}^A)_{\mbox{\scriptsize HD}} + \delta_a (\bar X_{\dot a}^A)_{\mbox{\scriptsize Yukawa}}  \Big)\Big\rangle\bigg|_{\mbox{\scriptsize fields = 0}}\qquad\nonumber\\
&& = - i \int d^8x\, \bar\theta \gamma^\mu \theta\, C(R)_k{}^m \Big\langle \Big\{-\frac{1}{8} \bar D^2 \Big[F\Big(-\frac{\nabla^2 \bar\nabla^2}{16\Lambda^2}\Big) \phi^{+} \cdot e^{2{\cal F}(V)} e^{2\bm{V}}\Big]^k\qquad\nonumber\\
&&\qquad\qquad\qquad\qquad\qquad\qquad\qquad\qquad\qquad\qquad\quad + \frac{1}{2} \bm{g}\,\lambda_0^{ijk} \phi_i \phi_j  \Big\}\,  \frac{D^2\partial_\mu}{2\partial^4} \phi_m \Big\rangle\bigg|_{\mbox{\scriptsize fields = 0}}.\qquad\quad
\end{eqnarray}

\noindent
The expression in the curly brackets coincides with the derivative of the action $S_{\mbox{\scriptsize matter}}$ with respect to $\phi_k$. Therefore, with the help of the identity (\ref{Auxiliary_A_Identity}) it is possible to rewrite Eq. (\ref{Variation_Matter_Total}) in the form

\begin{eqnarray}
&& - i \int d^8x\, \bar\theta \gamma^\mu \theta\, \Big\langle \frac{\delta S_{\mbox{\scriptsize matter}}}{\delta \phi_k}\, C(R)_k{}^m  \frac{D^2\partial_\mu}{2\partial^4} \phi_m \Big\rangle\bigg|_{\mbox{\scriptsize fields = 0}}\nonumber\\
&&\qquad\qquad\qquad\qquad\qquad\qquad = - \int d^8x\, \bar\theta \gamma^\mu \theta\, C(R)_m{}^m  \frac{D^2 \bar D^2 \partial_\mu}{4\partial^4} \delta^8_{xy}\Big|_{y\to x} = 0.\qquad
\end{eqnarray}

\noindent
Note that the last expression vanishes, because (after the Wick rotation) it is proportional to

\begin{equation}
\int \frac{d^4Q}{(2\pi)^4} \frac{Q_\mu}{Q^4} = 0.
\end{equation}

Thus, we see that formally the variation (\ref{We_Variate}) vanishes. This agrees with the argumentation of Ref. \cite{Stepanyantz:2019ihw}, where it was demonstrated that the higher order corrections appear to be 0 if the calculations are made formally. Certainly, in fact, the higher order corrections do not vanish, and the correct calculation should also take singular contributions into account. For this purpose, it is also necessary to write down total derivatives with respect to the loop momenta which appear in calculating the variation of the expression (\ref{Matter_Derivative_Variation}).

\section{Total derivatives generating matter singular contributions to the $\beta$-function}
\label{Appendix_X_Matter_Total_Derivatives}

\subsection{A simple example illustrating how the total derivatives appear}
\hspace*{\parindent}\label{Appendix_Toy_Expression}

Formally calculating the variation $\delta_a$ of various expressions we in fact omit integrals of total derivatives. In this section we describe a simple example which illustrates how these total derivatives appear. For this purpose we consider the expression

\begin{equation}
\int d^8x\, \phi^{*i} \big(e^{2{\cal F}(V)} e^{2\bm{V}}\big)_i{}^j \phi_j.
\end{equation}

\noindent
It is evidently invariant under the background gauge transformations, but in calculating the variation $\delta_a$ we obtain some terms containing $x^\mu$,

\begin{eqnarray}\label{Variation_Toy_Model}
&& 0 = \delta_a\int d^8x\, \phi^{*i} \big(e^{2{\cal F}(V)} e^{2\bm{V}}\big)_i{}^j \phi_j = i a^A_\mu \int d^8x\, \Big(- x^\mu\phi^{*i} (T^A)_i{}^k \big(e^{2{\cal F}(V)} e^{2\bm{V}} \big)_k{}^j \phi_j  \qquad\nonumber\\
&& + \phi^{*i}  [T^A,\, x^\mu\, e^{2{\cal F}(V)} e^{2\bm{V}}]_i{}^j\phi_j + \phi^{*i} \big(e^{2{\cal F}(V)} e^{2\bm{V}}\big)_i{}^k (T^A)_k{}^j x^\mu \phi_j\Big).
\end{eqnarray}

\noindent
In the momentum representation the coordinate operator is proportional to the derivative with respect to the momentum. Taking this into account we see that the expression (\ref{Variation_Toy_Model}) is given by the sum of integrals of total derivatives,

\begin{eqnarray}\label{Variation_Example}
&& 0 = a^A_\mu \int d^4\theta \int \frac{d^4q}{(2\pi)^4} \frac{d^4p}{(2\pi)^4}
\left\{\frac{\partial}{\partial q_\mu} \Big(\phi^{*i}(-q-p,\theta) \left(e^{2{\cal F}(V)} e^{2\bm{V}}(p,\theta)\right)_i{}^k (T^A)_k{}^j \phi_j(q,\theta)\Big)\right.\qquad\nonumber\\
&&\left. + \frac{\partial}{\partial p_\mu} \Big(\phi^{*i}(-q-p,\theta) \left[T^A,\,e^{2{\cal F}(V)} e^{2\bm{V}}(p,\theta)\right]_i{}^j  \phi_j(q,\theta)\Big) \right\},
\end{eqnarray}

\noindent
where

\begin{equation}
e^{2{\cal F}(V)} e^{2\bm{V}}(p,\theta) = \int d^4x\, e^{ip_\alpha x^\alpha} e^{2{\cal F}(V)} e^{2\bm{V}}.
\end{equation}

\noindent
Note that using the notation  $\delta^8_{yx}(q)$ introduced in Eq. (\ref{Delta_Function}) the expression (\ref{Variation_Example}) can be equivalently presented in the form

\begin{eqnarray}
&& a^A_\mu \int \frac{d^4q}{(2\pi)^4} \frac{\partial}{\partial q_\mu} \int d^8x\, d^8y\, \delta^8_{yx}(q)\,
\Big[\phi^{*i} \big(e^{2{\cal F}(V)} e^{2\bm{V}}\big)_i{}^k\Big]_x (T^A)_k{}^j \phi_{j,y}\qquad\nonumber\\
&& + a_\mu^A \int \frac{d^4p}{(2\pi)^4} \frac{\partial}{\partial p_\mu} \int d^8x\, d^8y\, \delta^8_{xy}(p)\, \phi^{*i}_{\ ,y}
\Big[T^A,\, \big(e^{2{\cal F}(V)} e^{2\bm{V}}\big)_x\Big]_i{}^j \phi_{j,y}.\qquad
\end{eqnarray}

\subsection{Terms quadratic in the matter superfields}
\hspace*{\parindent}

In this paper we are interested in the singular contributions corresponding to the matter superfields. This means that they are produced by total derivatives acting on $Q_\mu/Q^4$, where $Q_\mu$ is the (Euclidian) momentum of a matter propagator. The matter propagator proportional to

\begin{equation}
\frac{\bar D^2 D^2}{4\partial^2} \delta^8_{xy}
\end{equation}

\noindent
can give only $Q^{-2}$. Moreover, the product of two such propagators is also proportional to $Q^{-2}$ due to Eq. (\ref{Chiral_Identity}). That is why the terms proportional to $Q_\mu/Q^4$ (where $Q_\mu$ is a momentum of the {\it matter} superfield) can appear only if the operator $(\gamma^\mu)^{\dot a b}\theta_{b} \bar D_{\dot a} D^2/\partial^2$ acts on the matter propagator (from the right) or stands between two matter propagators with the same momenta,\footnote{Note that such propagators can appear when a supergraph can be made disconnected by two cuts of the matter line. Certainly, in this case these propagators are separated by a subdiagram (or subdiagrams) contributing to the two-point Green function of the matter superfields.}

\begin{equation}\label{Origin_Of_Singularities}
\frac{\bar D^2 D^2}{4\partial^2}\cdot (\gamma^\mu)^{\dot a b}\theta_{b}\, \frac{\bar D_{\dot a} D^2}{\partial^2}\cdot \frac{\bar D^2 D^2}{4\partial^2} = 8i\, \frac{\bar D^2 D^2 \partial_\mu}{\partial^4}.
\end{equation}

\noindent
This implies that the total derivatives coming from the variation of the term containing $(\bar X_{\dot a}^A)_{\mbox{\scriptsize HD}}$ in the expression (\ref{Matter_Derivative_Variation}) do not contain singular contributions of the considered type (i.e. corresponding to the cuts of the matter propagators) and can be excluded from the consideration. (Note that singularities corresponding to the cuts of the quantum gauge superfield propagators are not considered in this paper.)

Next, we consider the term in Eq. (\ref{Matter_Derivative_Variation}) which contains $(\bar X_{\dot a}^A)_{\mbox{\scriptsize WZ}}$ given by Eq. (\ref{X_WZ}). This term is similar to the toy expression considered in the previous section. That is why, repeating the calculations of Sect. \ref{Appendix_Toy_Expression} we obtain that the matter singularities come from

\begin{eqnarray}\label{WZ_Total_Derivative_Result}
&& \frac{\partial}{\partial a_\mu^A}  \delta_a \int d^8x\, (\gamma_\mu)^{\dot a b} \theta_b\, \Big\langle (\bar X_{\dot a}^A)_{\mbox{\scriptsize WZ}} \Big\rangle
\ \to\  \int \frac{d^4q}{(2\pi)^4} \frac{\partial}{\partial q^\mu} \int d^8x\, d^8y\, \delta^8_{yx}(q)  (\gamma^\mu)^{\dot a b} \theta_b \qquad\nonumber\\
&& \times \Big\langle \Big[\Big(F\Big(-\frac{\nabla^2 \bar\nabla^2}{16\Lambda^2}\Big) \phi^{+} \cdot  e^{2{\cal F}(V)} e^{2\bm{V}}\Big)^i\,\Big]_x C\big(R\big)_i{}^j  \Big(\frac{\bar D_{\dot a} D^2}{16\partial^2} \phi_j\Big)_y \Big\rangle\bigg|_{\mbox{\scriptsize fields = 0}},
\end{eqnarray}

\noindent
where the arrow points that we consider only terms containing the derivative with respect to the momentum of the matter superfield $\phi_j$.

\subsection{Total derivatives induced by the Yukawa interaction}
\hspace*{\parindent}

Now, let us find the total derivative terms corresponding to the Yukawa interaction. They are obtained from the third term in Eq. (\ref{Matter_Derivative_Variation}), where $(\bar X_{\dot a}^A)_{\mbox{\scriptsize Yukawa}}$ is given by Eq. (\ref{X_Yukawa}), after making the infinitesimal transformation (\ref{Infinitesimal_Background_Gauge_Transformations}) with the parameter (\ref{Parameters}). In this section we will consider only total derivative contributions, because all other terms have already been calculated in Sect. \ref{Appendix_X_Matter_Variation}. Making the transformations similar to the ones described in Sect. \ref{Appendix_Toy_Expression} and using Eq. (\ref{Yukawa_Gauge_Invariance}) we obtain

\begin{eqnarray}\label{Yukawa_Original}
&&\hspace*{-7mm} \frac{\partial}{\partial a_\mu^A}  \delta_a \int d^8x\, (\gamma_\mu)^{\dot a b} \theta_b\, \Big\langle (\bar X_{\dot a}^A)_{\mbox{\scriptsize Yukawa}} \Big\rangle\ \to\  - \frac{4}{3} \lambda_0^{ijk} C(R)_k{}^m \int \frac{d^4q}{(2\pi)^4} \frac{d^4k}{(2\pi)^4} \frac{\partial}{\partial k^\mu} \int d^8x\,d^8y\,d^8z\, \nonumber\\
&&\hspace*{-7mm} \times \bm{g}_y\, \delta^8_{yz}(q) \delta^8_{xz}(k) (\gamma^\mu)^{\dot a b} \theta_b\, \Big\langle  \Big(\frac{D^2}{8\partial^2} \phi_m\Big)_x\, \Big(\frac{\bar D_{\dot a} D^2}{16\partial^2} \phi_i\Big)_y\, \phi_{j,z} \Big\rangle - \frac{4}{3} \lambda_0^{ijk} (T^A)_k{}^m (T^A)_i{}^n \int \frac{d^4q}{(2\pi)^4} \frac{d^4k}{(2\pi)^4} \nonumber\\
&&\hspace*{-7mm} \times  \frac{\partial}{\partial q^\mu} \int d^8x\,d^8y\,d^8z\, \bm{g}_y\,\delta^8_{yz}(q) \delta^8_{xz}(k) (\gamma^\mu)^{\dot a b} \theta_b\, \Big\langle  \Big(\frac{D^2}{8\partial^2} \phi_m\Big)_x\, \Big(\frac{\bar D_{\dot a} D^2}{16\partial^2} \phi_n\Big)_y\, \phi_{j,z} \Big\rangle\bigg|_{\mbox{\scriptsize fields = 0}}.
\end{eqnarray}

\noindent
Note that due to the presence of

\begin{equation}
\delta^8_{yz}(q)\, \delta^8_{xz}(k) = \delta^4(\theta_y-\theta_z) \delta^4(\theta_x-\theta_z) \exp\big(iq_\alpha y^\alpha\big) \exp\big(ik_\alpha x^\alpha\big) \exp\big(-i(q_\alpha+k_\alpha) z^\alpha \big)
\end{equation}

\noindent
the momenta of the superfields $\phi_{m,x}$, $\big(\bm{g} \bar D_{\dot a} D^2/(16\partial^2) \phi_i\big)_y$, and $\phi_{j,z}$ are $k_\mu$, $q_\mu$, and $-k_\mu-q_\mu$, respectively. Integrating by parts with respect to the derivative $\bar D_{\dot a}$, using Eq. (\ref{Chiral_Identity}) and the identity $\bar D_{\dot a} \bar D^{\dot c}  = - \delta_{\dot a}^{\dot c} \bar D^2/2$, it is possible to present the expression under consideration in the form

\begin{eqnarray}\label{Yukawa_Total_Derivative}
&&\hspace*{-5mm} \frac{4}{3} \lambda_0^{ijk} C(R)_k{}^m \int \frac{d^4q}{(2\pi)^4} \frac{d^4k}{(2\pi)^4} \frac{\partial}{\partial k^\mu} \int d^8x\,d^8y\,d^8z\, \bm{g}_y\, \delta^8_{yz}(q)\, \delta^8_{xz}(k) (\gamma^\mu)^{\dot a b} \theta_b\, \Big\langle  \Big(\frac{\bar D_{\dot a} D^2}{8\partial^2} \phi_m\Big)_x\,\nonumber\\
&&\hspace*{-5mm} \times  \Big(\frac{D^2}{16\partial^2} \phi_i\Big)_y\, \phi_{j,z} \Big\rangle + \frac{4}{3} \lambda_0^{ijk} (T^A)_k{}^m (T^A)_i{}^n \int \frac{d^4q}{(2\pi)^4} \frac{d^4k}{(2\pi)^4} \frac{\partial}{\partial q^\mu} \int d^8x\,d^8y\,d^8z\, \bm{g}_y\, \delta^8_{yz}(q)\, \delta^8_{xz}(k)\nonumber\\
&&\hspace*{-5mm}\times (\gamma^\mu)^{\dot a b} \theta_b\, \Big\langle  \Big(\frac{\bar D_{\dot c} D^2}{8\partial^2} \phi_m\Big)_x\, \Big(\frac{D^2}{16\partial^2} \phi_n\Big)_y\, \Big( \frac{\bar D_{\dot a} \bar D^{\dot c} D^2}{8\partial^2}\phi_j\Big)_z \Big\rangle\bigg|_{\mbox{\scriptsize fields = 0}}.
\end{eqnarray}

\noindent
In the second term only a part of $\lambda_0^{ijk} (T^A)_k{}^m (T^A)_i{}^n$ symmetric in $mj$ survives. Really, let us present the integral in the second term as a sum of two equal expressions. We leave the first one unchanged, while in the second one we integrate by parts with respect to the derivative $\bar D_{\dot a}$. Then the integral in the second term of Eq. (\ref{Yukawa_Total_Derivative}) will be written as

\begin{eqnarray}\label{Second_Term_Integral}
&&\hspace*{-4mm} \int \frac{d^4q}{(2\pi)^4} \frac{d^4k}{(2\pi)^4} \frac{\partial}{\partial q^\mu} \int d^8x\,d^8y\,d^8z\, \bm{g}_y\, \delta^8_{yz}(q)\, \delta^8_{xz}(k) (\gamma^\mu)^{\dot a b} \theta_b\, \Big\langle  \Big(\frac{\bar D_{\dot c} D^2}{8\partial^2} \phi_m\Big)_x\, \Big(\frac{D^2}{16\partial^2} \phi_n\Big)_y\,\nonumber\\
&&\hspace*{-4mm} \times \Big(\frac{\bar D_{\dot a} \bar D^{\dot c} D^2}{16\partial^2}\phi_j\Big)_z + \Big(\frac{\bar D_{\dot a}  \bar D_{\dot c} D^2}{8\partial^2} \phi_m\Big)_x\, \Big(\frac{D^2}{16\partial^2} \phi_n\Big)_y\, \Big(\frac{\bar D^{\dot c} D^2}{16\partial^2}\phi_j\Big)_z - \Big(\frac{\bar D_{\dot c} D^2}{8\partial^2} \phi_m\Big)_x\, \Big(\frac{\bar D_{\dot a} D^2}{16\partial^2} \phi_n\Big)_y\, \nonumber\\
&&\hspace*{-4mm} \times \Big(\frac{\bar D^{\dot c} D^2}{16\partial^2}\phi_j\Big)_z \Big\rangle.
\end{eqnarray}

\noindent
Let us note that the sequence of two changes of the integrations variables $k_\mu\to k_\mu-q_\mu$; $k_\mu\to - k_\mu$ is equivalent to the permutation of the points $x$ and $z$ in the expression inside the angular brackets. This implies that the expression (\ref{Second_Term_Integral}) is symmetric in $mj$, because in our notation

\begin{equation}
\bar\psi^{\dot a} \bar\xi_{\dot a} = - \bar\xi_{\dot a} \bar\psi^{\dot a} = \bar\xi^{\dot a} \bar\psi_{\dot a}
\end{equation}

\noindent
for any left anticommuting spinors $\bar\psi$ and $\bar\xi$. Therefore, in Eq. (\ref{Yukawa_Total_Derivative}) $\lambda_0^{ijk} (T^A)_k{}^m (T^A)_i{}^n$ can be replaced by its symmetric part

\begin{equation}
\frac{1}{2}\Big(\lambda_0^{ijk} (T^A)_k{}^m + \lambda_0^{imk} (T^A)_k{}^j\Big)(T^A)_i{}^n = - \frac{1}{2}\lambda_0^{jmk} (T^A)_k{}^i (T^A)_i{}^n = - \frac{1}{2}\lambda_0^{jmk} C(R)_k{}^n,
\end{equation}

\noindent
where we took Eq. (\ref{Yukawa_Gauge_Invariance}) into account. Then the expression (\ref{Yukawa_Total_Derivative}) takes the form

\begin{eqnarray}
&& \frac{4}{3} \lambda_0^{ijk} C(R)_k{}^m \int \frac{d^4q}{(2\pi)^4} \frac{d^4k}{(2\pi)^4} \frac{\partial}{\partial k^\mu} \int d^8x\,d^8y\,d^8z\,\bm{g}_y\, \delta^8_{yz}(q) \delta^8_{xz}(k) (\gamma^\mu)^{\dot a b} \theta_b \Big\langle  \Big(\frac{\bar D_{\dot a} D^2}{8\partial^2} \phi_m\Big)_x\,\qquad\nonumber\\
&& \times  \Big(\frac{D^2}{16\partial^2} \phi_i\Big)_y\, \phi_{j,z} \Big\rangle\bigg|_{\mbox{\scriptsize fields = 0}} - \frac{2}{3} \lambda_0^{jmk} C(R)_k{}^n \int \frac{d^4q}{(2\pi)^4} \frac{d^4k}{(2\pi)^4} \frac{\partial}{\partial q^\mu} \int d^8x\,d^8y\,d^8z\, \bm{g}_y\,\delta^8_{yz}(q) \qquad\nonumber\\
&&\times  \delta^8_{xz}(k) (\gamma^\mu)^{\dot a b} \theta_b\, \Big\langle  \Big(\frac{\bar D_{\dot a} D^2}{8\partial^2} \phi_m\Big)_x\, \Big(\frac{D^2}{16\partial^2} \phi_n\Big)_y\, \phi_{j,z} \Big\rangle\bigg|_{\mbox{\scriptsize fields = 0}}.\qquad
\end{eqnarray}

\noindent
It is convenient to rename the integration variables in the first term, $q_\mu \leftrightarrow k_\mu$ and $x \leftrightarrow y$. Next, we replace $\bm{g}_x$ by $\bm{g}_y$. This is possible, because due to the $\delta$-functions $\theta_x = \theta_y$, while the momentum of superfield $\bm{g}$ is negligibly small. Really, in the expression $\Delta_{\mbox{\scriptsize matter}}$ the derivative with respect to $\mbox{\sl g}$ is multiplied to $(v^B)^2$. This implies that the corresponding momentum is of the order $1/R\to 0$. Therefore, all terms containing space-time derivatives of $\bm{g}$ are suppressed by powers of $1/R$ and should be omitted.
After the procedure described above the considered expression will be written as

\begin{eqnarray}\label{Q_Derivative}
&& \lambda_0^{jmk} C(R)_k{}^n \int \frac{d^4q}{(2\pi)^4} \frac{d^4k}{(2\pi)^4} \frac{\partial}{\partial q^\mu} \int d^8x\,d^8y\,d^8z\,  \bm{g}_y\,\delta^8_{yz}(q)\, \delta^8_{xz}(k) \nonumber\\
&&\qquad\qquad\qquad\qquad\qquad\qquad \times\, (\gamma^\mu)^{\dot a b} \theta_b\, \Big\langle  \Big(\frac{D^2}{4\partial^2} \phi_m\Big)_x\, \Big(\frac{\bar D_{\dot a} D^2}{16\partial^2} \phi_n\Big)_y\, \phi_{j,z} \Big\rangle\bigg|_{\mbox{\scriptsize fields = 0}}.\qquad
\end{eqnarray}

\begin{figure}[ht]
\begin{picture}(0,4)
\put(2,0.6){\includegraphics[scale=0.25]{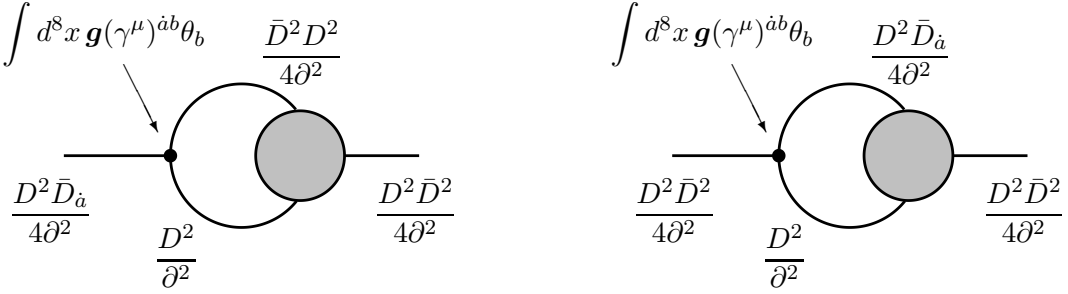}}
\put(1.2,3.1){${\displaystyle \int d^8x\,\bm{g} (\gamma^\mu)^{\dot a b} \theta_b}$}\put(2.8,2.8){\vector(1,-2){0.45}}
\put(1.3,0.7){${\displaystyle \frac{D^2 \bar D_{\dot a}}{4\partial^2}}$} \put(6.1,0.7){${\displaystyle \frac{D^2\bar D^2}{4\partial^2}}$}
\put(3.2,0.1){${\displaystyle \frac{D^2}{\partial^2}}$}  \put(4.6,2.8){${\displaystyle \frac{\bar D^2 D^2}{4\partial^2}}$}
\put(10,0.6){\includegraphics[scale=0.25]{yukawa_effective_diagram.eps}}
\put(9.2,3.1){${\displaystyle \int d^8x\,\bm{g} (\gamma^\mu)^{\dot a b} \theta_b}$}\put(10.8,2.8){\vector(1,-2){0.45}}
\put(9.5,0.7){${\displaystyle \frac{D^2\bar D^2}{4\partial^2}}$} \put(14.1,0.7){${\displaystyle \frac{D^2\bar D^2}{4\partial^2}}$}
\put(11.2,0.1){${\displaystyle \frac{D^2}{\partial^2}}$} \put(12.6,2.8){${\displaystyle \frac{D^2 \bar D_{\dot a}}{4\partial^2}}$}
\end{picture}
\caption{This figure illustrates how singularities can appear in the case of the existence of coinciding momenta. In the left diagram the external momentum is $q_\mu$, while in the right one the external momentum is $-(q+k)_\mu$. This implies that they can give terms proportional to $q_\mu/q^4$ and $(q+k)_\mu/(q+k)^4$, respectively.}\label{Figure_Yukawa_Effective_Diagram}
\end{figure}

\begin{figure}[ht]
\begin{picture}(0,4)
\put(6,0.6){\includegraphics[scale=0.25]{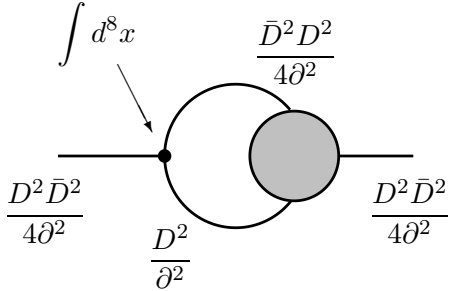}}
\put(6.0,3.1){${\displaystyle \int d^8x}$}\put(6.8,2.8){\vector(1,-2){0.45}}
\put(5.3,0.7){${\displaystyle \frac{D^2 \bar D^2}{4\partial^2}}$} \put(10.1,0.7){${\displaystyle \frac{D^2\bar D^2}{4\partial^2}}$}
\put(7.2,0.1){${\displaystyle \frac{D^2}{\partial^2}}$}  \put(8.6,2.8){${\displaystyle \frac{\bar D^2 D^2}{4\partial^2}}$}
\end{picture}
\caption{Usual subdiagrams corresponding to the insertion of the two-point Green function of the matter superfields with a Yukawa vertex in the left point.}\label{Figure_Usual_Subdiagrams}
\end{figure}

We are interested in singular terms which nontrivially contribute to this integral. Such terms can be proportional either to $q^\mu/q^4$ or $(q+k)^\mu/(q+k)^4$, where $q^\mu$ corresponds to $\phi_{n,y}$ and $-(q+k)^\mu$ corresponds to $\phi_{j,z}$. They appear when a supergraph contains coinciding momenta $q_\mu$ or $(q+k)_\mu$, respectively. The structure of the corresponding subdiagrams is illustrated in Fig. \ref{Figure_Yukawa_Effective_Diagram}. The coinciding momenta correspond to external lines of the subdiagrams, and the grey disks stand for the sums of all relevant subgraphs. For comparison, in Fig. \ref{Figure_Usual_Subdiagrams} we present a usual subdiagram of the considered type which gives quantum corrections to the two-point Green function of the matter superfields and contains a Yukawa vertex in the left point. Evidently, the diagram in Fig. \ref{Figure_Usual_Subdiagrams} is proportional to

\begin{equation}
\frac{D^2 \bar D^2}{4\partial^2} \cdot \frac{D^2 \bar D^2}{4\partial^2} = - \frac{D^2 \bar D^2}{\partial^2}
\end{equation}

\noindent
and does not produce singularities.

Comparing the diagrams in Figs. \ref{Figure_Yukawa_Effective_Diagram} and \ref{Figure_Usual_Subdiagrams} we see that the left superdiagram in Fig. \ref{Figure_Yukawa_Effective_Diagram} (with the coinciding momenta $q_\mu$) is proportional to\footnote{The indices are different, because in general the left hand side is not proportional to $q^\mu$.}

\begin{equation}\label{Singularity_Of_Left_Supergraph}
\frac{D^2 \bar D_{\dot a}}{4\partial^2} \cdot \bm{g} (\gamma^\mu)^{\dot a b} \theta_b \cdot \frac{D^2 \bar D^2}{4\partial^2} \sim \frac{q^\alpha}{q^4}.
\end{equation}

\noindent
Note that, as we have already mentioned, the momentum of the superfield $\bm{g}$ is of the order $1/R$ and should be neglected, because the integrals of double total derivatives appear only for the vanishing external momentum $p$ corresponding to the limit $R\to\infty$. Hence, calculating the integrals one should always assume that the radius of the sphere $S^3_\varepsilon$ is much larger than $1/R$. That is why instead of $q^2 (q+p)^2$ in Eq. (\ref{Singularity_Of_Left_Supergraph}) we simply write $q^4$.

The right subdiagram in Fig. \ref{Figure_Yukawa_Effective_Diagram} corresponds to the coinciding momenta $-(q+k)_\mu$ and appears to be proportional to

\begin{equation}
\frac{D^2 \bar D^2}{4\partial^2} \cdot \bm{g} \theta_b D^b \cdot \frac{D^2 \bar D^2}{4\partial^2} = 0.
\end{equation}

\noindent
This implies that the singularities corresponding to the coinciding momenta $-(q+k)_\mu$ are absent. Therefore, all singular contributions will be obtained from the integral over the infinitely small sphere $S^3_\varepsilon$ surrounding the point $Q_\mu=0$. Taking this into account we calculate the integrals over $d^4k$ and $d^8z$ in Eq. (\ref{Q_Derivative}) and rewrite the expression (\ref{Yukawa_Original}) in the form

\begin{equation}\label{Yukawa_Total_Derivative_Result}
\lambda_0^{jmk} C(R)_k{}^n \int \frac{d^4q}{(2\pi)^4} \frac{\partial}{\partial q^\mu} \int d^8x\,d^8y\, \delta^8_{yx}(q)\, (\gamma^\mu)^{\dot a b} \theta_b\, \Big\langle \Big(\frac{\bar D_{\dot a} D^2}{16\partial^2} \phi_n\Big)_y\, \Big(\bm{g} \phi_j \frac{D^2}{4\partial^2} \phi_m\Big)_x \Big\rangle\bigg|_{\mbox{\scriptsize fields = 0}}.
\end{equation}

\noindent
Note that here the superfield $\bm{g}$ is taken in the point $x$, while originally it was taken in the point $y$. This is possible, because $\delta^8_{yx}(q)$ includes $\delta^4(\theta_x-\theta_y)$, while the momentum of $\bm{g}$ (of the order $1/R$) is negligibly small.

\subsection{Result for the total derivatives}
\hspace*{\parindent}

Summing up Eqs. (\ref{WZ_Total_Derivative_Result}) and (\ref{Yukawa_Total_Derivative_Result}) and taking into account that the term containing $(\bar X_{\dot a}^A)_{\mbox{\scriptsize HD}}$ does not produce singular contributions, we can present the expression (\ref{We_Variate}) as the integral of a total derivative,

\begin{eqnarray}
&& \frac{\partial}{\partial a_\mu^A}  \delta_a \int d^8x\, (\gamma_\mu)^{\dot a b} \theta_b \Big\langle (\bar X_{\dot a}^A)_{\mbox{\scriptsize WZ}} + (\bar X_{\dot a}^A)_{\mbox{\scriptsize HD}} + (\bar X_{\dot a}^A)_{\mbox{\scriptsize Yukawa}} \Big\rangle\bigg|_{\mbox{\scriptsize fields = 0}}\nonumber\\
&& = \int \frac{d^4q}{(2\pi)^4} \frac{\partial}{\partial q^\mu} \int d^8x\, d^8y\, (\gamma^\mu)^{\dot a b} \theta_b\, \delta^8_{yx}(q)\, \Big\langle \bigg\{\Big[F\Big(-\frac{\nabla^2 \bar\nabla^2}{16\Lambda^2}\Big) \phi^{+}\cdot  e^{2{\cal F}(V)} e^{2\bm{V}}\Big]^k\nonumber\\
&&\qquad\qquad\qquad\qquad\qquad\quad + \bm{g} \lambda_0^{kmn} \phi_m\, \frac{D^2}{4\partial^2} \phi_n \bigg\}_x C(R)_k{}^j  \Big(\frac{\bar D_{\dot a} D^2}{16\partial^2} \phi_j\Big)_y \Big\rangle\bigg|_{\mbox{\scriptsize fields = 0}}.\qquad\quad
\end{eqnarray}

\noindent
It can be equivalently rewritten in terms of the derivative of $S_{\phi_0}$ with respect to $\phi_{0i}$,

\begin{eqnarray}
&& \int \frac{d^4q}{(2\pi)^4} \frac{\partial}{\partial q^\mu} \int d^8x\, d^8y\, \delta^8_{yx}(q)  (\gamma^\mu)^{\dot a b} \theta_b\, \Big\langle
\frac{\delta S_{\phi_0}}{\delta \phi_{0i,x}} C(R)_i{}^j  \Big(\frac{\bar D_{\dot a} D^2}{4\partial^2} \phi_j\Big)_y \Big\rangle\bigg|_{\mbox{\scriptsize fields = 0}}.\qquad
\end{eqnarray}

\end{document}